\pdfoutput=1
\documentclass[cernpreprint, atlasdraft=true, texlive=2016, UKenglish, texmf]{atlasdoc}
 
\usepackage[subcaption]{atlaspackage}
\usepackage{atlasbiblatex}
 
\usepackage{atlascomment}
 
\usepackage{atlasphysics}
 
\graphicspath{{logos/}{figures_final/}}
 
\usepackage{ANA-HIGG-2020-23-PAPER-defs}
 
\usepackage{PMGRefs-defs}
 
\usepackage{multirow}
 
\AtlasTitle{Measurement of Higgs boson decay into $b$-quarks in associated production with a top-quark pair in $pp$ collisions at $\sqrt{s}=13$~TeV with the ATLAS detector}

\AtlasAbstract{
The associated production of a Higgs boson and a top-quark pair is measured in events characterised by the presence of one or two electrons or muons. The Higgs boson decay into a $b$-quark pair is used. The analysed data, corresponding to an integrated luminosity of 139~$\mathrm{fb^{-1}}$, were collected in proton--proton collisions at the Large Hadron Collider between 2015 and 2018 at a centre-of-mass energy of $\sqrt{s}=13$~TeV. The measured signal strength, defined as the ratio of the measured signal yield to that predicted by the Standard Model, is $0.35^{+0.36}_{-0.34}$. This result is compatible with the Standard Model prediction and corresponds to an observed (expected) significance of 1.0 (2.7) standard deviations. The signal strength is also measured differentially in bins of the Higgs boson transverse momentum in the simplified template cross-section framework, including a bin for specially selected boosted Higgs bosons with transverse momentum above 300~GeV.
 
}

\AtlasRefCode{HIGG-2020-23}
 
\PreprintIdNumber{CERN-EP-2021-202}

\arXivId{2111.06712}
 
\HepDataRecord{114360}
 
\AtlasJournal{JHEP}
\AtlasJournalRef{\JHEP 06 (2022) 097}
\AtlasDOI{10.1007/JHEP06(2022)097}

\hypersetup{pdftitle={ATLAS document},pdfauthor={The ATLAS Collaboration}}
 
\begin{document}
 
\maketitle

\section{Introduction}
The Higgs boson~\cite{Englert:1964et,Higgs:1964pj,Guralnik:1964eu} was
discovered by the ATLAS and CMS
collaborations~\cite{HIGG-2012-27,CMS-HIG-12-028} in 2012, with a mass
of around 125~\GeV~\cite{HIGG-2014-14}. Since then, the analysis
of proton--proton ($pp$) collision data at centre-of-mass energies of
7~\TeV, 8~\TeV\ and 13~\TeV\ delivered by the Large Hadron Collider
(LHC)~\cite{Evans:2008zzb} has led to more detailed measurements of
its properties and couplings, testing the predictions of the Standard
Model (SM). Of particular interest is the Yukawa coupling to the top
quark, the heaviest elementary particle in the SM, which could be very sensitive
to effects of physics beyond the SM (BSM)~\cite{Englert:2014uua}.
 
Both the ATLAS and CMS collaborations have observed the
interaction of the Higgs boson with third-generation fermions of the
SM. The coupling to $\tau$-leptons was measured in the observation of
\htautau decays~\cite{HIGG-2015-07,HIGG-2017-07,CMS-HIG-16-043}, while the
observation of the decay of the Higgs boson into $b$-quark pairs
provided direct evidence for the Yukawa coupling to down-type
quarks~\cite{HIGG-2018-04,CMS-HIG-18-016}. The interaction of the
Higgs boson with the top quark was indirectly constrained (assuming no BSM phenomena in the loops) from measurements of gluon--gluon fusion Higgs production and decay to $\gamma\gamma$~\cite{HIGG-2015-07}, before being directly measured in the observation of
Higgs boson production in association with a pair of top quarks
(\ttH)~\cite{HIGG-2018-13,CMS-HIG-17-035}.
 
The \ttH production mode, in which the top quark couples to the Higgs boson at tree level,
is the most favourable to extract direct information on the top-quark Yukawa coupling without assumptions about
the potential presence of BSM
physics~\cite{Ng:1983jm,Kunszt:1984ri,Dawson:2002tg,Beenakker:2001rj}.
 
\begin{figure}[tp]
\centering
\captionsetup[subfigure]{belowskip=5pt}
\begin{subfigure}[]{0.31\linewidth}
\includegraphics[width=\linewidth]{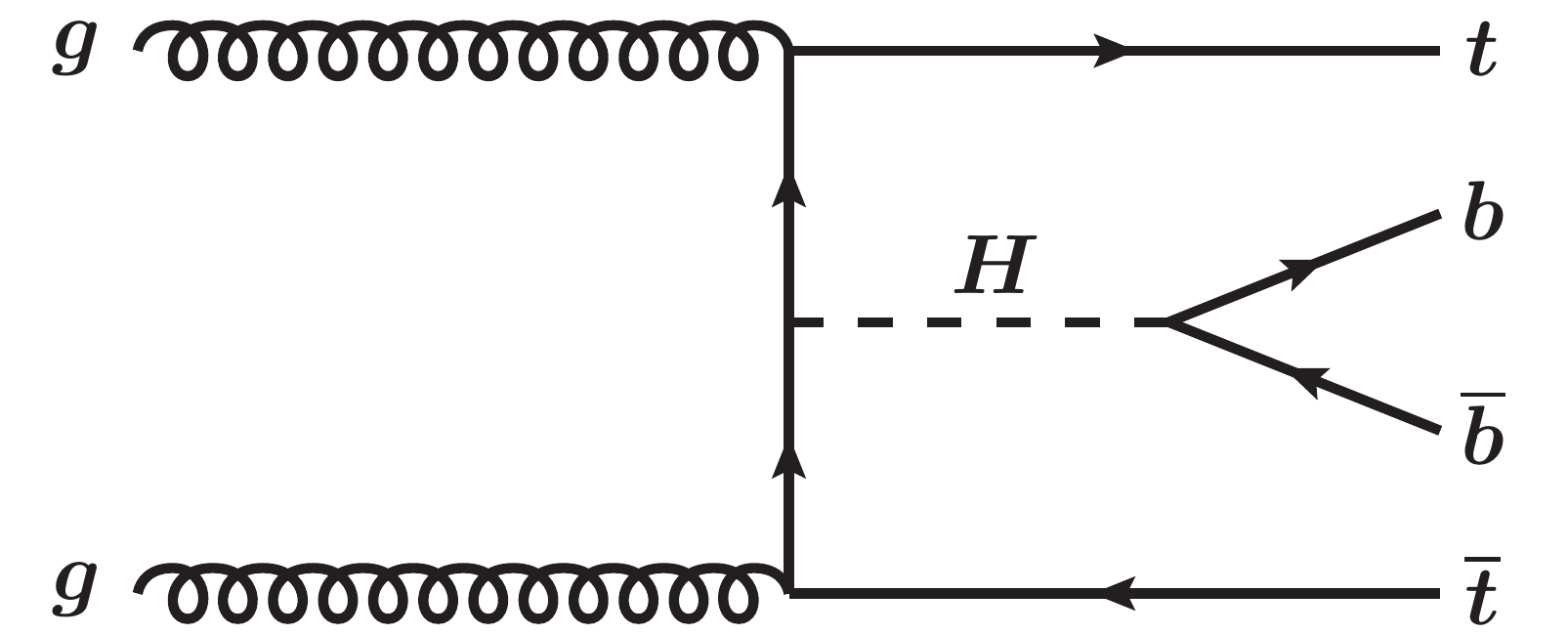}
\caption{\ttHbb \tchan}
\label{fig:intro:feyman_ttH1}
\end{subfigure}
\hspace{1cm}
\begin{subfigure}[]{0.31\linewidth}
\includegraphics[width=\linewidth]{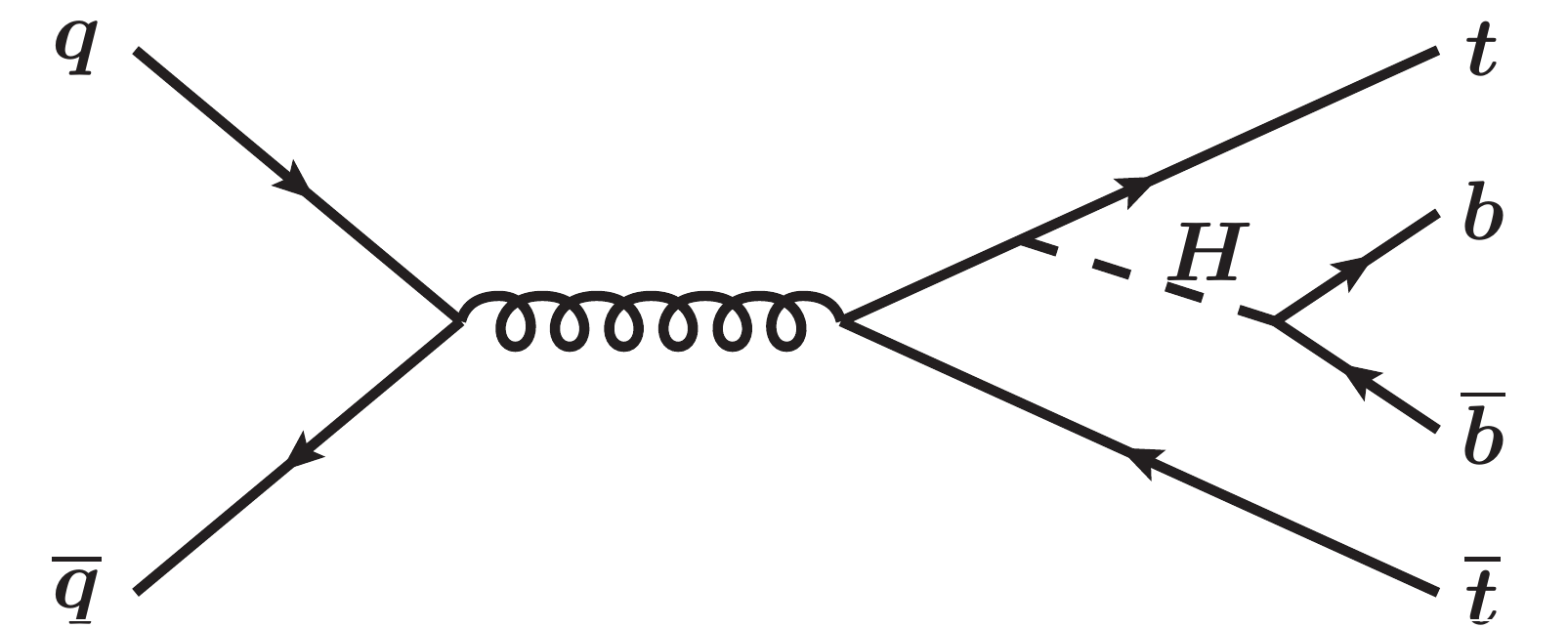}
\caption{\ttHbb \schan}
\label{fig:intro:feyman_ttH2}
\end{subfigure}
\hspace{1cm}
\begin{subfigure}[]{0.31\linewidth}
\includegraphics[width=\linewidth]{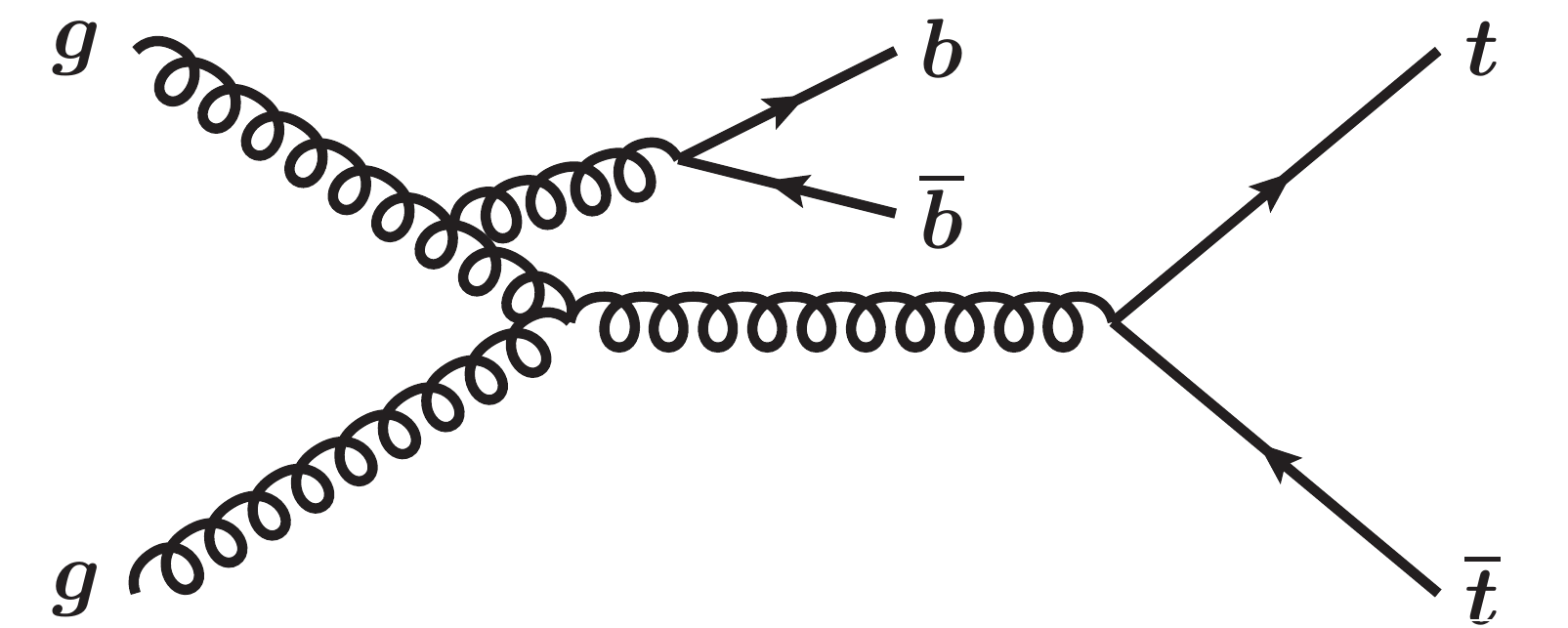}
\caption{\ttbb}
\label{fig:intro:feyman_ttbb}
\end{subfigure}
\caption{
Representative tree-level Feynman diagrams for the production of a Higgs boson in association with a top-quark pair (\ttH) in \subref{fig:intro:feyman_ttH1} the t-channel and \subref{fig:intro:feyman_ttH2} the s-channel and the subsequent decay of the Higgs boson into \bbbar, and for \subref{fig:intro:feyman_ttbb} the main \ttbb background.
\label{fig:intro:feynman}
}
\end{figure}

Although this production mode contributes only around 1\% of the total
Higgs boson production cross-section at the LHC~\cite{deFlorian:2016spz}, the top
quarks in the final state offer a distinctive signature and allow
access to many Higgs boson decay modes. The decay into two $b$-quarks (\hbb) is
the most probable, with a SM branching fraction of about
58\%~\cite{deFlorian:2016spz}. Furthermore, in the \hbb\
decay mode the reconstruction of the Higgs boson kinematics is
possible, which allows the extraction of additional information about the
structure of the top--Higgs interaction~\cite{Ellis:2013yxa,Boudjema:2015nda,Buckley:2016prl,AmorDosSantos:2017ayi}. This analysis therefore
aims to select events with a Higgs boson produced in association with a pair of top quarks and decaying into a pair of $b$-quarks $\left(\ttHbb\right)$, in which one or both top quarks decay
semileptonically, producing an electron or a muon, collectively referred to as leptons ($\ell$).\footnote{
Electrons and muons from the decay of a
$\tau$-lepton itself originating from a $W$ boson are
included.}
With many final-state particles, the main challenges are
the low efficiency to reconstruct and identify all of them, the large
combinatorial ambiguities when trying to match the observed objects to the decay products
of the Higgs boson and top quarks,
and the large background of \ttjets\ processes (in particular when these jets originate from $b$- or $c$-quarks), which have a much larger production cross-section than the signal.
Representative Feynman diagrams for the \ttHbb signal and dominant \ttbb background are shown in Figure~\ref{fig:intro:feynman}.
 
The ATLAS Collaboration searched for \ttHbb production at
$\sqs = 8$~\TeV\ in final states with at least
one~\cite{HIGG-2013-27} or no lepton~\cite{HIGG-2015-05}, and at
$\sqs = 13$~\TeV\ with at least
one lepton in the final state with data collected in 2015 and 2016,
corresponding to an integrated luminosity of
36.1~\ifb~\cite{HIGG-2017-03}. A combined signal strength (defined as the ratio of the measured signal yield to that predicted by the Standard Model) of
$0.84^{+0.64}_{-0.61}$ was measured, with an observed (expected)
significance of 1.4 (1.6) standard deviations.  This result was
combined with analyses of Higgs boson decays into massive vector bosons,
$\tau$-leptons, or photons to reach observation of the \ttH production
mode~\cite{HIGG-2018-13}. The CMS Collaboration searched for the same
processes using 35.9~\ifb\ of data collected at $\sqs=13$~\TeV\ in 2016, in events with at least one
lepton~\cite{CMS-HIG-17-026} or no lepton~\cite{CMS-HIG-17-022}, and
measured a signal strength of $0.72 \pm 0.45$ and $0.9 \pm 1.5$, respectively.
These results also contributed to the observation of the \ttH production
mode~\cite{CMS-HIG-17-035}.
 
In this paper, a measurement of the \ttH production cross-section in
the \hbb channel is performed using the LHC Run~2 $pp$ collision
data collected by the ATLAS detector, corresponding to an integrated
luminosity of 139~\ifb\ at $\sqs=13$~\TeV. 
Events with either one lepton or two leptons are analysed separately in exclusive
single-lepton or dilepton categories depending on the number of leptons, the number of jets and
the number of jets identified as originating from $b$-hadrons ($b$-jets).
In the single-lepton channel (but not in the dilepton channel because the expected number of events is small),
a specific category, referred to as `boosted' in the following, is designed to select events in which
the Higgs boson is produced with high transverse momentum (\pt).
The non-boosted categories, where the Higgs boson decay products are less collimated, are referred to as `resolved'.
Machine-learning algorithms are used to classify events into signal-rich
categories, which are analysed together with the signal-depleted ones
in a combined profile likelihood fit. The output distributions of these multivariate
algorithms are used as the main discriminant to extract the signal. This signal extraction fit simultaneously determines the event yields for the signal and for the most important background
components, while constraining the overall background model within the
assigned systematic uncertainties.  In addition, making use of the
possibility to reconstruct the Higgs boson kinematics in the \hbb\
channel, the cross-section is measured as a function of the Higgs
boson `truth' transverse momentum \ptHhat, as obtained from Monte Carlo simulation, in the simplified template
cross-sections (STXS) formalism~\cite{deFlorian:2016spz}, which aims
to reduce the theory dependencies that are folded into the measurements.

Several aspects of the analysis have been improved relative to the previous version~\cite{HIGG-2017-03}.
The full Run~2 dataset is analysed, with improved reconstruction algorithms and detector calibrations, which in turn enhance the performance
of the $b$-tagging algorithm. This allows the use of tighter selection criteria to reject events in poorly modelled regions of phase space,
and to define analysis regions differential in Higgs boson \pT.
An enhanced model is adopted for the
\ttjets\ background, with updated event generator versions, a higher-order precision prediction for the \ttbb\ process,
an increased number of simulated events, and an improved assessment of uncertainties.
Finally, a new deep neural network improves the performance of the boosted
channel.
 
This paper is organised as follows. The ATLAS detector is described
in Section~\ref{sec:Detector}. The signal and background modelling is
presented in Section~\ref{sec:Modelling}. Section~\ref{sec:Selection}
summarises the selection criteria for reconstructed objects and
events, while Section~\ref{sec:Regions} describes the analysis
strategy. Systematic uncertainties are discussed in
Section~\ref{sec:Systematics}. Results are presented in
Section~\ref{sec:Results}, and the conclusions in
Section~\ref{sec:Conclusion}.


\section{ATLAS detector}
\label{sec:Detector}
\newcommand{\AtlasCoordFootnote}{
ATLAS uses a right-handed coordinate system with its origin at the nominal interaction point (IP)
in the centre of the detector and the \(z\)-axis along the beam pipe.
The \(x\)-axis points from the IP to the centre of the LHC ring,
and the \(y\)-axis points upwards.
Cylindrical coordinates \((r,\phi)\) are used in the transverse plane,
\(\phi\) being the azimuthal angle around the \(z\)-axis.
The rapidity is defined as
$y=(1/2) \ln[(E+p_z)/(E-p_z)]$ where $E$
is the energy and $p_z$ is the longitudinal component of the
momentum along the beam pipe.
The pseudorapidity is defined in terms of the polar angle \(\theta\) as \(\eta = -\ln \tan(\theta/2)\).
Angular distance is measured in units of \(\Delta R \equiv \sqrt{(\Delta\eta)^{2} + (\Delta\phi)^{2}}\).}
 
The ATLAS experiment~\cite{PERF-2007-01,ATLAS-TDR-19,PIX-2018-001} at the LHC is a multipurpose particle detector
with a forward--backward symmetric cylindrical geometry and a near \(4\pi\) coverage in
solid angle.\footnote{\AtlasCoordFootnote}
It consists of an inner tracking detector surrounded by a thin superconducting solenoid
providing a \SI{2}{\tesla} axial magnetic field, electromagnetic and hadron calorimeters, and a muon spectrometer.
The inner tracking detector (ID) covers the pseudorapidity range \(|\eta| < 2.5\).
It consists of silicon pixel, silicon microstrip, and transition radiation tracking detectors.
Lead/liquid-argon (LAr) sampling calorimeters provide electromagnetic (EM) energy measurements
with high granularity.
A steel/scintillator-tile hadron calorimeter covers the central pseudorapidity range \(|\eta| < 1.7\).
The endcap and forward regions are instrumented with LAr calorimeters
for both the EM and hadronic energy measurements up to \(|\eta| = 4.9\).
The muon spectrometer surrounds the calorimeters and is based on
three large air-core toroidal superconducting magnets with eight coils each.
The field integral of the toroids ranges between \num{2} and \SI{6}{\tesla\metre}
across most of the detector.
The muon spectrometer includes a system of precision chambers for tracking and fast detectors for triggering.
A two-level trigger system is used to select events.
The first-level trigger is implemented in hardware and uses a subset of the detector information
to accept events at a rate below \SI{100}{\kHz}.
This is followed by a software-based trigger that
reduces the accepted event rate to \SI{1}{\kHz} on average
depending on the data-taking conditions~\cite{TRIG-2016-01}.
An extensive software suite~\cite{ATL-SOFT-PUB-2021-001} is used in the reconstruction and analysis of real and simulated data, in detector operations, and in the trigger and data acquisition systems of the experiment.
 
 
\section{Signal and background modelling}
\label{sec:Modelling}
Simulated event samples are used to model the \ttH\ signal and the background processes.
The numbers of selected events and the distribution shapes of variables used to discriminate between signal and background enter into the signal extraction fit, which also constrains the modelling of the background processes.
The Monte Carlo (MC) samples were produced using either the full ATLAS detector simulation~\cite{SOFT-2010-01} based on \GEANT~\cite{Agostinelli:2002hh} or a faster simulation where the full \GEANT simulation of the calorimeter response is replaced by a detailed parameterisation of the shower shapes~\cite{SOFT-2010-01}. For the observables used in the analysis, the two simulations were found to give similar modelling. To simulate the effects of multiple interactions in the same and neighbouring bunch crossings (pile-up), additional interactions were generated using \PYTHIA[8.186]~\cite{Sjostrand:2014zea} with a set of tuned parameters called the A3 tune~\cite{ATL-PHYS-PUB-2016-017} and overlaid onto the simulated hard-scatter event. Simulated events are reweighted to match the pile-up conditions observed in the full Run~2 dataset, with a mean number of $pp$ interactions per bunch crossing of 34. All simulated events are processed through the same reconstruction algorithms and analysis chain as the data.
 
In all samples where the parton shower (PS), hadronisation, and multi-parton interactions (MPI) were generated with either
\PYTHIA[8] or \HERWIG[7]~\cite{Bahr:2008pv,Bellm:2015jjp}, the decays of $b$- and $c$-hadrons were simulated using the \EVTGEN[1.6.0] program~\cite{Lange:2001uf}. The $b$-quark mass was set to $m_b=4.80$~\GeV\ ($4.50$~\GeV) for samples using \PYTHIA[8] (\HERWIG[7]).
For \PYTHIA[8], the A14 tune~\cite{ATL-PHYS-PUB-2014-021} and the \NNPDF[2.3lo] parton distribution function (PDF) set~\cite{Ball:2012cx} were used.
For \HERWIG[7], the H7UE tune~\cite{Bellm:2015jjp} was used with the \MMHT[lo] PDF set \cite{Harland-Lang:2014zoa}. The Higgs boson mass was set to $m_H=125.0$~\GeV, and the top-quark mass to $m_t$ = 172.5\,\GeV.
The precision of the matrix element (ME) generators is next-to-leading order (NLO) in quantum chromodynamics (QCD) for most samples. Some samples are normalised to higher precision in QCD (next-to-next-to-leading order, NNLO, or next-to-next-to-leading logarithm, NNLL) or with electroweak (EW) corrections.
A summary of all generated samples is presented in Table~\ref{tab:modelling:samples2}, which includes both the samples used for nominal predictions and other samples used to assess systematic uncertainties. Further details are provided in the following subsections.

\begin{table}[tp]
\renewcommand{\arraystretch}{1.2}
\centering
\footnotesize
\begin{tabular}{ccccc}
\toprule \midrule
Process & ME generator 		& ME PDF 					&	 PS 			& Normalisation   		\\ \midrule \midrule
\multicolumn{5}{l}{Higgs boson} \\ 
\midrule \midrule
\ttH 	& \POWHEGBOX[v2] 	& \NNPDF[3.0nlo] 				& \PYTHIA[8.230]		& NLO+NLO (EW)
~\cite{deFlorian:2016spz}					\\
\tabsyst & \POWHEGBOX[v2] 	& \NNPDF[3.0nlo] 				& \HERWIG[7.04]	& NLO+NLO (EW)
~\cite{deFlorian:2016spz}					\\
\tabsyst & \MGNLO[2.6.0] 	& \NNPDF[3.0nlo] 				& \PYTHIA[8.230]		& NLO+NLO (EW)
~\cite{deFlorian:2016spz}					\\
\tHjb 	& \MGNLO[2.6.2] 	& \NNPDF[3.0nlo]nf4		& \PYTHIA[8.230]		& --						\\
\tWH 	& \MGNLO[2.6.2][DR]	& \NNPDF[3.0nlo] 				& \PYTHIA[8.235]		& --						\\ \midrule \midrule
\multicolumn{5}{l}{\ttjets\ and single-top} \\ \midrule \midrule
\ttbar	& \POWHEGBOX[v2] 	& \NNPDF[3.0nlo] 				& \PYTHIA[8.230]		& NNLO+NNLL
\cite{Beneke:2011mq,Cacciari:2011hy,Baernreuther:2012ws,Czakon:2012zr,Czakon:2012pz,Czakon:2013goa,Czakon:2011xx}			\\
\tabsyst & \POWHEGBOX[v2] 	& \NNPDF[3.0nlo] 				& \HERWIG[7.04]	& NNLO+NNLL
\cite{Beneke:2011mq,Cacciari:2011hy,Baernreuther:2012ws,Czakon:2012zr,Czakon:2012pz,Czakon:2013goa,Czakon:2011xx}		   	\\
\tabsyst & \MGNLO[2.6.0] 	& \NNPDF[3.0nlo] 				& \PYTHIA[8.230]		& NNLO+NNLL
\cite{Beneke:2011mq,Cacciari:2011hy,Baernreuther:2012ws,Czakon:2012zr,Czakon:2012pz,Czakon:2013goa,Czakon:2011xx}			\\
\ttbb 	& \POWHEGBOXRES 	& \NNPDF[3.0nlo]nf4 	& \PYTHIA[8.230]		& --						\\
\tabsyst & \SHERPA[2.2.1] 	& \NNPDF[3.0nnlo]nf4 	& \SHERPA				& --						\\
\tW 	& \POWHEGBOX[v2][DR]	& \NNPDF[3.0nlo] 			& \PYTHIA[8.230]		& NLO+NNLL
~\cite{Kidonakis:2010ux,Kidonakis:2013zqa} 				   	\\
\tabsyst & \POWHEGBOX[v2][DS]	& \NNPDF[3.0nlo] 			& \PYTHIA[8.230]		& NLO+NNLL
~\cite{Kidonakis:2010ux,Kidonakis:2013zqa} 						\\
\tabsyst & \POWHEGBOX[v2][DR]	& \NNPDF[3.0nlo] 			& \HERWIG[7.04]	& NLO+NNLL
~\cite{Kidonakis:2010ux,Kidonakis:2013zqa} 				   		\\
\tabsyst & \MGNLO[2.6.2][DR]	& \CT[10nlo] 		& \PYTHIA[8.230]		& NLO+NNLL
~\cite{Kidonakis:2010ux,Kidonakis:2013zqa} 					   	\\
\tchan 	& \POWHEGBOX[v2] 	& \NNPDF[3.0nlo]nf4		& \PYTHIA[8.230]		& NLO
~\cite{Aliev:2010zk,Kant:2014oha} 						\\
\tabsyst & \POWHEGBOX[v2] 	& \NNPDF[3.0nlo]nf4 	& \HERWIG[7.04]	& NLO
~\cite{Aliev:2010zk,Kant:2014oha} 						\\
\tabsyst & \MGNLO[2.6.2] 	& \NNPDF[3.0nlo]nf4		& \PYTHIA[8.230]		& NLO
~\cite{Aliev:2010zk,Kant:2014oha} 						\\
\schan 	& \POWHEGBOX[v2] 	& \NNPDF[3.0nlo] 				& \PYTHIA[8.230]		& NLO
~\cite{Aliev:2010zk,Kant:2014oha} 						\\
\tabsyst & \POWHEGBOX[v2] 	& \NNPDF[3.0nlo] 				& \HERWIG[7.04]	& NLO
~\cite{Aliev:2010zk,Kant:2014oha} 						\\
\tabsyst & \MGNLO[2.6.2] 	& \NNPDF[3.0nlo] 				& \PYTHIA[8.230]		& NLO
~\cite{Aliev:2010zk,Kant:2014oha}  					\\ \midrule \midrule
\multicolumn{5}{l}{Other} \\ \midrule \midrule
\Wjets 	& \SHERPA[2.2.1] (NLO [2j], LO [4j]) 	& \NNPDF[3.0nnlo] 	& \SHERPA		& NNLO
~\cite{Anastasiou:2003ds}						\\
\Zjets 	& \SHERPA[2.2.1] (NLO [2j], LO [4j])	& \NNPDF[3.0nnlo]	& \SHERPA		& NNLO
~\cite{Anastasiou:2003ds}						\\
$VV$ (had.)  		& \SHERPA[2.2.1] 				& \NNPDF[3.0nnlo] 	& \SHERPA		& --						\\
$VV$ (lep.)  	 	& \SHERPA[2.2.2] 				& \NNPDF[3.0nnlo] 	& \SHERPA		& --						\\
$VV$ (lep.) + jj	& \SHERPA[2.2.2] (LO [EW])		& \NNPDF[3.0nnlo] 	& \SHERPA		& --						\\
\ttW 	& \MGNLO[2.3.3] 	& \NNPDF[3.0nlo] 				& \PYTHIA[8.210]		& NLO+NLO (EW)
~\cite{deFlorian:2016spz}						\\
\tabsyst & \SHERPA[2.0.0] (LO [2j]) 	& \NNPDF[3.0nnlo] 	& \SHERPA					& NLO+NLO (EW)
~\cite{deFlorian:2016spz}						\\
\ttbar{}$\ell\ell$ 	& \MGNLO[2.3.3] 	& \NNPDF[3.0nlo] 	& \PYTHIA[8.210]		& NLO+NLO (EW)
~\cite{deFlorian:2016spz}						\\
\tabsyst & \SHERPA[2.0.0] (LO [1j]) 	& \NNPDF[3.0nnlo] 	& \SHERPA					& NLO+NLO (EW)
~\cite{deFlorian:2016spz}						\\
\ttZ($qq,\nu\nu$) 	& \MGNLO[2.3.3] 	& \NNPDF[3.0nlo] 	& \PYTHIA[8.210]		& NLO+NLO (EW)
~\cite{deFlorian:2016spz}						\\
\tabsyst & \SHERPA[2.0.0] (LO [2j]) 	& \NNPDF[3.0nnlo] 	& \SHERPA					& NLO+NLO (EW)
~\cite{deFlorian:2016spz}						\\
\tttt 	& \MGNLO[2.3.3]		& \NNPDF[3.1nlo]	& \PYTHIA[8.230]		& NLO+NLO (EW)
~\cite{Frederix:2017wme}						\\
\tZq 	& \MGNLO[2.3.3] (LO)	& \CTEQ[6L1]		 			& \PYTHIA[8.212]		& --						\\
\tWZ 	& \MGNLO[2.3.3][DR]	& \NNPDF[3.0nlo] 			& \PYTHIA[8.230]		& --						\\ \midrule \bottomrule
\end{tabular}
\caption{Table summarising the generator set-ups for samples used in this analysis.
The first row for each sample details the nominal settings used for this process in the analysis. Any additional rows describe samples which are used to evaluate the modelling and performance of the analysis. For overlap between \ttbar\ and \tW-like diagrams, samples using the diagram removal scheme~\cite{Frixione:2008yi} are listed as [DR] and samples using the diagram subtraction scheme~\cite{Frixione:2008yi,ATL-PHYS-PUB-2016-020} are listed as [DS].
The precision of the ME generator is NLO in QCD if no additional information is provided in parentheses. The higher-order cross-section used to normalise these samples is listed in the last column and refers to the order of QCD processes if no additional information is provided. If no information is present in this column, no higher-order $K$-factor is applied to this process. For the $VV$ \SHERPA samples, `lep.' (`had.') means that both bosons decay leptonically (one decays leptonically and one hadronically).
}
\label{tab:modelling:samples2}
\end{table}

\subsection{Signal modelling}
 
In \ttH\ events the production and decays were modelled in the five-flavour scheme using the
\POWHEGBOX~\cite{Nason:2004rx,Frixione:2007vw,Alioli:2010xd,Hartanto:2015uka,Frixione:2007nw}
generator at NLO in QCD with the \NNPDF[3.0nlo]~\cite{Ball:2014uwa} PDF set.
The \hdamp\ parameter\footnote{The \hdamp\ parameter
controls the \pt\ of the first additional emission beyond the leading-order Feynman diagram
in the PS and therefore regulates the  high-\pt\ emission against which the \ttH\ system recoils.}
was set to $0.75 \times (m_t + m_{\bar{t}} + m_H)=352.5$~\GeV,
and the functional form of the renormalisation and factorisation scales were both set to $\sqrt[3]{m_\text{T}(t)\cdot m_\text{T}(\bar{t}) \cdot m_\text{T}(H)}$ (where $m_\text{T} = \sqrt{m^2 + \pt^2}$ is the transverse mass of a particle).
The events were showered by \PYTHIA[8] and all Higgs boson decay modes are considered.
The samples are normalised to the fixed-order cross-section calculation, $\sigma_{\ttH}=507^{+35}_{-50}$\,fb, which includes NLO QCD and EW corrections~\cite{deFlorian:2016spz} for a Higgs boson mass of 125~\GeV.

\subsection{\texorpdfstring{$\ttjets$}{ttbar + jets} background}
 
Simulated \ttjets\ events are categorised according to the flavour of additional jets in the event, using the procedure described in Ref.~\cite{HIGG-2013-27}. For this purpose, jets are reconstructed from stable particles (mean lifetime $\tau > 3\times10^{-11}$\,~s) using the \antikt\ algorithm~\cite{Cacciari:2008gp,Fastjet}, and the number of $b$- or $c$-hadrons within $\Delta R = 0.4$ of the jet axis is considered (with $\pt > 5$~\GeV\ for the leading $b$- or $c$-hadron around the jet), excluding particles produced by the top-quark decay. Events are labelled as \ttbin\ if at least one $b$-flavour jet is identified, or else as \ttcin\ if at least one $c$-flavour jet is identified, and otherwise as \ttlight. Where necessary, the \ttbin events are further separated into \ttbexcl (where exactly one jet is matched to at least one $b$-hadron) and \ttbbin (all remaining events).

To accurately model the dominant \ttbin\ background, a sample with \ttbb\ MEs was produced at NLO QCD accuracy in the four-flavour scheme with the \POWHEGBOXRES~\cite{Jezo:2018yaf}
generator and \OPENLOOPS~\cite{Cascioli:2011va,Denner:2016kdg}, using a pre-release of the implementation of this process in \POWHEGBOXRES provided by the authors~\cite{ttbbPowheg}, with the \NNPDF[3.0nlo]nf4~\cite{Ball:2014uwa} PDF set, and using \PYTHIA[8] for the PS and hadronisation.
The factorisation scale was set to $0.5 \times \Sigma_{i=t,\bar{t},b,\bar{b},j}m_{\mathrm{T}}(i)$ (where $j$ stands for extra partons), the renormalisation scale was set to $\sqrt[4]{m_{\text{T}}(t)\cdot m_{\text{T}}(\bar{t})\cdot m_{\text{T}}(b)\cdot m_{\text{T}}(\bar{b})}$, and the \hdamp\ parameter was set to $0.5 \times \Sigma_{i=t,\bar{t},b,\bar{b}}m_{\mathrm{T}}(i)$.
The mass of the two $b$-quarks produced in the ME in association with the two top quarks was set to the same value as the mass of the $b$-quarks from the top-quark decays, $m_b=4.95$~\GeV.
 
Inclusive \ttjets\ events were also generated with \ttbar\ MEs in the five-flavour scheme using the \POWHEGBOX[v2] generator at NLO in QCD, using \PYTHIA[8] for the PS and hadronisation. Here, the \hdamp\ parameter was set to 
1.5~$m_t$~\cite{ATL-PHYS-PUB-2016-020}, and the functional form of the renormalisation and factorisation scales was set to $m_\text{T}(t)$.\footnote{This scale is calculated in the \ttbar\ rest-frame and hence the \pt\ values of the top quark and top antiquark are equivalent.} The \ttcin\ and \ttlight\ events using this prediction are combined with the previously described \ttbin sample to form the nominal \ttbar model, while the \ttbin\ events from this five-flavour scheme are used only to assign modelling uncertainties.

\subsection{Other backgrounds}
 
The QCD \Vjets\ processes (i.e.\ \Wjets\ and \Zjets) were simulated with the \SHERPA[2.2.1] generator~\cite{Bothmann:2019yzt}.
In this set-up, NLO-accurate MEs for up to two partons, and leading-order (LO) accurate MEs for up to four partons were calculated with the Comix~\cite{Gleisberg:2008fv} and \OPENLOOPS libraries.
They were matched with the \SHERPA parton shower~\cite{Schumann:2007mg} by using the \MEPSatNLO prescription~\cite{Hoeche:2011fd,Hoeche:2012yf,Catani:2001cc,Hoeche:2009rj} with the set of tuned parameters developed by the \SHERPA authors and based on the \NNPDF[3.0nnlo] set of PDFs~\cite{Ball:2014uwa}.
A data-driven correction to the normalisation was derived for \Zjets\ predictions containing at least two heavy-flavour jets (where heavy-flavour means jets originating from $b$-hadrons and $c$-hadrons). Events are selected, with objects passing the selection discussed in Section~\ref{sec:Selection}, in dedicated control regions in data defined by requiring at least two jets and two opposite-electric-charge same-flavour leptons ($e^+e^-$ or $\mu^+\mu^-$) with an invariant mass inside the \Zboson-boson mass window 83--99~\GeV. 
A 25\% yield increase is found to improve the modelling of \Zjets\ events with at least two heavy-flavour jets, consistent with Ref.~\cite{STDM-2017-38}.
 
Diboson events were simulated with the \SHERPA[2.2.1] and~2.2.2 generators.
In this set-up multiple MEs were matched and merged with the \SHERPA PS based on Catani--Seymour dipole factorisation~\cite{Gleisberg:2008fv,Schumann:2007mg} using the \MEPSatNLO\ prescription.
For semileptonically and fully leptonically decaying diboson event samples, as well as
loop-induced diboson event samples, the virtual QCD correction for MEs at NLO accuracy was
provided by the \OPENLOOPS library.
For EW $VVjj$ production, the calculation was performed in the $G_\mu$ scheme, ensuring an optimal description
of pure EW interactions at the EW scale~\cite{Denner:2005fg,Dittmaier:2009cr,Andersen:2014efa}.
All samples were generated using the \NNPDF[3.0nnlo]\ PDF set, along with the dedicated set of tuned PS parameters developed by the \SHERPA authors.

The associated production of a single top quark and a Higgs boson is treated as background.
Samples for two subprocesses, \tHjb\ and \tWH\ production, were generated using the \MGNLO[2.6.2] generator at NLO in QCD.
The functional form of the renormalisation and factorisation scale was set to $0.5 \times \sum_i m_\text{T}(i)$, where the sum runs over all the particles generated from the ME calculation.
For \tHjb (\tWH), events were generated in the four-flavour (five-flavour) scheme using the \NNPDF[3.0nlo]nf4 (\NNPDF[3.0nlo]) PDF set,
and the diagram removal scheme was employed to handle the interference with \ttH\ in the \tWH\ sample~\cite{Frixione:2008yi,Demartin:2016axk}.
 
Single-top \tchan, \schan, and \tW production were modelled using the \POWHEGBOX[v2]~\cite{Frederix:2012dh,Alioli:2009je,Re:2010bp} generator at NLO in QCD.
For \tchan production, events were generated in the four-flavour scheme with the \NNPDF[3.0nlo]nf4 PDF set, and the functional form of the renormalisation and factorisation scales was set to $m_\text{T}(b)$ following the recommendation of Ref.~\cite{Frederix:2012dh}.
For \schan and $tW$ production, events were generated in the five-flavour scheme with the \NNPDF[3.0nlo]\ PDF set, and the functional form of the renormalisation and factorisation scales was set to the top-quark mass.
For \tW production, the diagram removal scheme~\cite{Frixione:2008yi} was employed to handle the interference with \ttbar\ production~\cite{ATL-PHYS-PUB-2016-020}. An additional sample which applies the diagram subtraction scheme~\cite{Frixione:2008yi,ATL-PHYS-PUB-2016-020} was used to assess an uncertainty in the modelling of this interference.
 
The production and decay of a top-quark pair in association with a vector boson~(i.e. \ttW\ or \ttZ), referred to collectively as \ttV, was modelled using the \MGNLO[2.3.3] generator at NLO in QCD.
The functional form of the renormalisation and factorisation scale was set to $0.5 \times \sum_i m_\text{T}(i)$, where the sum runs over all the particles generated from the ME calculation.
 
For events with four top quarks (\tttt) the production and decays were modelled using the \MGNLO[2.3.3]
generator at NLO in QCD with the \NNPDF[3.1nlo]~\cite{Ball:2014uwa} PDF set.
The functional form of the renormalisation and factorisation scales was set to $0.25 \times \sum_i m_\text{T}(i)$, where the sum runs over all the particles generated from the ME calculation, following the recommendation of Ref.~\cite{Frederix:2017wme}.
 
The \tZq events were generated using the \MGNLO[2.3.3] generator in the four-flavour scheme at LO in QCD, with the \CTEQ[6L1]~\cite{Pumplin:2002vw} PDF set.
Following the recommendations of Ref.~\cite{Frederix:2012dh}, the renormalisation and factorisation scales were set to $4 \times m_\text{T}(b)$, where the $b$-quark is the one coming from the gluon splitting.
 
The \tWZ sample was produced using the \MGNLO[2.3.3] generator at NLO in QCD with the \NNPDF[3.0nlo] PDF set. The renormalisation and factorisation scales were set to the top-quark mass.
The diagram removal scheme was employed to handle the interference between \tWZ and \ttZ.
 
In the single-lepton channel, a negligible contribution is found from events with a jet or a photon misidentified as a lepton, or from events with a non-prompt lepton. In the dilepton channel, the contribution from events with non-prompt leptons is estimated from simulated events (from \ttbar, \ttZ, \ttW, diboson, \Zjets, \Wjets, \tW, \schan and \tchan single top production) where at least one of the reconstructed leptons does not match a true lepton in the event record. These events are collectively referred to as fakes.


\section{Object and event selection}
\label{sec:Selection}
Events are selected from $pp$ collisions at $\sqrt{s}=13$~\TeV\ recorded
by the ATLAS detector between 2015 and 2018, corresponding to an
integrated luminosity of 139~\ifb. Only events
for which the LHC beams were in stable-collision mode and all relevant subsystems were operational are
considered~\cite{DAPR-2018-01}. Events are required to have at least one primary vertex with two
or more tracks with $\pt > 0.5$~\GeV. If more than
one vertex is found, the hard-scattering primary vertex is selected as the
one with the highest sum of squared transverse momenta of associated
tracks~\cite{PERF-2015-01}.
 
Events were recorded using single-lepton triggers with either a low
\pt\ threshold and a lepton isolation requirement, or a higher
threshold but a looser identification criterion and without any
isolation requirement.  The lowest \pt\ threshold in the single-muon trigger
was 20 (26)~\GeV~\cite{TRIG-2018-01} for
data taken in 2015 (2016--2018), while in the single-electron trigger it was
24 (26)~\GeV~\cite{TRIG-2018-05}).
 
Electrons are reconstructed from tracks in the ID associated with
topological clusters of energy depositions in the
calorimeter~\cite{EGAM-2018-01} and are required to have
$\pt>10$~\GeV\ and $|\eta|<2.47$.  Candidates in the calorimeter
barrel--endcap transition region ($1.37 <|\eta| < 1.52$) are
excluded. Electrons must satisfy the \textit{Medium} likelihood
identification criterion. Muon candidates are
identified by matching ID tracks to full tracks or track segments
reconstructed in the muon spectrometer, using the \textit{Loose}
identification criterion~\cite{MUON-2018-03}. Muons are required to
have $\pt>10$~\GeV\ and $|\eta|<2.5$. Lepton tracks must match the
primary vertex of the event, i.e.\ they have to satisfy $|z_0\sin(\theta)|<0.5$~mm and
$|d_0/\sigma(d_0)| < 5\,(3)$ for electrons (muons), where 
$z_0$ is the longitudinal impact parameter relative to the primary vertex
and $d_0$ (with uncertainty $\sigma(d_0)$) is the transverse impact parameter
relative to the beam line.
 
Jets are reconstructed from noise-suppressed topological clusters of
calorimeter energy depositions~\cite{PERF-2014-07} calibrated at the
electromagnetic scale~\cite{JETM-2018-05}, using the \antikt\
algorithm with a radius parameter of
0.4. These are referred to as small-$R$ jets. The average energy contribution from pile-up is subtracted
according to the jet area and jets are calibrated as described in
Ref.~\cite{JETM-2018-05} with a series of simulation-based corrections and in situ techniques. Jets are required to satisfy $\pt > 25$~\GeV\
and $|\eta|<2.5$.
The effect of pile-up is reduced by an algorithm requiring that the
calorimeter-based jets are consistent with originating from the primary
vertex using tracking information~\cite{PERF-2014-03}.
 
Jets containing $b$-hadrons are identified
($b$-tagged) with the MV2c10 multivariate
algorithm~\cite{FTAG-2018-01}, which combines information about the
transverse impact parameters of displaced tracks and the topological properties
of secondary and tertiary decay vertices reconstructed within the
jet. Four working points, defined by different thresholds for the
MV2c10 discriminant, are used in this analysis, corresponding to
efficiencies 
ranging from 85\% to 60\% for $b$-jets with $\pt>20$~\GeV\
as determined in simulated \ttbar events. The corresponding rejection rates are in
the range 2--22 for $c$-jets (containing $c$-hadrons and no $b$-hadrons) and 27--1150 for light-jets
. Jets are then assigned a pseudo-continuous $b$-tagging value~\cite{FTAG-2018-01} according to the tightest working point they satisfy. Correction factors are applied to the simulated events to
compensate for differences between data and simulation in the
$b$-tagging efficiency for $b$-, $c$-, and
light-jets~\cite{FTAG-2018-01,ATLAS-CONF-2018-001,ATLAS-CONF-2018-006}.
 
Hadronically decaying $\tau$-leptons (\tauhad) are distinguished from
jets using their track multiplicity and a multivariate discriminant
based on calorimetric shower shapes and tracking
information~\cite{ATLAS-CONF-2017-029}.  They are required to have
$\pt>25$~\GeV\ and $|\eta| < 2.5$, and to pass the \textit{Medium}
$\tau$-identification working point.
 
An overlap removal procedure is applied to prevent double-counting of
objects. The closest jet within
$\Delta R_{y}=\sqrt{(\Delta y)^2+(\Delta \phi)^2}=0.2$ of a selected
electron is removed. 
If the nearest jet surviving that selection is within
$\Delta R_{y} = 0.4$ of the electron, the electron is discarded.
Muons are usually removed if they are separated from the nearest jet by
$\Delta R_{y} < 0.4$, since this reduces the background from heavy-flavour
decays inside jets.  However, if this jet has fewer than three
associated tracks, the muon is kept and the jet is removed instead;
this avoids an inefficiency for high-energy muons undergoing
significant energy loss in the calorimeter.  A \tauhad{} candidate is
rejected if it is separated by $\Delta R_{y} < 0.2$ from any selected
electron or muon. No overlap removal is performed between jets and \tauhad{} candidates.
 
The missing transverse momentum (with magnitude \met) is reconstructed
as the negative vector sum of the \pt of all the selected electrons,
muons, \tauhad\ and jets described above, with an extra `soft term'
built from additional tracks associated with the primary vertex, to make
it resilient to pile-up contamination~\cite{PERF-2016-07}. The missing
transverse momentum is not used for event selection but is included in
the inputs to the multivariate discriminants that are built in the
most sensitive analysis categories (see Section~\ref{sec:Regions:mva}).
 
For the boosted category, the small-$R$ jets are
reclustered~\cite{Nachman:2014kla} using the \antikt\ algorithm with a
radius parameter of 1.0, resulting in a collection of large-$R$
jets (referred to as RC jets). These RC jets are required to have a reconstructed invariant
mass higher than 50~\GeV, $\pt > 200$~\GeV\ and at least two
small-$R$ constituent jets. RC jets are used as input to a deep neural network (DNN),
described in Section~\ref{sec:Regions:mva}, to identify high-\pt\ (boosted) top-quark
and Higgs boson candidates decaying into collimated hadronic final states.
The event is flagged as
containing a boosted Higgs boson candidate if one of the RC jets has a high
probability of originating from a Higgs boson, as estimated by the DNN.

Events are required to have exactly one lepton in the single-lepton
channels and exactly two leptons with opposite electric
charge in the dilepton channel. At least one of the leptons must have
$\pt>27$~\GeV\ and match a corresponding object at trigger level. In the $ee$
and $\mu\mu$ channels, the dilepton invariant mass must be above 15~\GeV\
and outside the \Zboson-boson mass window 83--99~\GeV.  To
maintain orthogonality with other \ttH\ channels~\cite{HIGG-2017-02},
events are vetoed if they contain one or more (two or more) \tauhad{}
candidates in the dilepton (single-lepton) channel. Leptons are
further required to satisfy additional identification and isolation
criteria to increase background rejection: electrons (muons) must
pass the \textit{Tight} (\textit{Medium})
identification criterion and the \textit{Gradient}
(\textit{FixedCutTightTrackOnly}) isolation
criteria~\cite{EGAM-2018-01,PERF-2015-10}.


\section{Analysis strategy}
\label{sec:Regions}
In order to target the \ttHbb\ final state, events are categorised into mutually exclusive regions defined by the number of leptons, the number of jets, the number of $b$-tagged jets at different $b$-tagging efficiencies (60\%, 70\%, 77\%, or 85\%) and the number of boosted Higgs boson candidates (see Section~\ref{sec:Regions:analysis}).
 
The analysis regions with higher signal-to-background ratio 
are referred to as `signal regions' (SR). Higgs boson candidates are reconstructed using a DNN in the boosted category, and boosted decision trees (BDT) referred to as `reconstruction BDTs' in the resolved categories, aiming at associating the reconstructed jets to the final state partons.
Kinematic variables of these Higgs boson candidates, such as their transverse momentum \ptH, are computed and used together with several other variables as input to other BDTs. These, referred to as `classification BDTs', are then employed to separate signal from background, in each of the SRs (see Section~\ref{sec:Regions:mva}).
The remaining analysis regions, depleted in signal, 
are referred to as `control regions' (CR). They provide stringent constraints on the normalisation of the backgrounds and on systematic uncertainties when used in a combined fit with the signal regions.
 
\subsection{Analysis regions}
\label{sec:Regions:analysis}
Table~\ref{tab:regions:summary} defines the 16 regions into which the events are classified:
11 SRs (dilepton \SR{4}{4}, single-lepton \SR{6}{4} and \SRboosted, split according to the reconstructed \ptH into four, five and two regions, respectively), and five CRs.
 
In the single-lepton channel, events are assigned to the boosted category if they contain at least four jets $b$-tagged at the 85\% working point, one boosted Higgs boson candidate, and at least two jets not belonging to the boosted Higgs boson candidate which are $b$-tagged at the 77\% working point.
The boosted Higgs boson candidate must satisfy $\pt>300$~\GeV, have an invariant mass in the range 100--140~\GeV, a DNN score
$P(H)>0.6$ (see Section~\ref{sec:Regions:mva}), and exactly two jet constituents $b$-tagged at the 85\% working point. If more than one candidate is identified, the one with a mass closest to the Higgs boson mass is chosen.
The selected RC jet is used to determine the kinematic properties of the boosted Higgs boson candidate (reconstructed 
\ptH, $m_{\bbbar}$, etc.).
All other selected events belong to the resolved categories.

The analysis regions are further split according to the reconstructed
\ptH\ to allow the extraction of multiple signal parameters, sensitive
to new physics effects. The \ptH\ ranges are the same as used to
define STXS bins with `truth' \ptHhat (where the `truth' \ptHhat is
taken from the MC event record before the Higgs boson decays), and
were chosen to minimise the correlation among signal strengths in
different STXS bins. The split goes as follows:
\begin{itemize}
\item The single-lepton resolved signal region, \SR{6}{4}, is split in five
reconstructed \ptH regions: 0--120\,\GeV, 120--200\,\GeV,
200--300\,\GeV, 300--450\,\GeV\ and $\geq$\,450\,\GeV.
\item The dilepton signal region, \SR{4}{4}, is split in a similar manner, with the two
highest \ptH\ bins merged because only a small number of events are
expected.
\item The boosted signal region, \SRboosted, is split into
two reconstructed \ptH regions: 300--450\,\GeV\ and
$\geq$\,450\,\GeV.
\item Control regions are inclusive in reconstructed \ptH to
maintain the constraints on the background composition.
\end{itemize}
 
After these selections are applied, \ttbar\ + heavy-flavour jets dominate the background composition and the \ttH\ selection efficiency is 1.2\%. Although all Higgs boson decay modes are considered, \hbb events account for at least $94\%$ of \ttH\ events selected in the signal regions. In the SRs the shape and normalisation of a multivariate discriminant distribution is used in the statistical analysis, except in the highest reconstructed \ptH bin of the single-lepton resolved analysis, where only the event yield is used.
 
In the dilepton CRs, only the event yield is used to correct the amount of \ttcin\ predicted from the inclusive \ttjets sample. In the single-lepton channel the distribution shape and normalisation of the average $\Delta R$ for all possible combinations of $b$-tagged jet pairs, \drbbav, is used in the CRs to help better constrain the background contributions and correct their shape.
Control regions have different ratios of \ttbin\ to \ttcin\ events: regions labelled `hi', referring to a higher $b$-tagging probability, are enriched in \ttbin, while in regions labelled `lo', the proportion of \ttcin\ events is increased.
The different proportions of \ttbin\ and \ttcin\ in the control regions allow the signal extraction fit to better constrain the relative fractions of these processes in the signal regions.

\begin{table}[tp]
\centering \small
\renewcommand{\arraystretch}{1.4}
\begin{tabular}{l@{\hspace{3pt}}lc@{\hspace{3pt}}c@{\hspace{3pt}}c@{\hspace{3pt}}c>{\hspace*{-5pt}}cc@{\hspace{3pt}}c@{\hspace{3pt}}c@{\hspace{3pt}}c}
\toprule
\hline
\multicolumn{2}{l}{\multirow{2}*{Region}} & \multicolumn{4}{c}{Dilepton}  && \multicolumn{4}{c}{Single-lepton} \\
& & \SR{4}{4} & \CR[hi]{4}{3} & \CR[lo]{4}{3} & \CR[hi]{3}{3} && \SR{6}{4} & \CR[hi]{5}{4} & \CR[lo]{5}{4} & \SRboosted \\
\midrule \midrule 
\multicolumn{2}{l}{\#leptons} & \multicolumn{4}{c}{$=2$} && \multicolumn{4}{c}{$=1$}  \\
\cline{3-6} \cline{8-11}
\multicolumn{2}{l}{\#jets} &  \multicolumn{3}{c|}{$\geq4$} & $=3$ && $\geq6$  & \multicolumn{2}{|c|}{$=5$} & $\geq4$ \\
\cline{1-6} \cline{8-11}
\multirow{4}{*}{\#$b$-tag}         & @85\%  & \multicolumn{4}{c}{--} && \multicolumn{4}{c}{$\geq4$} \\
\cline{3-6} \cline{8-11}
& @77\% & \multicolumn{4}{c}{--} && \multicolumn{3}{c|}{--} & $\ge2^\dagger$ \\
\cline{3-6} \cline{8-11}
& @70\% & $\geq4$ & \multicolumn{3}{|c}{$=3$} && \multicolumn{3}{c|}{$\geq4$} & -- \\
\cline{3-6} \cline{8-11}
& @60\% & -- & $=3$ & $<3$ & $=3$ && -- & $\geq4$ & $<4$ & -- \\
\cline{1-6} \cline{8-11}
\multicolumn{2}{l}{\#boosted cand.}  & \multicolumn{4}{c}{--} &&  \multicolumn{3}{c|}{0} & $\geq1$\\
\cline{1-6} \cline{8-11}
\cline{1-6} \cline{8-11}
\multicolumn{2}{l}{Fit input} & BDT & \multicolumn{3}{|c}{Yield} && BDT/Yield &  \multicolumn{2}{|c|}{\drbbav} & BDT \\
\hline
\bottomrule
\end{tabular}
\caption{Definition of the analysis regions, split according to the number of leptons, jets, and $b$-tagged jets using different working points, and the number of boosted Higgs candidates. For \SRboosted, $b$-tagged jets flagged with $^\dagger$ are extra $b$-jets not part of the boosted Higgs boson candidate. All SRs are further split in reconstructed \ptH as described in the text. The last row specifies the type of input to the fit used in each region: normalisation only (Yield) or shape and normalisation of the classification BDT or \drbbav\ distribution. In the highest bin ($\ptH\geq 450$~\GeV) of the single-lepton resolved analysis, only the event yield is used.}
\label{tab:regions:summary}
\end{table}
 
\subsection{Multivariate analysis}
\label{sec:Regions:mva}
Multivariate classifiers are used in two parts of this analysis: identifying Higgs boson candidate objects and classifying \ttH\ signal events. In all SRs of the resolved categories, the multivariate classifiers are constructed analogously to the reconstruction and classification BDTs used in the previous analysis~\cite{HIGG-2017-03} and trained with TMVA~\cite{TMVA}.
The training for the reconstruction BDTs is identical to this previous analysis, matching reconstructed jets to the partons emitted from top-quark and Higgs boson decays. For this purpose, $W$-boson, top-quark and Higgs boson candidates are built from combinations of jets and leptons. The $b$-tagging information is used to discard combinations containing jet–parton assignments inconsistent with the correct parton candidate flavour. The combination of jets with the highest reconstruction BDT score is selected, allowing the computation of kinematic properties of the Higgs boson candidate (reconstructed \ptH, $m_{\bbbar}$, etc.) by summing the four-momenta of the two jets associated with the Higgs boson candidate in this combination.
 
In the boosted channel the Higgs boson candidate is found using a DNN with a
three-node softmax output layer trained with Keras~\cite{chollet2015keras}
and a TensorFlow backend~\cite{tensorflow2015-whitepaper} on a sample of
RC jets from signal events. The DNN quantifies the probability
that an RC jet is originated from a Higgs boson (labelled $P(H)$ and shown in Figure~\ref{fig:regions:boosted_pHiggs}), top quark 
or any other process (mostly multijet production
). The most
important DNN input variables for identifying a Higgs boson candidate jet are built from the
small-$R$ jet constituent masses and pseudo-continuous $b$-tagging values, while substructure variables~\cite{PERF-2012-02} also contribute. The Higgs boson jets are correctly identified 76\% of the time, while top-quark jets are identified in 67\% of the cases, in signal events.
 
The performance of the reconstruction in the three channels is reported in Figure~\ref{fig:RecoMigrationAll}, which shows the migration matrix for Higgs boson candidates between reconstructed \ptH bins and `truth' \ptHhat bins, for all reconstructed Higgs boson candidates. When applying the reconstruction BDT or the DNN (which are trained with \ttH\ events to match $b$-jets to a Higgs boson candidate) to the \ttbin\ sample, the two selected jets do not originate from the top-quark decay products about half (two-thirds) of the time in the lower (higher) \ptH\ bin. The number of events with both jets originating from top-quark decays is negligible.
 
\begin{figure}[tp]
\centering
\begin{subfigure}[]{0.4\linewidth}
\vspace{0pt}
\includegraphics[width=\linewidth]{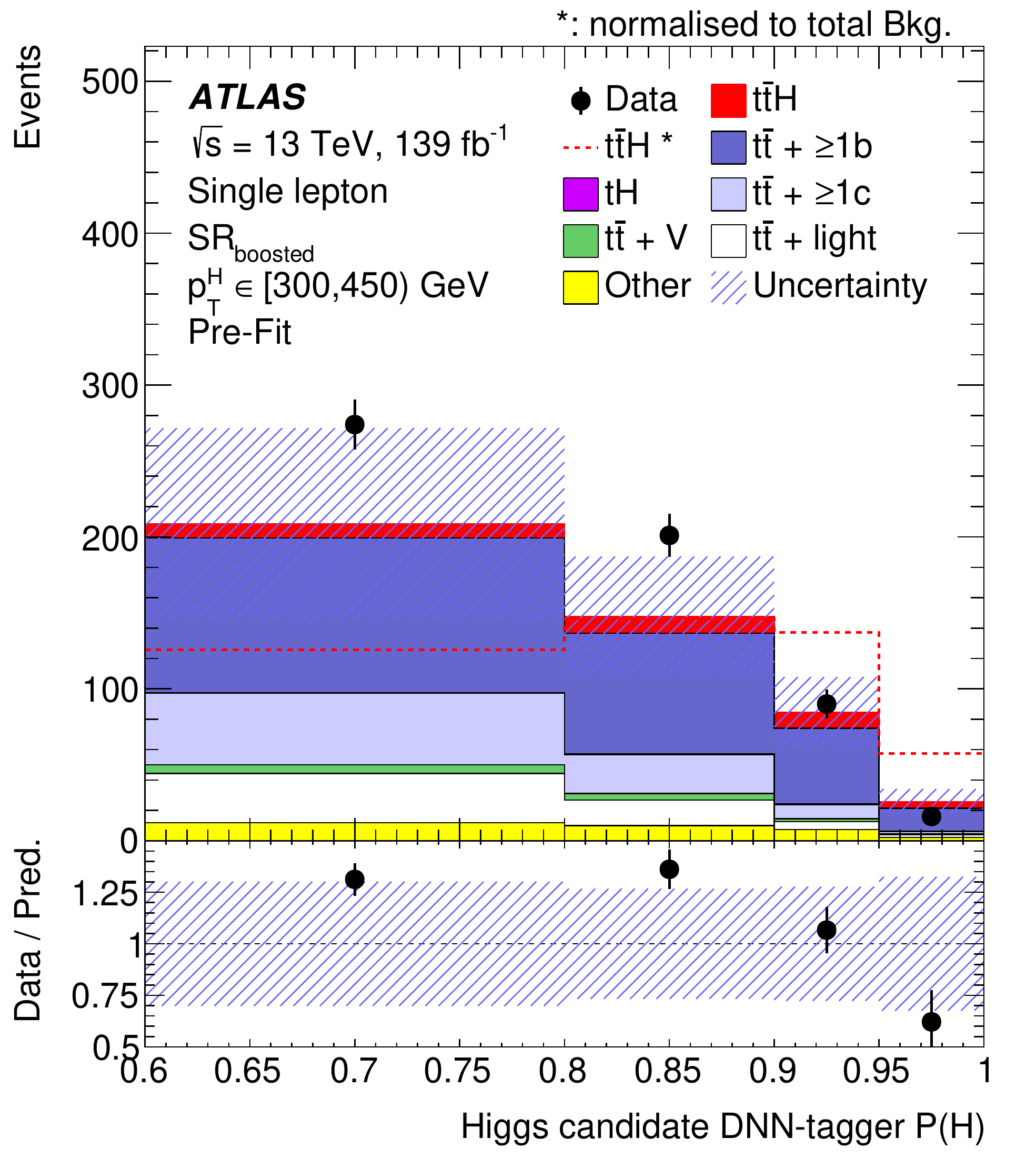}
\caption{}
\end{subfigure}
\hspace{0.5cm}
\begin{subfigure}[]{0.4\linewidth}
\vspace{0pt}
\includegraphics[width=\linewidth]{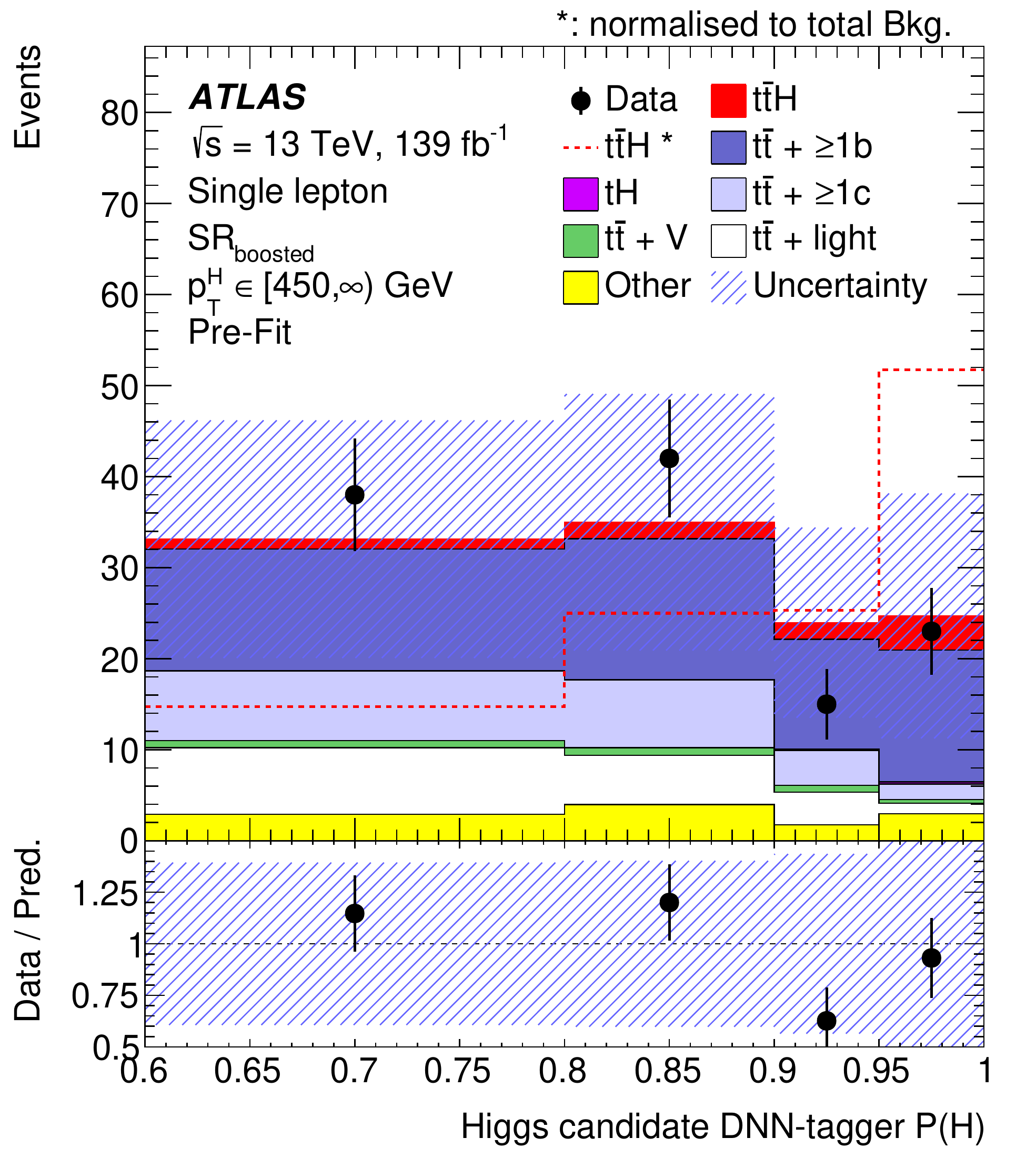}
\caption{}
\end{subfigure}
\caption{Comparison between data and prediction for the DNN $P(H)$ output for the Higgs boson candidate prior to any fit to the data in the single-lepton boosted channel for (a) $300\le \ptH<450$~\GeV\ and (b) $\ptH\ge 450$~\GeV. The dashed line shows the \ttH\ signal distribution normalised to the total background prediction. The uncertainty band includes all uncertainties and their correlations.}
\label{fig:regions:boosted_pHiggs}
\end{figure}

\begin{figure}[tp]
\centering
\includegraphics[width=.8\textwidth]{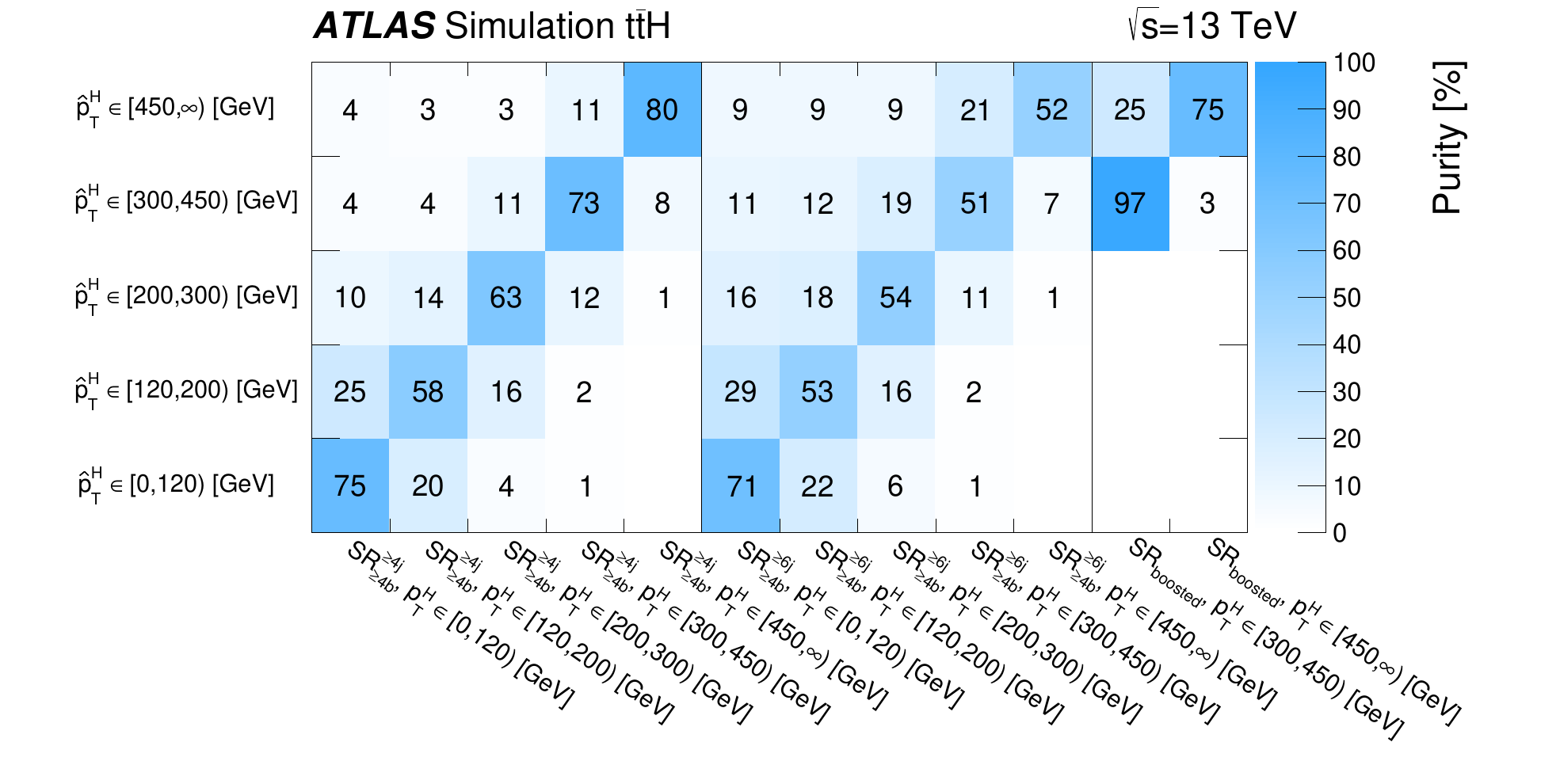}
\caption{Performance of the Higgs boson reconstruction algorithms. For each row of `truth' \ptHhat, the matrix shows (in percentages) the fraction of all Higgs boson candidates with reconstructed \ptH in the various bins of the dilepton (left), single-lepton resolved (middle) and boosted (right) channels.}
\label{fig:RecoMigrationAll}
\end{figure}

The classification BDTs have been retrained since the previous analysis to profit from the improved background modelling and from the increased number of simulated events, using the signal and components of the nominal background model presented in this paper. The dilepton BDT is trained only against \ttbb events (as it constitutes most of the background), the single-lepton resolved BDT is trained against \ttjets\ events (because \ttcin and \ttlight events also contribute) and the single-lepton boosted BDT is trained against all background processes.
 
These BDTs are built by combining kinematic variables, such as invariant masses and angular separations of pairs of reconstructed jets and leptons, outputs of the reconstruction discriminants, as well as the pseudo-continuous $b$-tagging discriminant of selected jets. The complete lists of input variables can be found in Appendix~\ref{sec:Appendix}. The reconstruction discriminants provide their own output value as well as variables derived from the selected combination of jets with the highest reconstruction BDT score in the resolved channels. In the single-lepton resolved channel, a likelihood discriminant method that combines the signal and background probabilities of all possible jet combinations in each event is also used as input to the classification BDT~\cite{HIGG-2017-03}. In the boosted channel, information from the DNN is used as input to the classification BDT, including the Higgs probability, $P(H)$. 
Distributions of the output of these BDT classifiers serve as SR inputs to the signal extraction fit.
 
The most important variables entering the dilepton BDT include
the reconstruction BDT score for the Higgs boson candidate identified using Higgs boson information,
the average $\Delta\eta$ between $b$-tagged jets,
the minimum invariant mass of a $b$-tagged jet pair,
the $\Delta R$ between the Higgs candidate and the \ttbar\ candidate system,
and the number of $b$-tagged jet pairs with an invariant mass within 30\,\GeV\ of 125\,\GeV.
The most important variables entering the single-lepton resolved BDT include
the likelihood discriminant,
the average $\Delta R$ for all possible combinations of two $b$-tagged jets (\drbbav),
the invariant mass of the two $b$-tagged jets with the smallest $\Delta R$,
the reconstruction BDT score for the Higgs boson candidate identified using Higgs boson information,
and the $\Delta R$ between the two highest-\pt\ $b$-tagged jets.
The most important variables entering the single-lepton boosted BDT include
the DNN $P(H)$ output for the Higgs boson candidate,
the sum of $b$-tagging discriminants of small-$R$ jets from Higgs, hadronic top and leptonic top candidates,
the hadronic top candidate's invariant mass,
the small-$R$ jet multiplicity and, in the sum of $b$-tagging discriminants,
the fraction due to all jets not associated with the Higgs or hadronic top candidates.
 

\section{Systematic uncertainties}
\label{sec:Systematics}
Many sources of systematic uncertainty affect this analysis, impacting the categorisation of events as well as the shape and normalisation of the final discriminants used in the signal extraction fit.
All sources of experimental uncertainty considered, with the exception of the uncertainty in the luminosity, affect both the normalisations and shapes of distributions in all the simulated event samples.
Uncertainties related to the theoretical modelling of the signal and background processes affect both the normalisations and shapes of the distributions, with the exception of cross-section and normalisation uncertainties which only affect the overall yield of the considered sample. Nonetheless, the normalisation uncertainties modify the relative fractions of the different samples leading to a shape uncertainty in the distribution of the final discriminant for the total prediction in the different analysis regions.
 
A single independent nuisance parameter is assigned to each source of systematic uncertainty.
Some of the systematic uncertainties, in particular most of the experimental uncertainties, are decomposed into several independent sources, as specified in the following. Each individual source then has a correlated effect across all the channels, analysis regions, and signal and background samples.
Modelling uncertainties are typically broken down into components which target specific physics effects in the event generation, such as scale variations or changing the hadronisation model, and are uncorrelated between different samples.

\subsection{Experimental uncertainties}
 
The uncertainty in the combined 2015--2018 integrated luminosity is 1.7\%~\cite{ATLAS-CONF-2019-021}, obtained using the LUCID-2 detector~\cite{LUCID2} for the primary luminosity measurement. 
An uncertainty associated with the modelling of pile-up in the simulation is included to cover the difference between the predicted and measured inelastic cross-section values~\cite{STDM-2015-05}.
 
The jet energy scale uncertainty is derived by combining information from test-beam data, LHC collision data and simulation, and the jet energy resolution uncertainty is obtained by combining dijet \pt-balance measurements and simulation~\cite{JETM-2018-05}. Additional considerations related to jet flavour, pile-up corrections, $\eta$ dependence and high-\pt jets are included. These uncertainties are further propagated into the single-lepton boosted analysis by applying the reclustering described in Section~\ref{sec:Selection} with systematically varied inputs. A total of 40 independent contributions are considered. While the uncertainties are not large, varying between 1\% and 5\% per jet (depending on the jet \pt), the effects are amplified by the large number of jets considered in the final state. The efficiency to identify and remove jets from pile-up is measured with $Z \to \mu^+\mu^-$ events in data using techniques similar to those used in Ref.~\cite{PERF-2014-03}. All small-$R$ jet constituent uncertainties are propagated to RC jets.
 
The efficiency to correctly tag \bjets is measured using dileptonic \ttbar\ events~\cite{FTAG-2018-01}. The mis-tag rate for \cjets is measured using single-lepton \ttbar\ events, exploiting the \cjets from the hadronic \Wboson-boson decays using techniques similar to those in Ref.~\cite{ATLAS-CONF-2018-001}. The mis-tag rate for \ljets is measured using a negative-tag method similar to that in Ref.~\cite{ATLAS-CONF-2018-006} applied to \Zjets events. The uncertainty in tagging \bjets is 2\%--10\% depending on the working point and jet \pt. The uncertainty in mis-tagging \cjets (\ljets) is 10\%--25\% (15\%--50\%) depending on the working point and jet \pt. For the calibration of the four working points used in this analysis, a large number of uncertainty components are considered, 
yielding 45, 20, and 20 uncorrelated sources of uncertainty for $b$-, $c$- and light-jets, respectively.
 
Uncertainties associated with leptons arise from the trigger, reconstruction, identification, and isolation, as well as the lepton momentum scale and resolution. Efficiencies are measured and scale and resolution calibrations are performed using leptons in $Z\to \ell^+\ell^-$ and $J/\psi \to \ell^+\ell^-$ events~\cite{EGAM-2018-01,PERF-2015-10}. Systematic uncertainties in these measurements account for 22 independent sources but have only a small impact on the final result.
 
All uncertainties related to the energy scales or resolution of the reconstructed objects are propagated to the calculation of the missing transverse momentum. Three additional uncertainties associated with the scale and resolution of the soft term are also included. As the missing transverse momentum is only used for event reconstruction and not for event selection, these uncertainties have a minimal impact.
 
\subsection{Theoretical modelling uncertainties}
\label{sec:Systematics:modelling}

Uncertainties in the predicted SM \ttH\ signal cross-section are evaluated with a particular focus on the impact on STXS bins.
An uncertainty of $\pm3.6\%$ from varying the PDF and \alphas\ in the fixed-order calculation is applied~\cite{deFlorian:2016spz,Raitio:1978pt,Beenakker:2002nc,Dawson:2003zu,Yu:2014cka,Frixione:2015zaa}.
The effect of PDF variations on the shape of the distributions considered in this analysis is found to be negligible.
Uncertainties in the Higgs boson branching fractions are also considered, and amount to 2.2\% for the $b\bar{b}$ decay mode~\cite{deFlorian:2016spz}.
An uncertainty related to the amount of initial-state radiation (ISR) is estimated by simultaneously varying the renormalisation and factorisation scales in the ME and $\alphas^{\mathrm{ISR}}$ in the PS~\cite{ATL-PHYS-PUB-2020-023}, while an uncertainty related to final-state radiation (FSR) is estimated by varying $\alphas^{\mathrm{FSR}}$ in the PS.
The nominal \POWHEGBOX\!+\,\PYTHIA[8] sample is also compared with the \POWHEGBOX\!+\,\HERWIG[7] sample to assess an uncertainty related to the choice of PS and hadronisation model, and with the \MGNLO\!+\,\PYTHIA[8] sample to assess an uncertainty arising from changing the NLO matching procedure (sample details are given in Table~\ref{tab:modelling:samples2}).
Uncertainties due to missing higher-order terms in the perturbative QCD calculations affecting the total cross-section and event migration between STXS bins are estimated by varying the renormalisation and factorisation scales independently by a factor of two, as well as evaluating the ISR and FSR uncertainties. The largest effect was found to originate from the ISR uncertainty, corresponding to a 9.2\% variation of the total cross-section, leading to an uncertainty of 10\%--17\% in bin migrations estimated using the Stewart--Tackmann procedure~\cite{Stewart:2011cf}.
All signal uncertainties are correlated across STXS bins, with the exception of bin migration uncertainties.

The systematic uncertainties affecting the \ttjets\ background modelling are summarised in Table~\ref{tab:systs:ttjets}. An uncertainty of 6\% is assumed for the inclusive \ttbar\ production cross-section predicted at NNLO+NNLL, including effects from varying the factorisation and renormalisation scales, the PDFs, $\alphas$, and the top-quark mass~\cite{Beneke:2011mq,Cacciari:2011hy,Baernreuther:2012ws,Czakon:2012zr,Czakon:2012pz,Czakon:2013goa,Czakon:2011xx}. This uncertainty is applied to \ttlight\ samples only, since this component is dominant in \ttbar\ production in the full phase-space. An uncertainty of 100\% in the normalisation of \ttcin\ events is applied, motivated by the fitted value of this normalisation in the previous analysis~\cite{HIGG-2017-03}. The normalisation of \ttbin\ is allowed to float freely in the signal extraction fit.
The \ttbin, \ttcin\ and \ttlight\ processes are affected by different types of uncertainties: \ttlight\ profits from relatively precise measurements in data; \ttbin\ and \ttcin\ can have similar or different diagrams depending on the precision of the ME and the flavour scheme used for the PDF, and the different masses of the $c$- and $b$-quarks contribute to additional differences between these two processes. For these reasons, all uncertainties in the \ttjets\ background modelling are assigned independent nuisance parameters for the \ttbin, \ttcin\ and \ttlight\ processes. The effect of PDF uncertainties was found to be negligible.
 
Systematic uncertainties in the acceptance and the distribution shapes are extracted from comparisons between the nominal prediction and different samples or settings. The fraction of \ttbin\ events in the selected phase-space in all alternative samples is reweighted to match the fraction in the nominal sample. This is to allow the normalisation of \ttbin\ to be driven solely by the free-floating parameter in the signal extraction fit to data. The systematic uncertainties related to varying the amount of ISR, the amount of FSR, the PS and hadronisation model, and the NLO matching procedure are estimated using the same procedure as for \ttH, comparing the nominal prediction with alternative samples.
In the specific case of \ttbin, relative uncertainties are used to estimate the effect of changing the PS and hadronisation model or the NLO matching procedure by comparing predictions from the NLO \ttbar\ generators (see Table~\ref{tab:systs:ttjets}). These comparisons are made using predictions in which the additional $b$-quarks were generated at leading-log precision from gluon splitting. 
A check was performed, comparing the nominal \ttbin prediction with a smaller \ttbb sample produced with \SHERPA[2.2.1].
The size of the difference estimated from \ttbar\ samples was observed to be generally the same as or larger than the difference between the two \ttbb\ NLO predictions. 
The impact of these uncertainties on the final results is reported in Section~\ref{sec:Results}.
 
\begin{table}[tp]
\centering
\renewcommand{\arraystretch}{1.2}
\footnotesize
\begin{tabular}{ l ll l }
\toprule
\midrule
Uncertainty source     & \multicolumn{2}{l}{Description}							& Components         \\
\midrule
\ttbar\ cross-section  & \multicolumn{2}{l}{$\pm 6$\%}					& \ttlight\          \\
\ttbin\ normalisation  & \multicolumn{2}{l}{Free-floating} 						& \ttbin             \\
\ttcin\ normalisation  & \multicolumn{2}{l}{$\pm 100$\%}					& \ttcin             \\
\midrule
NLO matching           & \multicolumn{2}{l}{\MGNLO\!+\,\PYTHIA[8] vs \POWHEGBOX\!+\,\PYTHIA[8]} 		& All                \\
PS \& hadronisation    & \multicolumn{2}{l}{\POWHEGBOX\!+\,\HERWIG[7] vs \POWHEGBOX\!+\,\PYTHIA[8]} 		& All                \\ 
\multirow{2}{*}{ISR}
& \multirow{2}{*}{Varying $\alphas^{\mathrm{ISR}}$ (PS), $\muR$\,\&\,$\muF$ (ME)}
& in \POWHEGBOXRES\!+\,\PYTHIA[8]    		& \ttbin\            \\
& & in \POWHEGBOX\!+\,\PYTHIA[8]   	& \ttcin, \ttlight\  \\ 
\multirow{2}{*}{FSR}
& \multirow{2}{*}{Varying $\alphas^{\mathrm{FSR}}$ (PS)}
& in \POWHEGBOXRES\!+\,\PYTHIA[8]   		& \ttbin\            \\
& & in \POWHEGBOX\!+\,\PYTHIA[8]   	& \ttcin, \ttlight\  \\
\midrule
\ttbin\ fractions       & \multicolumn{2}{l}{\POWHEGBOX\!+\,\HERWIG[7] vs \POWHEGBOX\!+\,\PYTHIA[8]}     & \ttbexcl, \ttbbin   \\
\ptbb\ shape            & \multicolumn{2}{l}{Shape mismodelling measured from data}       & \ttbin   \\
\midrule
\bottomrule
\end{tabular}
\caption{Summary of the sources of systematic uncertainty for \ttjets\ modelling. The systematic uncertainties listed in the second section of the table are evaluated in such a way as to   have no impact on the normalisation of the three \ttbin, \ttcin, and \ttlight\ components in the phase-space selected in this analysis. The last column of the table indicates the \ttjets\ components to which a systematic uncertainty is assigned. All systematic uncertainty sources are treated as uncorrelated across the three components.
}
\label{tab:systs:ttjets}
\end{table}
 
Special consideration is given to the correlation of modelling uncertainties across different \ptH bins, in order to provide the fit with enough flexibility to cover background mismodelling without biasing the signal extraction. The \ttbin NLO matching uncertainty is shown to depend on \ptH and is therefore decorrelated across \pt\ bins in the SRs. The NLO matching uncertainty and the PS and hadronisation uncertainties for \ttbin are further decorrelated between the single-lepton and dilepton channels in order to avoid the transfer of constraints from the single-lepton resolved channel to the dilepton channel, which is less sensitive to the high-\ptH\ regime and produces less additional radiation.
The pre-fit distributions of the reconstructed \ptH\ are shown in Figure~\ref{fig:systematics:hpt}. An additional uncertainty is derived for the \ttbin\ sample to cover the mismodelling observed in this distribution.
After removing the overall normalisation difference by scaling the \ttbin\ background in the dilepton \SR{4}{4} (single-lepton \SR{6}{4}), a weight is computed in each reconstructed \ptH\ bin of the dilepton \SR{4}{4} (single-lepton \SR{6}{4}), which corrects the predicted \ttbin contribution so that the data and background model yields agree in each \ptH\ bin.
The derived weights are not applied to the nominal sample: instead they define the $+1\sigma$ variation of 
an additional uncertainty in the \ttbin\ sample in each reconstructed \ptH\ bin.
The weights derived in the single-lepton resolved channel are also applied in the boosted channel. This uncertainty enters the signal extraction fit as a single nuisance parameter (\ptbb\ shape), correlated across all channels, such that a pull of $+1\sigma$ corresponds to applying this weight, effectively correcting the reconstructed \ptH\ spectrum.

\begin{figure}[tp]
\centering
\begin{subfigure}[]{0.32\linewidth}
\vspace{0pt}
\includegraphics[width=\linewidth]{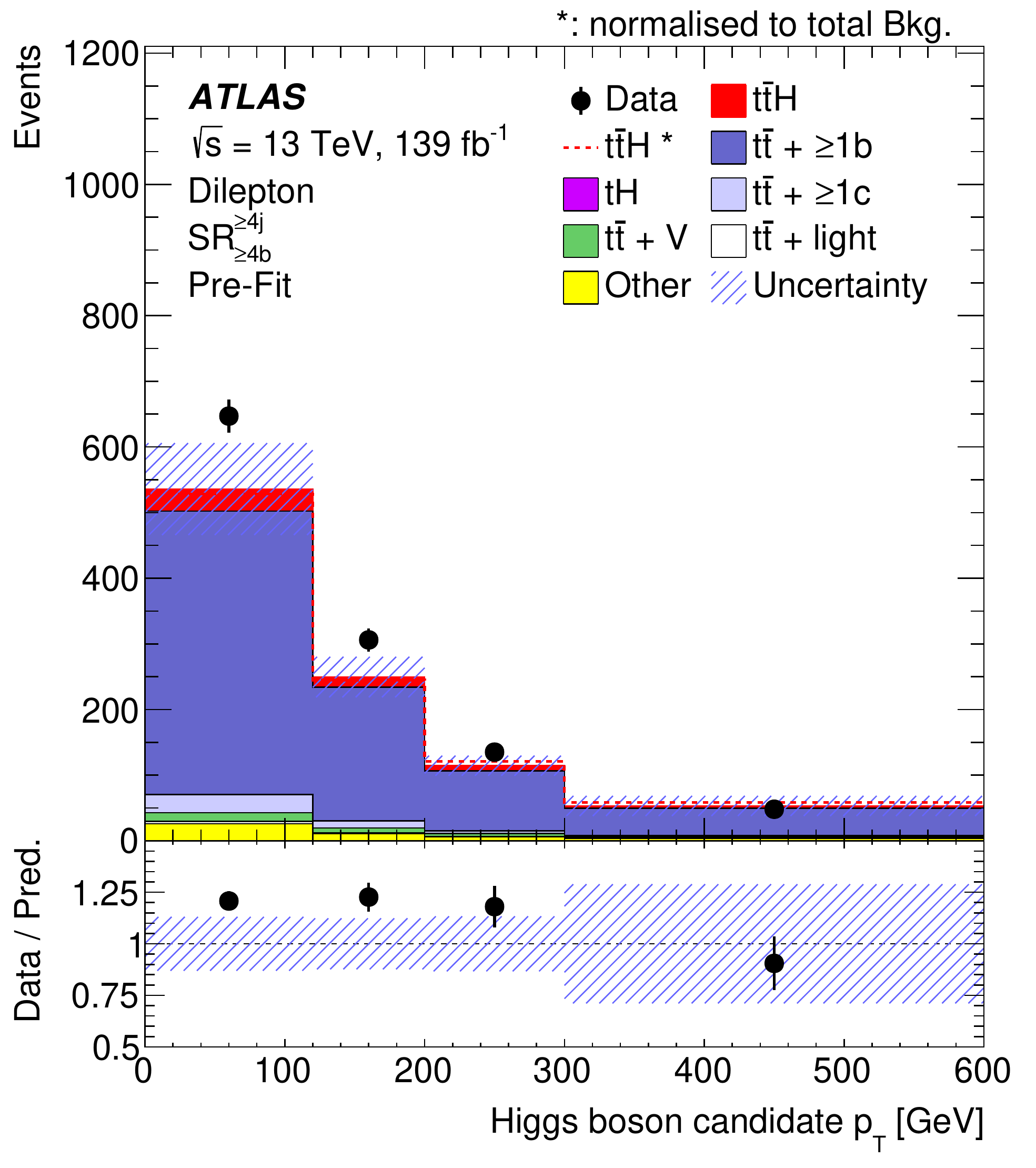}
\caption{}
\end{subfigure}
\begin{subfigure}[]{0.32\linewidth}
\vspace{0pt}
\includegraphics[width=\linewidth]{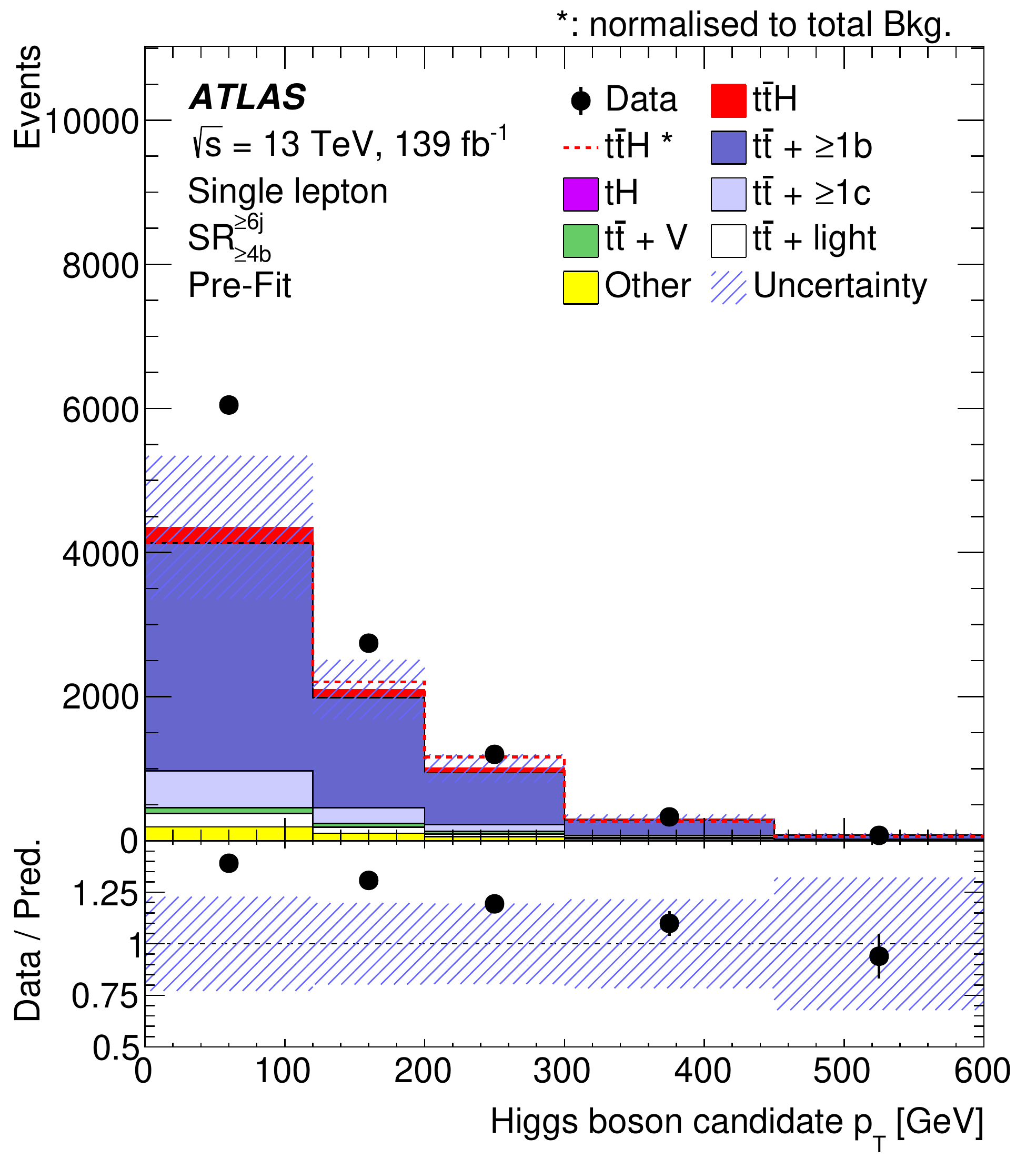}
\caption{}
\end{subfigure}
\begin{subfigure}[]{0.32\linewidth}
\vspace{0pt}
\includegraphics[width=\linewidth]{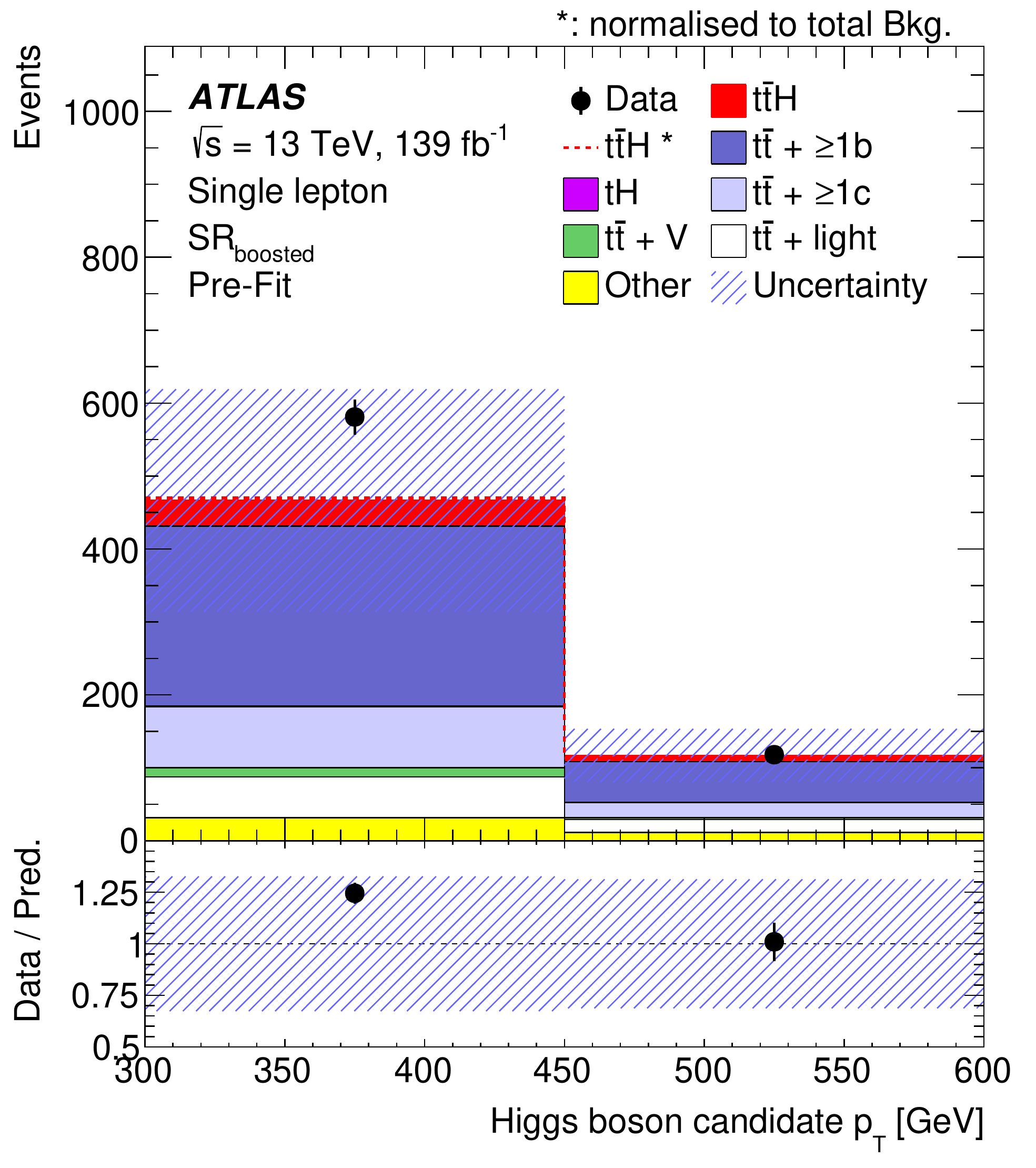}
\caption{}
\end{subfigure}
\caption{Pre-fit distributions of the reconstructed Higgs boson candidate \ptH\ for the (a) dilepton \SR{4}{4}, (b) single-lepton resolved \SR{6}{4} and (c) single-lepton boosted \SRboosted signal regions. The dashed line shows the \ttH\ signal distribution normalised to the total background prediction. The uncertainty band includes all uncertainties and their correlations, except for the uncertainty in the $k(\ttbin)$ normalisation factor which is not defined pre-fit. The last bin includes the overflow.}
\label{fig:systematics:hpt}
\end{figure}
 
To account for variations in the \ttbin\ subcomponent fractions found in different predictions, an additional nuisance parameter is introduced to cover the largest discrepancy between two models for the fraction of \ttbexcl\ and \ttbbin. The $1\sigma$ variation of this nuisance parameter corresponds to reducing the amount of \ttbbin\ by 13\% and increasing the amount of \ttbexcl\ by 22\%. This uncertainty is correlated across all regions, and impacts each region differently due to the varying compositions of \ttbin.

An uncertainty of 5\% is considered for the cross-sections of the three single-top production modes~\cite{Martin:2009iq,Martin:2009bu,Aliev:2010zk,Kant:2014oha}. Uncertainties associated with the PS and hadronisation model, and with the NLO matching scheme, are evaluated by comparing, for each process, the nominal \POWHEGBOX\!+\,\PYTHIA[8] sample with a sample produced using \POWHEGBOX\!+\,\HERWIG[7] and \MGNLO\!+\,\PYTHIA[8], respectively.
The uncertainty associated with the interference between $tW$ and \ttbar\ production at NLO~\cite{Frixione:2008yi} is assessed by comparing the nominal \POWHEGBOX\!+\,\PYTHIA[8] sample produced using the diagram removal scheme with the alternative sample produced with the same generator but using the diagram subtraction scheme.
 
An uncertainty of 40\% is assumed for the \Wjets\ cross-section, with an additional 30\% normalisation uncertainty used for $W$+ heavy-flavour jets, taken as uncorrelated between events with two and more than two heavy-flavour jets. These uncertainties are based on variations of the factorisation and renormalisation scales and of the \SHERPA\ matching parameters. An uncertainty of 35\% is applied to the \Zjets\ normalisation, uncorrelated across jet bins, to account for both the variations of the scales and \SHERPA\ matching parameters and the uncertainty in the extraction from data of the correction factor for the heavy-flavour component.
 
The uncertainty in the \ttV\ NLO cross-section prediction is 15\%~\cite{Campbell:2012dh}, split into PDF and scale uncertainties as for \ttH. An additional \ttV\ modelling uncertainty, related to the choice of PS and hadronisation model and NLO matching scheme is assessed by comparing the nominal \MGNLO\!+\,\PYTHIA[8] samples with alternative ones generated with \SHERPA.
 
The uncertainty in the diboson background is assumed to be 50\%, which includes uncertainties in the inclusive cross-section and additional jet production~\cite{ATL-PHYS-PUB-2016-002}. A 50\% normalisation uncertainty is considered for the four-top background, covering effects from varying the factorisation and renormalisation scales, the PDFs and \alphas~\cite{Alwall:2014hca}.
 
The small backgrounds from \tZq\ and $tWZ$ are each assigned cross-section uncertainties; for \tZq\ two uncertainties are used, 7.9\% accounting for factorisation and renormalisation scale variations and 0.9\% accounting for PDFs, and for $tWZ$ a single uncertainty of 50\% is used~\cite{Alwall:2014hca}.
Uncertainties in the associated production of a single top quark and a Higgs boson include factorisation and renormalisation scale variations as well as PDF uncertainties: they amount to $+6.5/-14.9$\% ($+6.5/-6.7$\%) and $\pm 3.7$\% ($\pm 6.3$\%) for \tHjb (\tWH), respectively~\cite{deFlorian:2016spz}.
Finally, a 25\% normalisation uncertainty is considered for the fake lepton background in the dilepton channel.


\section{Results}
\label{sec:Results}
The distributions of the classification BDT in the signal regions, and the event yield or the \drbbav\ distributions in the dilepton or single-lepton control regions, respectively, are combined in a profile likelihood fit to extract the signal, while simultaneously determining the yields and constraining the normalisation and shape of differential distributions of the most important background components.
The binning of these distributions is optimised to maximise the analysis sensitivity while keeping the total MC statistical uncertainty in each bin to a level adjusted to avoid biases due to fluctuations in the predicted number of events.
 
The statistical analysis is based on a binned likelihood function ${\cal L}(\mu,\theta)$ constructed as a product of Poisson probability terms over all bins considered in the analysis. The likelihood function depends on the signal-strength parameter $\mu$, defined as $\mu = \sigma/\sigma_{\textrm{SM}}$ (where $\sigma$ is the measured cross-section and $\sigma_{\textrm{SM}}$ is the Standard Model prediction), and a set of nuisance parameters $\theta$ which characterise the effects of systematic uncertainties in the signal and background expectations.
They are implemented in the likelihood function as Gaussian or Poisson priors, with the exception of the unconstrained normalisation factor $k(\ttbin)$ for the \ttbin\ background, of which no prior knowledge from theory or subsidiary measurements is assumed.
The statistical uncertainty in the prediction, which incorporates the statistical uncertainty arising from the limited number of simulated events, is included in the likelihood in the form of additional nuisance parameters, one for each of the considered bins.
In the statistical analysis, the number of events expected in a given bin depends on $\mu$ and $\theta$.
The nuisance parameters $\theta$ adjust the expectations for signal and background according to the corresponding systematic uncertainties, and their fitted values correspond to the amount that best fits the data.
The test statistic $t_\mu$ is defined as the profile likelihood ratio: $t_\mu=-2\ln({\cal L}(\mu,\hat{\hat{\theta}}_\mu)/{\cal L}(\hat{\mu},\hat{\theta}))$, where $\hat{\mu}$ and $\hat{\theta}$ are the values of the parameters that maximise the likelihood function, and $\hat{\hat{\theta}}_\mu$ are the values of the nuisance parameters that maximise the likelihood function for a given value of $\mu$~\cite{Cowan:2010js}.
This test statistic, as implemented in the RooStat framework~\cite{RooFit,RooStat}, is used to assess the compatibility of the observed data with the background-only hypothesis (i.e.\ for $\mu=0$) and to make statistical inferences about $\mu$. The uncertainty in the best-fit value of the signal strength is extracted by finding the values of $\mu$ that correspond to varying $t_\mu$ by one unit.
 
Tables~\ref{tab:yields_dil} and \ref{tab:yields_ljets} show the observed and predicted signal and background
event yields in all SRs and CRs before and after the inclusive fit to data. Post-fit values are summarised in Figure~\ref{fig:results:summary}, where the precision increases post-fit due to profiling and the uncertainties can be observed to increase as a function of \ptH, ranging from 2\% to 12\%.
The SR BDT distributions are
presented in Figures~\ref{fig:results:dilep_PostFit_BDT} and
\ref{fig:results:ljets_PostFit_BDT}, while the \drbbav\ distributions in the single-lepton resolved CRs are shown in Figure~\ref{fig:results:ljets_PostFit_dRbb}. All distributions are
compatible with the data.
The normalisation factor for the \ttbin\ background is found to be $k(\ttbin) = 1.28 \pm 0.08$.
The best-fit $\mu$ value is
\[
\mu = 0.35 \pm 0.20\,\textrm{(stat.)}\,^{+0.30}_{-0.28}\textrm{\,(syst.)} = 0.35^{+0.36}_{-0.34},
\]
corresponding to an observed (expected) significance of 1.0 (2.7) standard deviations with respect to the background-only hypothesis. This observed inclusive $\mu$ and its uncertainty, common to all channels, cannot affect the shape of the signal distributions but only their normalisation, leading to post-fit signal yield uncertainties very close to 100\%.

\begin{table}[tp]
\setlength{\extrarowheight}{3pt}
\centering

\scalebox{0.7}{
\begin{tabular}{l
r@{$\,\pm\,$}l
r@{$\,\pm\,$}l
r@{$\,\pm\,$}l
r@{$\,\pm\,$}l
r@{$\,\pm\,$}l
r@{$\,\pm\,$}l
r@{$\,\pm\,$}l
r@{$\,\pm\,$}l
}
\toprule
\hline
& \multicolumn{4}{c}{\SR[]{4}{4}, \ptH $\in$ [0,120) \GeV} & \multicolumn{4}{c}{\SR[]{4}{4}, \ptH $\in$ [120,200) \GeV} & \multicolumn{4}{c}{\SR[]{4}{4}, \ptH $\in$ [200,300) \GeV} & \multicolumn{4}{c}{\SR[]{4}{4}, \ptH $\in$ [300,$\infty$) \GeV}	 \\
\cmidrule(lr){2-5}  \cmidrule(lr){6-9}  \cmidrule(lr){10-13}  \cmidrule(lr){14-17}
& \multicolumn{2}{c}{Pre-fit} & \multicolumn{2}{c}{Post-fit} & \multicolumn{2}{c}{Pre-fit} & \multicolumn{2}{c}{Post-fit} & \multicolumn{2}{c}{Pre-fit} & \multicolumn{2}{c}{Post-fit} & \multicolumn{2}{c}{Pre-fit} & \multicolumn{2}{c}{Post-fit}  \\
\hline
\ttH	 & 33.6 & 4.1 & 12 & 12 & 15.6 & 1.8 & 5.5 & 5.3 & 7.71 & 0.89 & 2.7 & 2.6 & 3.72 & 0.44 & 1.3 & 1.3 \\
$tH$	 & 0.249 & 0.065 & 0.249 & 0.064 & 0.148 & 0.063 & 0.146 & 0.061 & 0.043 & 0.032 & 0.043 & 0.031 & 0.031 & 0.027 & 0.031 & 0.025 \\
\ttbin	 & 432 & 59 & 546 & 24 & 203 & 27 & 263 & 12 & 92 & 14 & 116.9 & 8.8 & 42 & 15 & 37.9 & 6.0 \\
\ttcin	 & 27 & 29 & 48.5 & 9.1 & 11 & 12 & 16.3 & 5.0 & 4.0 & 4.2 & 6.5 & 1.4 & 1.9 & 2.1 & 3.69 & 0.96 \\
\ttbar + $Z$	 & 12.5 & 2.0 & 12.6 & 2.0 & 7.4 & 1.6 & 7.6 & 1.6 & 4.18 & 0.72 & 4.15 & 0.70 & 2.05 & 0.45 & 2.06 & 0.44 \\
\ttbar + $W$	 & 0.75 & 0.31 & 0.75 & 0.31 & 0.38 & 0.12 & 0.40 & 0.11 & 0.27 & 0.12 & 0.27 & 0.11 & 0.124 & 0.068 & 0.127 & 0.068 \\
\ttbar + light	 & 3.6 & 4.9 & 4.8 & 6.2 & 0.97 & 0.96 & 0.92 & 0.74 & 0.46 & 0.65 & 0.41 & 0.47 & 0.22 & 0.31 & 0.22 & 0.25 \\
\tttt	 & 3.1 & 1.5 & 3.0 & 1.5 & 2.4 & 1.2 & 2.3 & 1.2 & 1.38 & 0.70 & 1.36 & 0.69 & 0.81 & 0.41 & 0.79 & 0.40 \\
Fakes	 & 3.7 & 1.1 & 3.7 & 1.1 & 1.33 & 0.51 & 1.33 & 0.51 & 0.40 & 0.23 & 0.40 & 0.23 & 0.57 & 0.30 & 0.57 & 0.30 \\
Other sources	 & 19.1 & 6.9 & 19.3 & 7.0 & 7.1 & 4.4 & 7.7 & 4.3 & 4.3 & 4.0 & 4.5 & 4.1 & 2.0 & 1.5 & 2.0 & 1.5 \\
\hline
Total	 & 536 & 71 & 651 & 21 & 249 & 32 & 305 & 11 & 114 & 16 & 137.2 & 8.1 & 53 & 15 & 48.7 & 5.7 \\
\hline
Data	 & \multicolumn{4}{c}{647} & \multicolumn{4}{c}{306} & \multicolumn{4}{c}{135} & \multicolumn{4}{c}{48} \\
\hline
\bottomrule
\end{tabular}
}
 
\vspace{.2cm}

\scalebox{0.7}{
\begin{tabular}{l
r@{$\,\pm\,$}l
r@{$\,\pm\,$}l
r@{$\,\pm\,$}l
r@{$\,\pm\,$}l
r@{$\,\pm\,$}l
r@{$\,\pm\,$}l
}
\toprule
\hline
& \multicolumn{4}{c}{\CR[hi]{3}{3}} & \multicolumn{4}{c}{\CR[hi]{4}{3}} & \multicolumn{4}{c}{\CR[lo]{4}{3}} \\
\cmidrule(lr){2-5}  \cmidrule(lr){6-9}  \cmidrule(lr){10-13}
& \multicolumn{2}{c}{Pre-fit} & \multicolumn{2}{c}{Post-fit} & \multicolumn{2}{c}{Pre-fit} & \multicolumn{2}{c}{Post-fit} & \multicolumn{2}{c}{Pre-fit} & \multicolumn{2}{c}{Post-fit}  \\
\hline
\ttH	 & 25.2 & 3.1 & 8.8 & 8.6 & 117 & 13 & 42 & 41 & 76.4 & 8.4 & 27 & 27 \\
$tH$	 & 1.26 & 0.15 & 1.23 & 0.15 & 2.06 & 0.39 & 2.02 & 0.38 & 1.19 & 0.53 & 1.12 & 0.50 \\
\ttbin	 & 1900 & 510 & 2010 & 130 & 2810 & 300 & 4070 & 210 & 1730 & 210 & 2550 & 160 \\
\ttcin	 & 350 & 360 & 550 & 130 & 700 & 710 & 1190 & 240 & 1500 & 1500 & 2550 & 470 \\
\ttbar + $Z$	 & 11.1 & 1.8 & 10.8 & 1.7 & 57.1 & 7.4 & 57.5 & 7.3 & 51.7 & 7.0 & 52.3 & 6.7 \\
\ttbar + $W$	 & 1.88 & 0.59 & 1.84 & 0.55 & 10.8 & 1.6 & 10.9 & 1.6 & 21.5 & 3.7 & 21.8 & 3.4 \\
\ttbar + light	 & 128 & 74 & 119 & 61 & 200 & 120 & 210 & 120 & 850 & 350 & 900 & 340 \\
\tttt	 & 0.047 & 0.026 & 0.044 & 0.024 & 12.3 & 6.1 & 12.0 & 6.1 & 8.9 & 4.5 & 8.8 & 4.4 \\
Fakes	 & 6.3 & 1.8 & 6.4 & 1.8 & 47 & 12 & 47 & 12 & 56 & 14 & 56 & 14 \\
Other sources	 & 125 & 35 & 125 & 34 & 211 & 62 & 211 & 62 & 251 & 73 & 257 & 73 \\
\hline
Total	 & 2540 & 630 & 2835 & 54 & 4160 & 810 & 5855 & 79 & 4500 & 1600 & 6431 & 83 \\
\hline
Data	 & \multicolumn{4}{c}{2827} & \multicolumn{4}{c}{5865} & \multicolumn{4}{c}{6429} \\
\hline
\bottomrule
\end{tabular}
}
\caption{Pre-fit and post-fit event yields in the dilepton channel (top) signal regions and (bottom) control regions. Post-fit yields are after the inclusive fit in all channels. All uncertainties are included, taking into account correlations in the post-fit case. The $k(\ttbin)$ uncertainty is not defined pre-fit and therefore only included in the post-fit uncertainties. For the \ttH{} signal, the pre-fit yield values correspond to the theoretical prediction and corresponding uncertainties, while the post-fit yield and uncertainties correspond to those in the inclusive signal-strength measurement. `Other sources' refers to \schan, \tchan, \tW, \tWZ, \tZq, \Zjets and diboson events.}
\label{tab:yields_dil}
\end{table}

\begin{table}[tp]
\setlength{\extrarowheight}{3pt}
\centering

\scalebox{0.65}{
\begin{tabular}{l
r@{$\,\pm\,$}l
r@{$\,\pm\,$}l
r@{$\,\pm\,$}l
r@{$\,\pm\,$}l
r@{$\,\pm\,$}l
r@{$\,\pm\,$}l
r@{$\,\pm\,$}l
r@{$\,\pm\,$}l
r@{$\,\pm\,$}l
r@{$\,\pm\,$}l
}
\toprule
\hline
& \multicolumn{4}{c}{\SR[]{6}{4}, \ptH $\in$ [0,120) \GeV} & \multicolumn{4}{c}{\SR[]{6}{4}, \ptH $\in$ [120,200) \GeV} & \multicolumn{4}{c}{\SR[]{6}{4}, \ptH $\in$ [200,300) \GeV} & \multicolumn{4}{c}{\SR[]{6}{4}, \ptH $\in$ [300,450) \GeV} & \multicolumn{4}{c}{\SR[]{6}{4}, \ptH $\in$ [450,$\infty$) \GeV} \\
\cmidrule(lr){2-5}  \cmidrule(lr){6-9}  \cmidrule(lr){10-13}  \cmidrule(lr){14-17}  \cmidrule(lr){18-21}
& \multicolumn{2}{c}{Pre-fit} & \multicolumn{2}{c}{Post-fit} & \multicolumn{2}{c}{Pre-fit} & \multicolumn{2}{c}{Post-fit} & \multicolumn{2}{c}{Pre-fit} & \multicolumn{2}{c}{Post-fit} & \multicolumn{2}{c}{Pre-fit} & \multicolumn{2}{c}{Post-fit} & \multicolumn{2}{c}{Pre-fit} & \multicolumn{2}{c}{Post-fit}  \\
\hline
\ttH	 & 213 & 29 & 76 & 73 & 113 & 15 & 40 & 39 & 59.9 & 7.8 & 21 & 21 & 13.9 & 2.0 & 4.8 & 4.7 & 3.09 & 0.49 & 1.0 & 1.0 \\
$tH$	 & 3.01 & 0.47 & 3.03 & 0.44 & 1.81 & 0.44 & 1.83 & 0.40 & 1.24 & 0.26 & 1.28 & 0.25 & 0.26 & 0.17 & 0.25 & 0.14 & 0.000 & 0.041 & 0.000 & 0.041 \\
\ttbin	 & 3160 & 490 & 4530 & 160 & 1530 & 210 & 2050 & 80 & 720 & 130 & 865 & 47 & 215 & 61 & 240 & 22 & 55 & 27 & 44.7 & 8.5 \\
\ttcin	 & 510 & 540 & 930 & 190 & 220 & 230 & 401 & 83 & 96 & 100 & 176 & 36 & 26 & 27 & 45 & 10 & 6.9 & 7.5 & 12.8 & 3.2 \\
\ttbar + $W$	 & 7.0 & 1.2 & 7.1 & 1.1 & 4.31 & 0.90 & 4.37 & 0.84 & 2.47 & 0.52 & 2.49 & 0.46 & 1.05 & 0.32 & 1.06 & 0.30 & 0.47 & 0.15 & 0.47 & 0.14 \\
\ttbar + $Z$	 & 77 & 11 & 77 & 10 & 44.6 & 6.6 & 45.6 & 6.3 & 30.1 & 4.9 & 30.8 & 4.8 & 11.5 & 2.4 & 11.6 & 2.3 & 2.05 & 0.64 & 2.11 & 0.64 \\
\ttbar + light	 & 180 & 120 & 220 & 130 & 85 & 58 & 97 & 58 & 37 & 23 & 43 & 25 & 10.5 & 7.7 & 11.6 & 7.8 & 2.2 & 2.1 & 2.7 & 2.3 \\
Single top $tW$	 & 71 & 40 & 76 & 42 & 40 & 26 & 42 & 27 & 17.9 & 7.6 & 18.3 & 7.7 & 8.5 & 7.9 & 9.5 & 9.0 & 6.0 & 5.3 & 6.0 & 5.4 \\
\tttt	 & 15.9 & 8.0 & 15.7 & 7.9 & 12.6 & 6.3 & 12.5 & 6.3 & 7.7 & 3.9 & 7.6 & 3.8 & 2.7 & 1.4 & 2.7 & 1.3 & 0.98 & 0.50 & 0.95 & 0.49 \\
Other top sources	 & 46 & 24 & 45 & 24 & 23 & 16 & 24 & 16 & 13 & 10 & 12.7 & 9.6 & 4.3 & 2.8 & 4.4 & 2.6 & 1.08 & 0.54 & 1.06 & 0.54 \\
$V$ \& $VV$ + jets	 & 60 & 24 & 60 & 23 & 29 & 11 & 29 & 11 & 19.7 & 8.3 & 19.7 & 7.9 & 7.8 & 3.4 & 7.8 & 3.2 & 1.90 & 0.88 & 1.84 & 0.82 \\
\hline
Total	 & 4350 & 810 & 6045 & 82 & 2100 & 360 & 2745 & 48 & 1000 & 180 & 1198 & 33 & 301 & 72 & 339 & 16 & 80 & 29 & 73.7 & 7.2 \\
\hline
Data	 & \multicolumn{4}{c}{6047} & \multicolumn{4}{c}{2742} & \multicolumn{4}{c}{1199} & \multicolumn{4}{c}{331} & \multicolumn{4}{c}{75} \\
\hline
\bottomrule
\end{tabular}
}

\vspace{.2cm}

\scalebox{0.7}{
\begin{tabular}{l
r@{$\,\pm\,$}l
r@{$\,\pm\,$}l
r@{$\,\pm\,$}l
r@{$\,\pm\,$}l
r@{$\,\pm\,$}l
r@{$\,\pm\,$}l
r@{$\,\pm\,$}l
r@{$\,\pm\,$}l
}
\toprule
\hline
& \multicolumn{4}{c}{\SRboosted, \ptH $\in$ [300,450) \GeV} & \multicolumn{4}{c}{\SRboosted, \ptH $\in$ [450,$\infty$) \GeV}	 & \multicolumn{4}{c}{\CR[lo]{5}{4}} & \multicolumn{4}{c}{\CR[hi]{5}{4}} \\
\cmidrule(lr){2-5}  \cmidrule(lr){6-9}  \cmidrule(lr){10-13}  \cmidrule(lr){14-17}
& \multicolumn{2}{c}{Pre-fit} & \multicolumn{2}{c}{Post-fit} & \multicolumn{2}{c}{Pre-fit} & \multicolumn{2}{c}{Post-fit} & \multicolumn{2}{c}{Pre-fit} & \multicolumn{2}{c}{Post-fit} & \multicolumn{2}{c}{Pre-fit} & \multicolumn{2}{c}{Post-fit}  \\
\hline
\ttH	 & 35.1 & 4.1 & 12 & 12 & 8.5 & 1.1 & 3.0 & 2.9 & 61.7 & 8.1 & 21 & 21 & 62.1 & 8.6 & 22 & 21 \\
$tH$	 & 1.31 & 0.31 & 1.31 & 0.29 & 0.42 & 0.10 & 0.42 & 0.10 & 3.15 & 0.41 & 3.15 & 0.40 & 3.16 & 0.42 & 3.18 & 0.42 \\
\ttbin	 & 246 & 49 & 295 & 23 & 55 & 24 & 52.5 & 9.3 & 1370 & 160 & 1581 & 89 & 1000 & 150 & 1118 & 51 \\
\ttcin	 & 84 & 90 & 156 & 35 & 21 & 23 & 38 & 10 & 390 & 410 & 650 & 140 & 56 & 59 & 93 & 23 \\
\ttbar + $W$	 & 1.86 & 0.39 & 1.88 & 0.35 & 0.55 & 0.18 & 0.56 & 0.17 & 2.53 & 0.53 & 2.58 & 0.45 & 0.54 & 0.13 & 0.53 & 0.12 \\
\ttbar + $Z$	 & 10.7 & 2.1 & 10.9 & 2.1 & 2.21 & 0.60 & 2.32 & 0.59 & 26.4 & 3.7 & 26.0 & 3.5 & 23.5 & 3.4 & 23.0 & 3.2 \\
\ttbar + light	 & 56 & 26 & 61 & 25 & 17 & 10 & 16.2 & 7.3 & 260 & 120 & 270 & 95 & 22 & 16 & 20 & 14 \\
Single top $tW$	 & 13.1 & 8.0 & 13.7 & 8.3 & 6.1 & 5.8 & 5.3 & 4.6 & 58 & 32 & 59 & 32 & 27 & 20 & 28 & 20 \\
\tttt	 & 1.76 & 0.89 & 1.74 & 0.88 & 0.42 & 0.22 & 0.41 & 0.21 & 0.33 & 0.17 & 0.31 & 0.16 & 0.24 & 0.13 & 0.23 & 0.12 \\
Other top sources	 & 4.3 & 3.2 & 4.4 & 3.1 & 0.80 & 0.78 & 0.82 & 0.77 & 41 & 16 & 41 & 16 & 27 & 11 & 26 & 10 \\
$V$ \& $VV$ + jets	 & 12.4 & 5.7 & 12.4 & 5.4 & 4.3 & 2.3 & 4.2 & 2.0 & 42 & 16 & 41 & 15 & 24.2 & 8.8 & 24.2 & 8.5 \\
\hline
Total	 & 470 & 120 & 571 & 22 & 117 & 38 & 123.8 & 9.7 & 2260 & 490 & 2694 & 53 & 1250 & 170 & 1359 & 36 \\
\hline
Data	 & \multicolumn{4}{c}{581} & \multicolumn{4}{c}{118} & \multicolumn{4}{c}{2696} & \multicolumn{4}{c}{1362} \\
\hline
\bottomrule
\end{tabular}
}
\caption{Pre-fit and post-fit event yields in the single-lepton (top) resolved signal regions and (bottom) boosted signal regions and control regions. Post-fit yields are after the inclusive fit in all channels. All uncertainties are included, taking into account correlations in the post-fit case. The $k(\ttbin)$ uncertainty is not defined pre-fit and therefore only included in the post-fit uncertainties. For the \ttH{} signal, the pre-fit yield values correspond to the theoretical prediction and corresponding uncertainties, while the post-fit yield and uncertainties correspond to those in the inclusive signal-strength measurement. `Other top sources' refers to \schan, \tchan, \tWZ and \tZq events.}
\label{tab:yields_ljets}
\end{table}

\begin{figure}[tp]
\centering
\begin{subfigure}[]{0.45\linewidth}
\vspace{0pt}
\includegraphics[width=\linewidth]{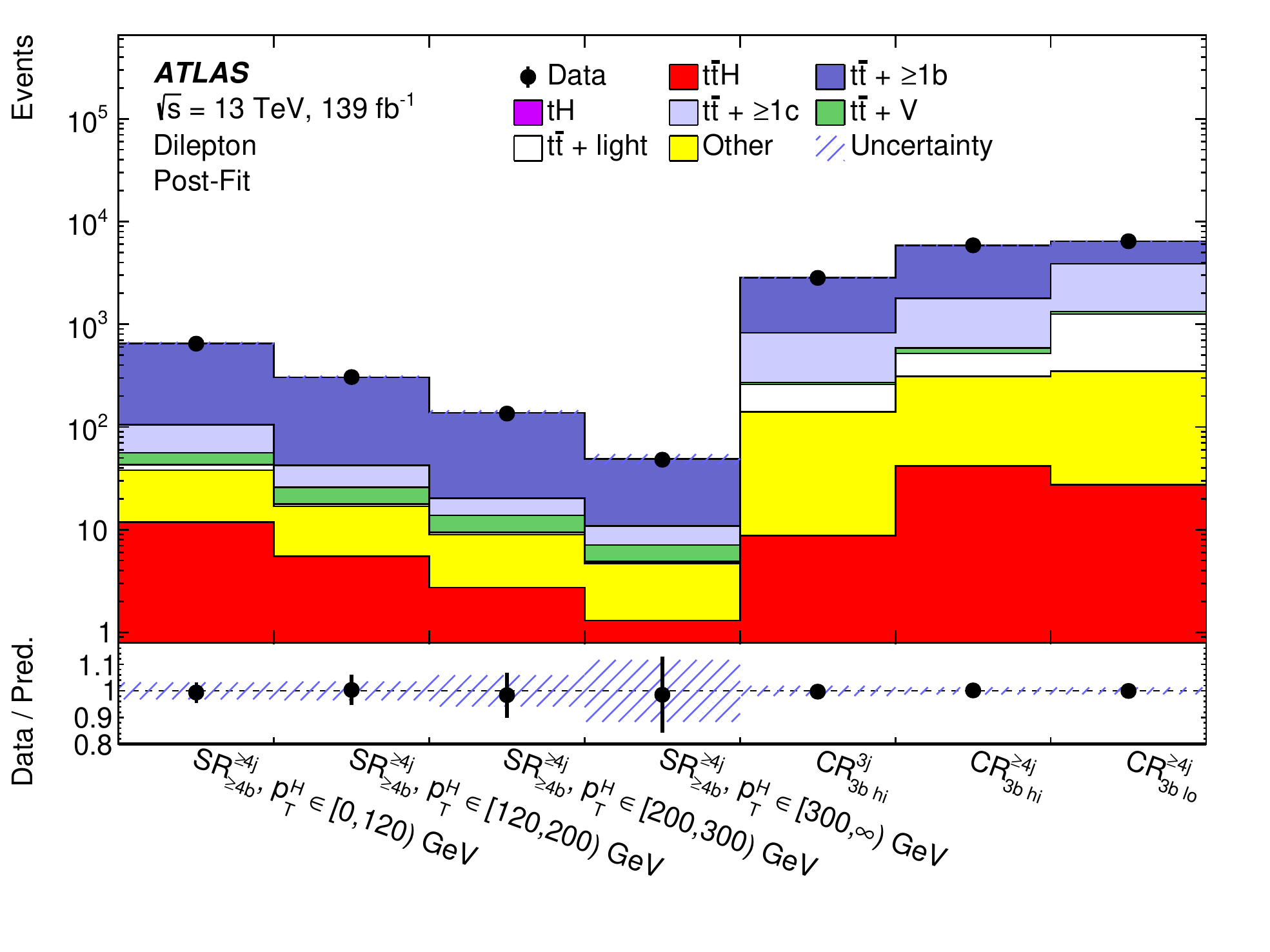}
\caption{}
\end{subfigure}
\hspace{0.5cm}
\begin{subfigure}[]{0.45\linewidth}
\vspace{0pt}
\includegraphics[width=\linewidth]{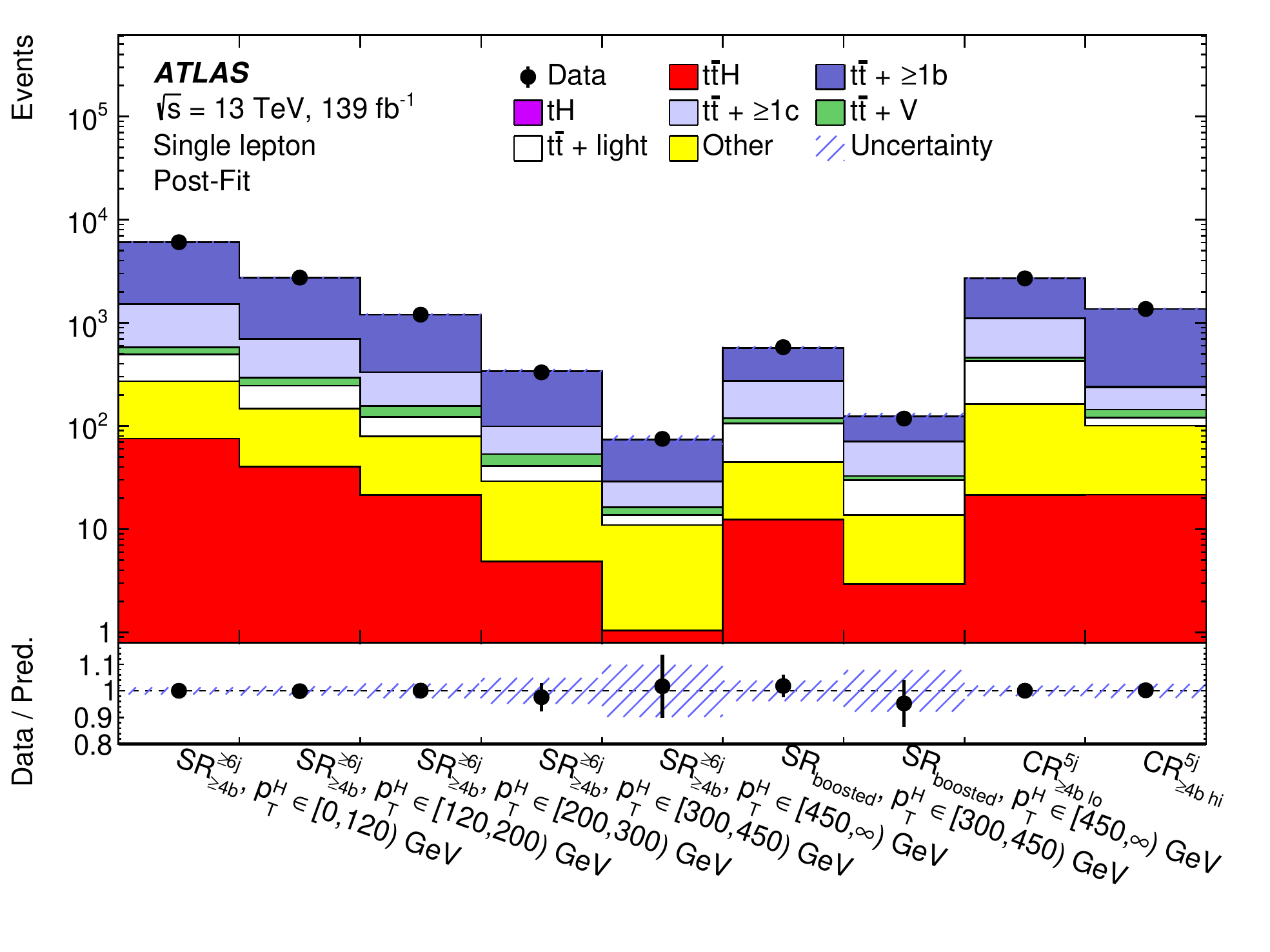}
\caption{}
\end{subfigure}
\caption{Comparison of predicted and observed event yields in each
of the control and signal regions in the (a) dilepton and (b)
single-lepton channels after the fit to the data.
The uncertainty band includes all uncertainties and their correlations.}
\label{fig:results:summary}
\end{figure}
 
\begin{figure}[tp]
\centering
\begin{subfigure}[]{0.24\linewidth}
\vspace{0pt}
\includegraphics[width=\linewidth]{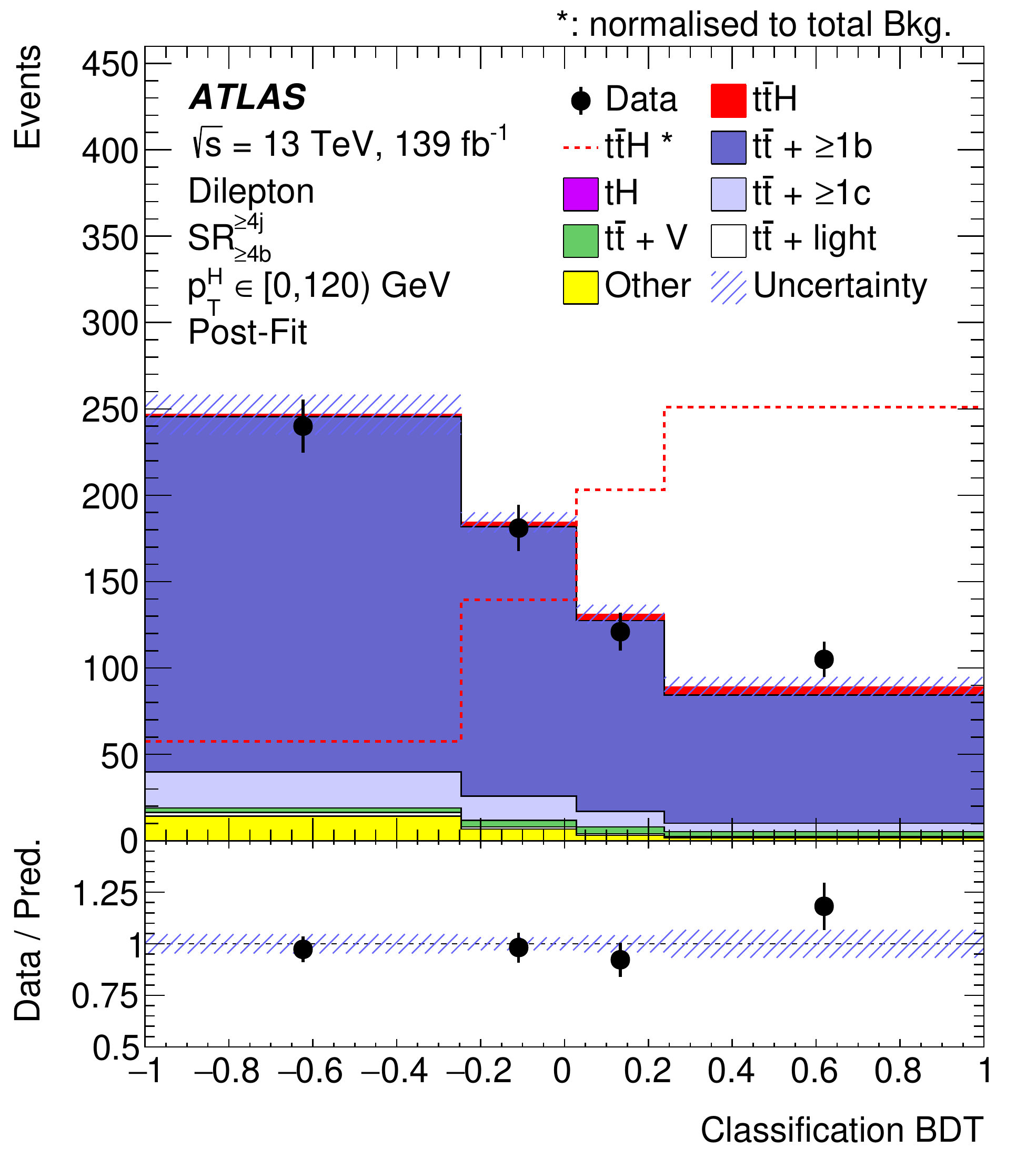}
\caption{}
\end{subfigure}
\begin{subfigure}[]{0.24\linewidth}
\vspace{0pt}
\includegraphics[width=\linewidth]{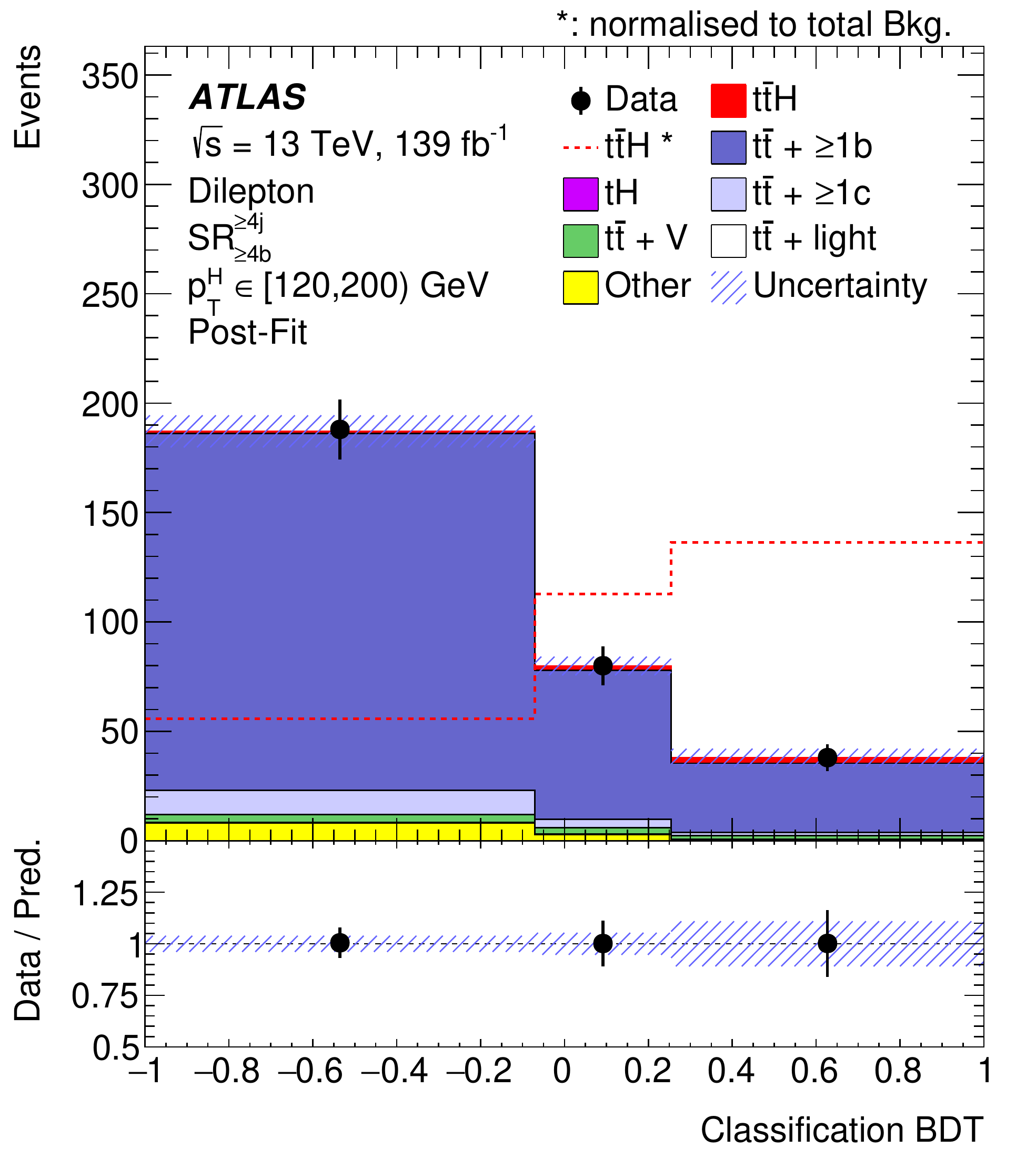}
\caption{}
\end{subfigure}
\begin{subfigure}[]{0.24\linewidth}
\vspace{0pt}
\includegraphics[width=\linewidth]{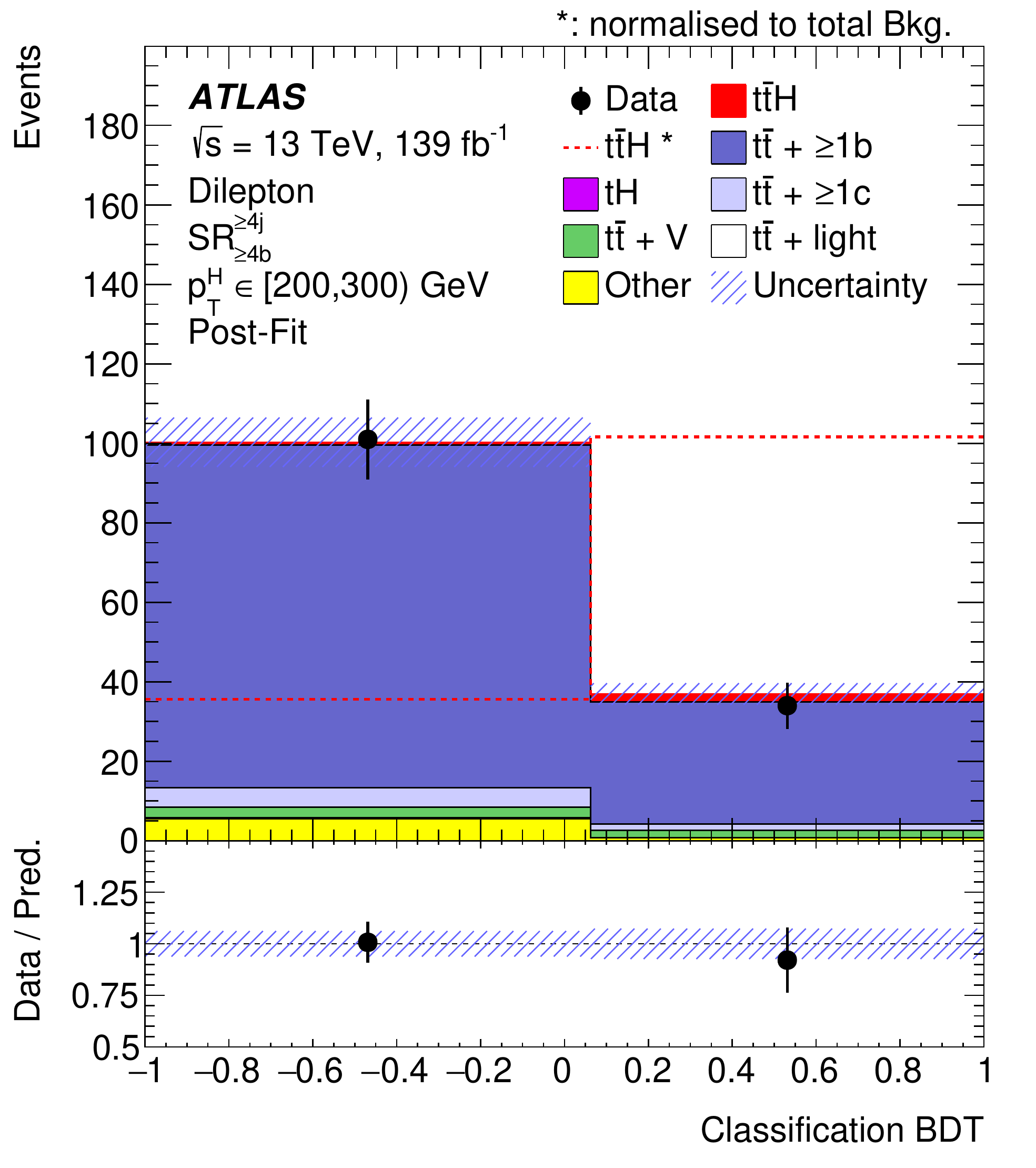}
\caption{}
\end{subfigure}
\begin{subfigure}[]{0.24\linewidth}
\vspace{0pt}
\includegraphics[width=\linewidth]{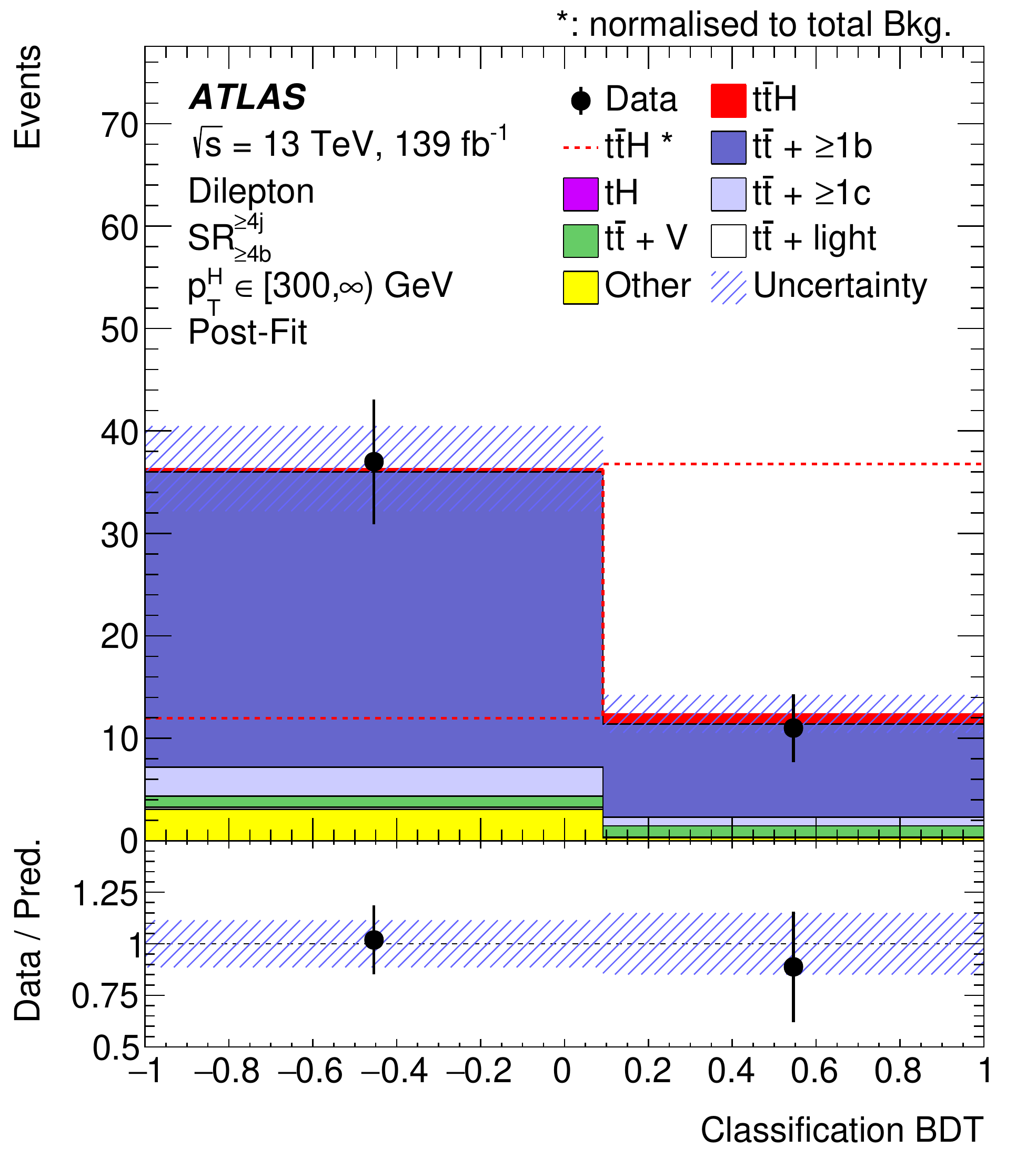}
\caption{}
\end{subfigure}
\caption{Comparison between data and prediction for the BDT
discriminant in the dilepton SRs after the inclusive fit
to the data, for (a) $0\le \ptH<120$~\GeV, (b) $120\le \ptH<200$~\GeV,
(c) $200\le \ptH<300$~\GeV\ and (d) $\ptH\ge 300$~\GeV.
The \ttH\ signal yield (solid red) is normalised to
the fitted $\mu$ value from the inclusive fit.
The dashed line shows the \ttH\
signal distribution normalised to the total background
prediction.
The uncertainty band includes all uncertainties and their correlations.}
\label{fig:results:dilep_PostFit_BDT}
\end{figure}

\begin{figure}[tp]
\centering
\begin{subfigure}[]{0.24\linewidth}
\vspace{0pt}
\includegraphics[width=\linewidth]{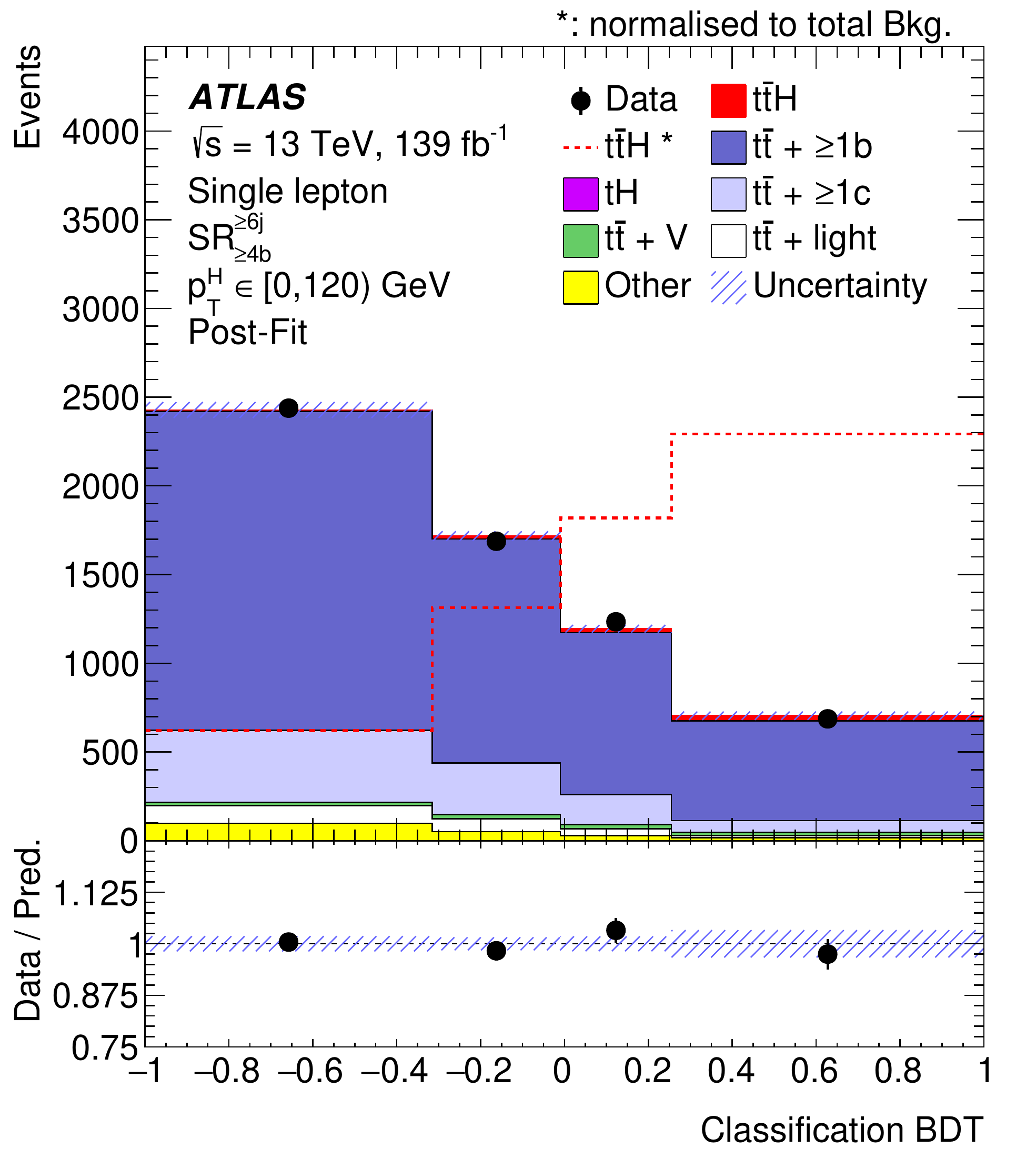}
\caption{}
\end{subfigure}
\begin{subfigure}[]{0.24\linewidth}
\vspace{0pt}
\includegraphics[width=\linewidth]{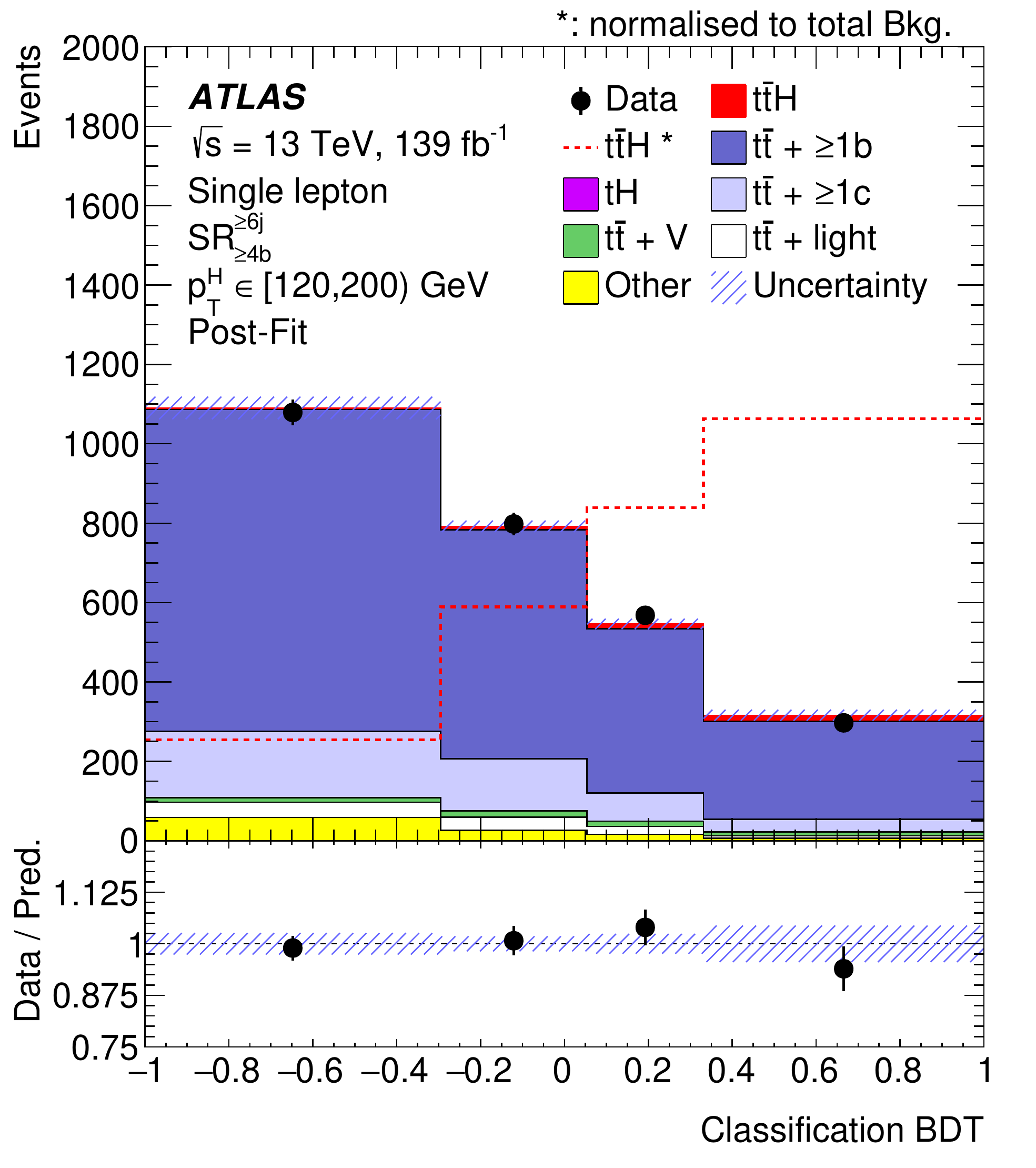}
\caption{}
\end{subfigure}
\begin{subfigure}[]{0.24\linewidth}
\vspace{0pt}
\includegraphics[width=\linewidth]{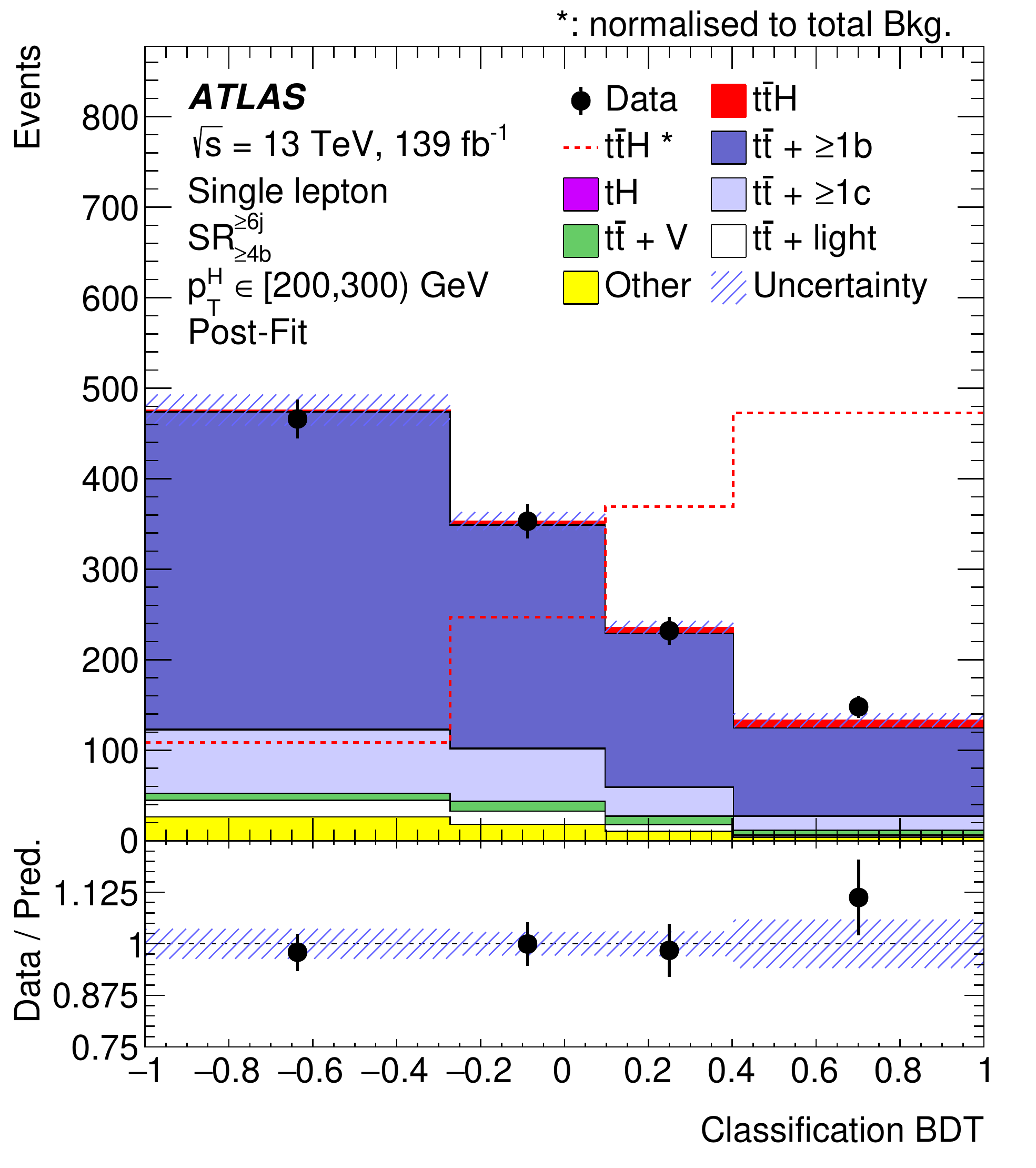}
\caption{}
\end{subfigure}
\begin{subfigure}[]{0.24\linewidth}
\vspace{0pt}
\includegraphics[width=\linewidth]{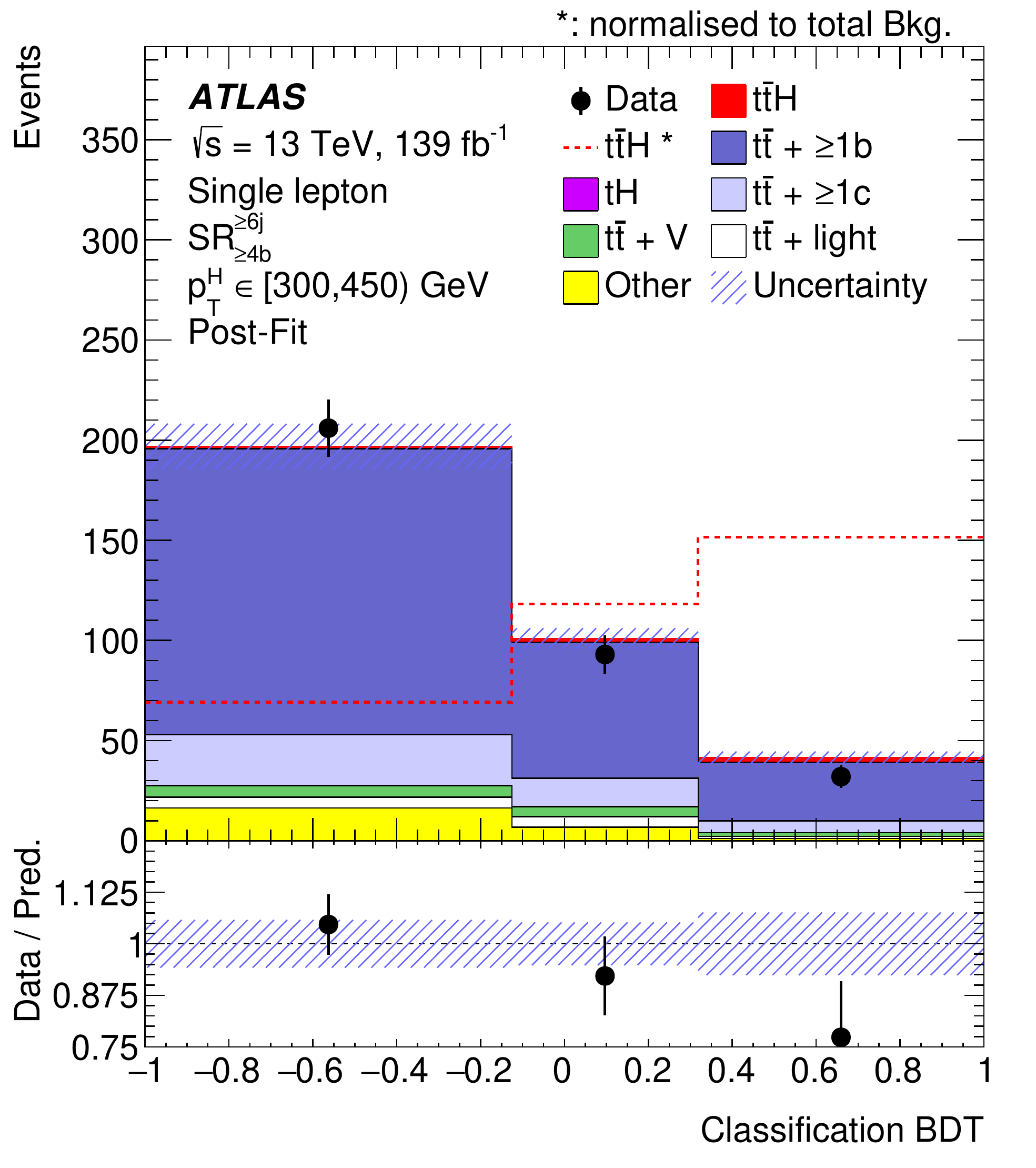}
\caption{}
\end{subfigure}
\begin{subfigure}[]{0.24\linewidth}
\vspace{0pt}
\includegraphics[width=\linewidth]{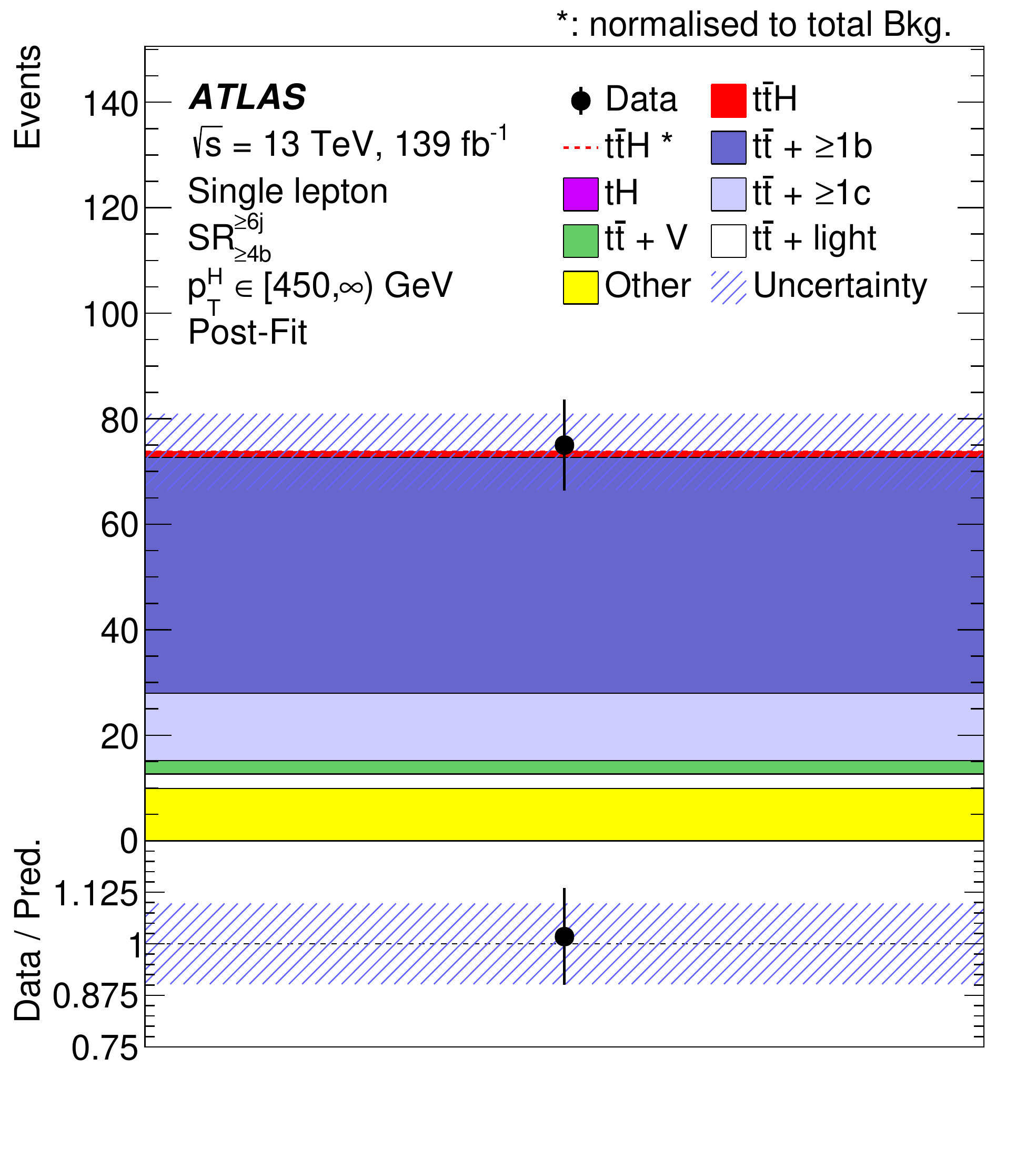}
\caption{}
\end{subfigure}
\begin{subfigure}[]{0.24\linewidth}
\vspace{0pt}
\includegraphics[width=\linewidth]{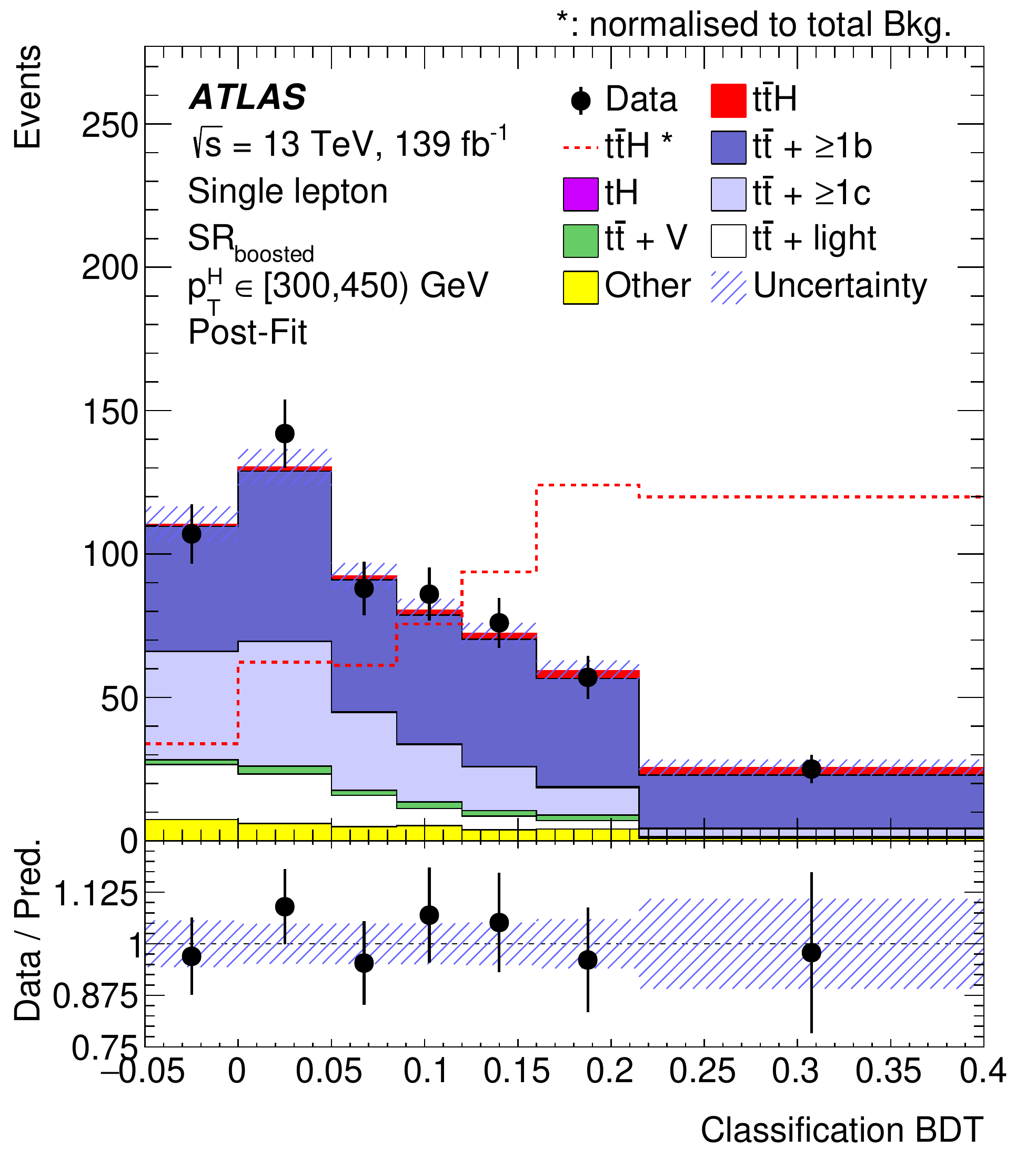}
\caption{}
\end{subfigure}
\begin{subfigure}[]{0.24\linewidth}
\vspace{0pt}
\includegraphics[width=\linewidth]{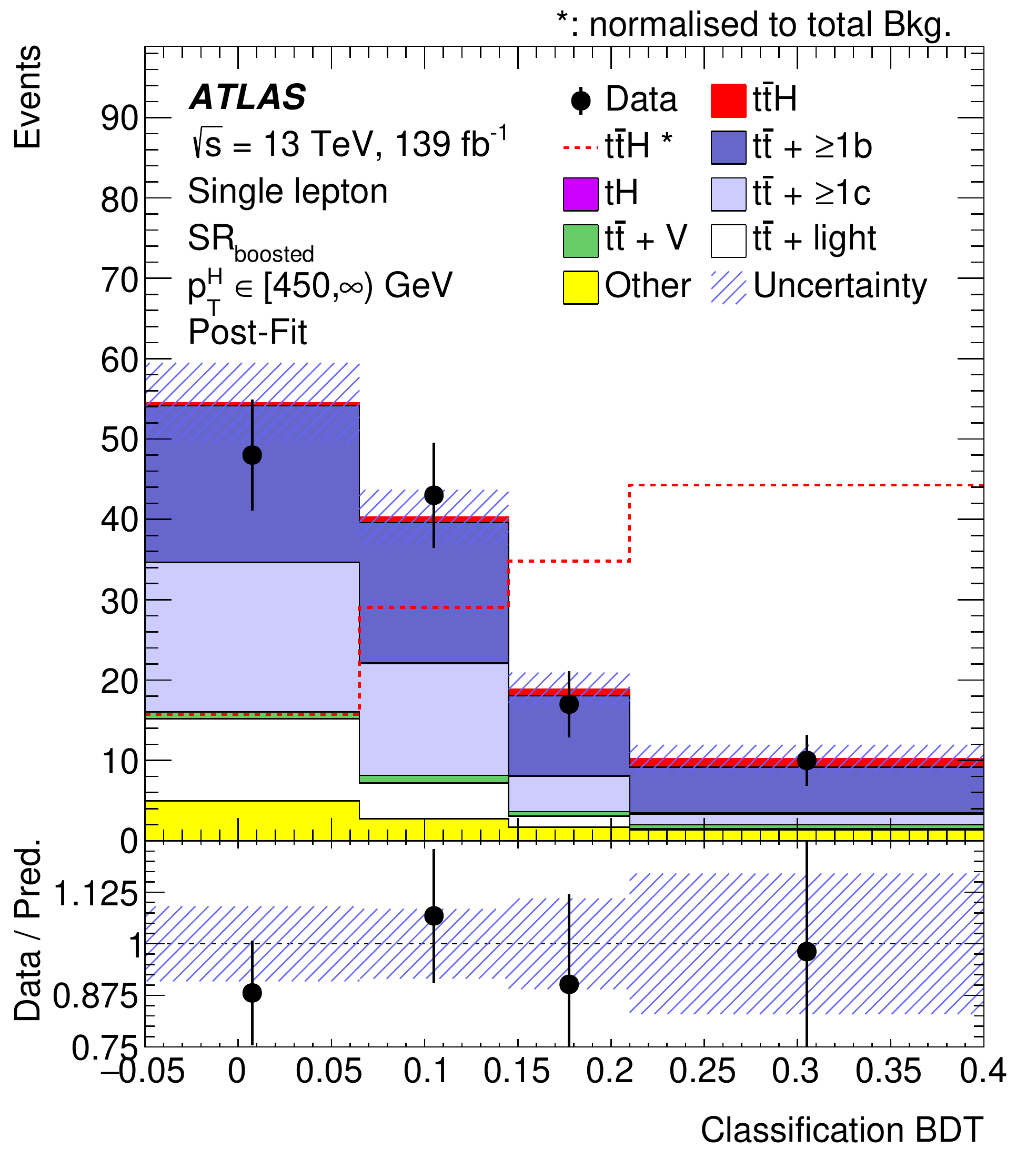}
\caption{}
\end{subfigure}
\caption{Comparison between data and prediction for the BDT
discriminant in the single-lepton SRs after the inclusive
fit to the data. The resolved channel is shown for (a) $0\le \ptH<120$~\GeV, (b) $120\le \ptH<200$~\GeV, (c) $200\le \ptH<300$~\GeV, (d) $300\le \ptH<450$~\GeV\ and (e) $\ptH\ge 450$~\GeV\ (yield only). The boosted channel is shown
for (f) $300\le \ptH<450$~\GeV\ and (g) $\ptH\ge 450$~\GeV. The \ttH\ signal yield
(solid red) is normalised to the fitted $\mu$ value from the inclusive fit. The
dashed line shows the \ttH\ signal distribution normalised to the
total background prediction.
The uncertainty band includes all uncertainties and their correlations.}
\label{fig:results:ljets_PostFit_BDT}
\end{figure}
 
\begin{figure}[tp]
\centering
\begin{subfigure}[]{0.32\linewidth}
\vspace{0pt}
\includegraphics[width=\linewidth]{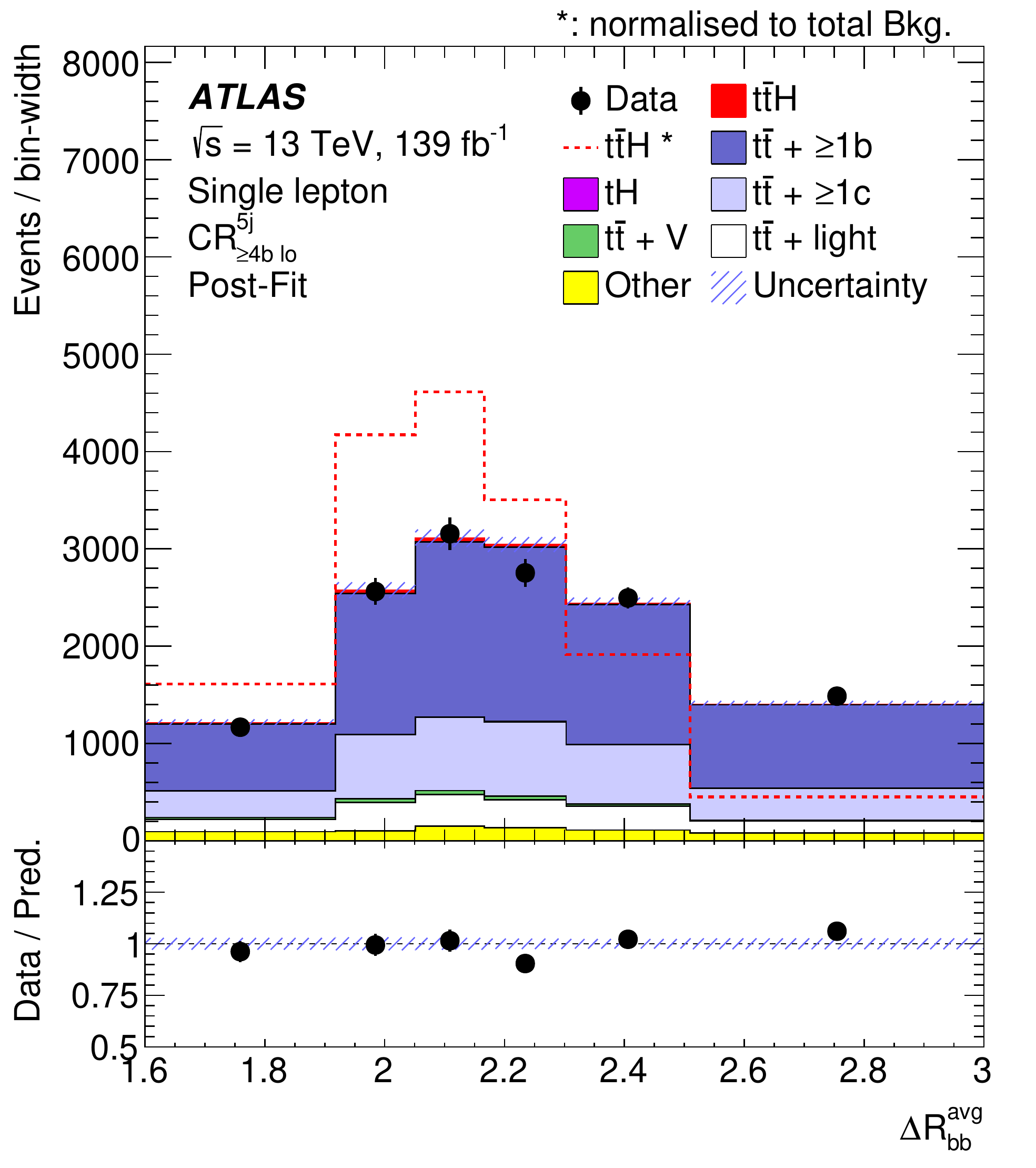}
\caption{}
\end{subfigure}
\hspace{0.5cm}
\begin{subfigure}[]{0.32\linewidth}
\vspace{0pt}
\includegraphics[width=\linewidth]{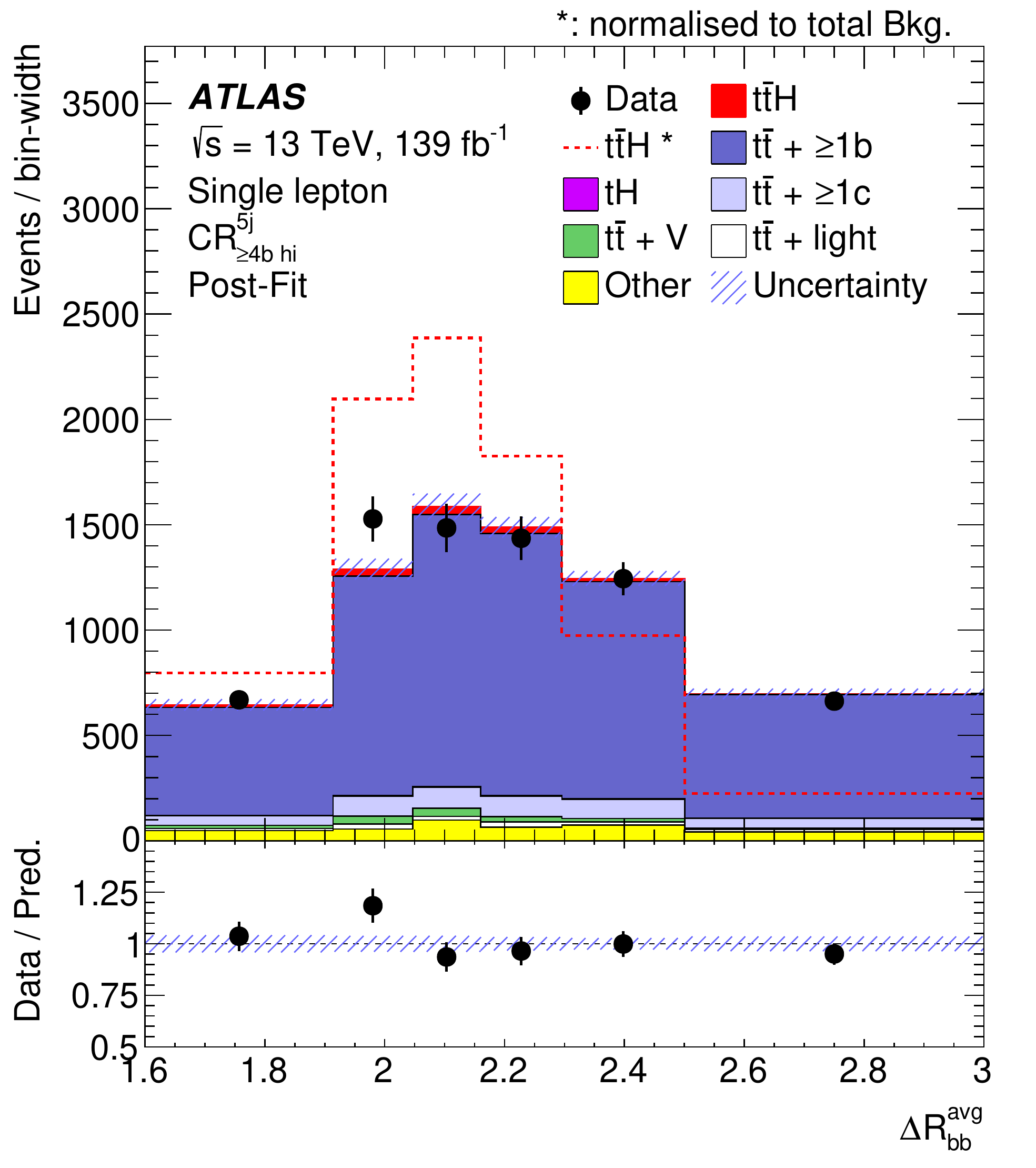}
\caption{}
\end{subfigure}
\caption{Comparison between data and prediction for \drbbav\ after the inclusive fit to the data in the single-lepton (a)
\CR[lo]{5}{4} and (b) \CR[hi]{5}{4} control regions. The \ttH\ signal yield
(solid red) is normalised to the fitted $\mu$ value from the inclusive fit. The dashed line shows the \ttH\ signal distribution normalised to the total background prediction.
The uncertainty band includes all uncertainties and their correlations. The first (last) bin includes the underflow (overflow).}
\label{fig:results:ljets_PostFit_dRbb}
\end{figure}

The statistical uncertainty is obtained by repeating the fit to data
after fixing all nuisance parameters to their post-fit values, with
the exception of the free normalisation factors in the fit:
$k(\ttbin)$ and $\mu$. The total systematic uncertainty is obtained
by subtracting the statistical variance from the total variance,
i.e. $\sigma_\text{syst} = \sqrt{\sigma_\text{tot}^2 - \sigma_\text{stat}^2}$.
The expected significance is computed from a
fit to a pseudo-dataset, built using the pulls from the nominal fit when
fixing $\mu=1$.
Figure~\ref{fig:results:logsb} shows the
event yield in data compared with the post-fit prediction for all events
entering the analysis selection, grouped and ordered by the
signal-to-background ratio of the corresponding final-discriminant
bins.
 
\begin{figure}[tp]
\centering
\includegraphics[width=0.6\textwidth]{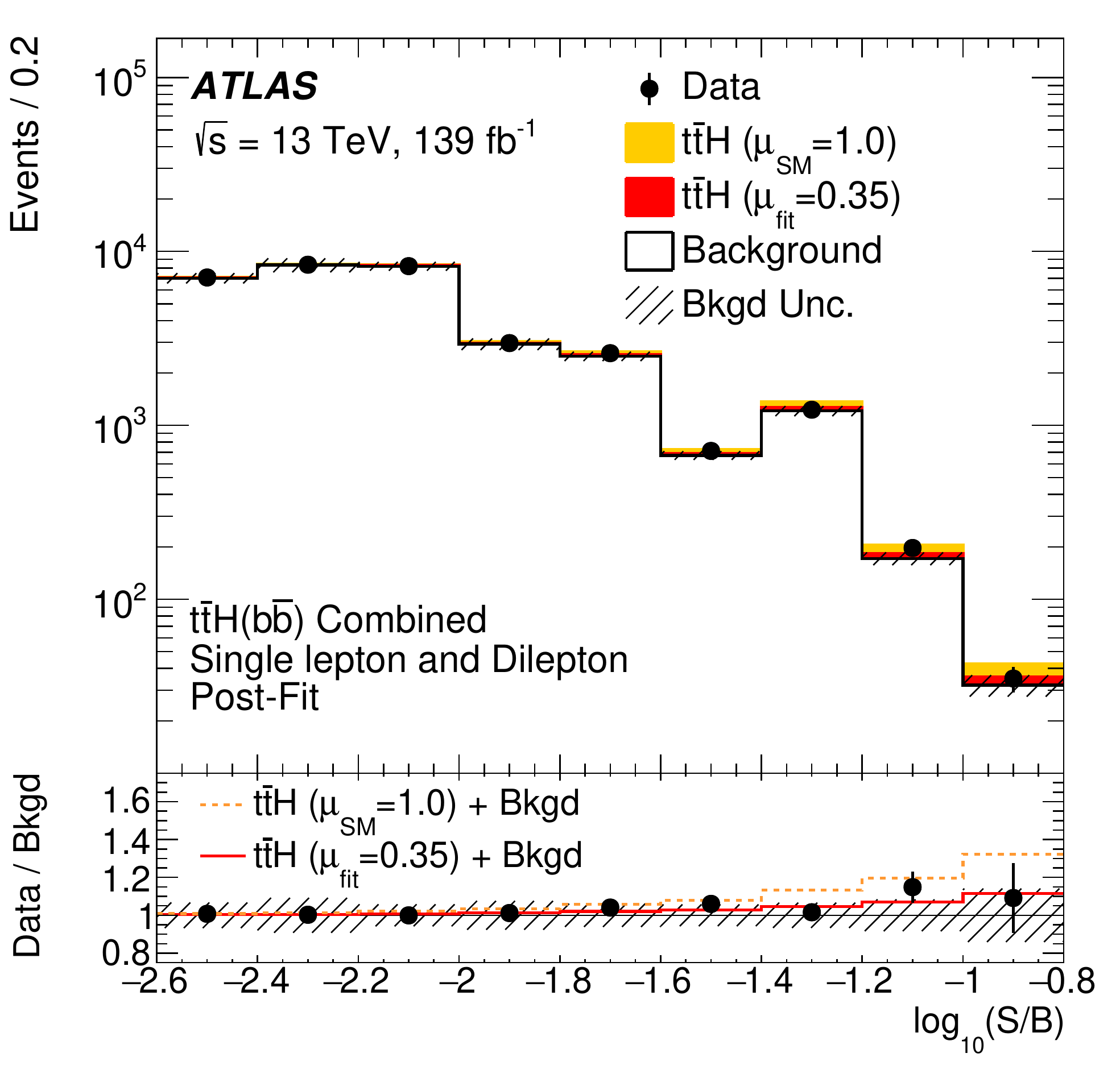}
\caption{Post-fit yields of signal ($S$) and total background ($B$)
as a function of $\log(S/B)$, compared with data.
Final-discriminant bins in all dilepton and single-lepton analysis
regions are combined into bins of $\log(S/B)$, with the signal
normalised to the SM prediction used for the computation of
$\log(S/B)$.
The signal is then shown normalised to the best-fit value and the SM prediction.
The lower frame reports the ratio of data to background, and this is compared with
the expected \ttH-signal-plus-background yield divided by the background-only yield for the
best-fit signal strength (solid red line) and the SM prediction (dashed orange line).
}
\label{fig:results:logsb}
\end{figure}
 
The global goodness of fit, including all input
variables to the classification BDTs and to a fit using the
saturated model~\cite{CousinsSaturated}, is 92\%, validating the good
post-fit modelling achieved.
Some of the most important variables used in the classification BDTs
are shown in Figure~\ref{fig:results:dilep_PostFit_inputs} for the dilepton channel,
Figure~\ref{fig:results:ljets_PostFit_inputs} for the single-lepton resolved channel
and Figure~\ref{fig:results:boosted_PostFit_pHiggs} for the single-lepton boosted channel.
The probability of obtaining a level of agreement worse than observed between the fitted predictions and data
was evaluated by calculating the $\chi^2$ for a given number of degrees of freedom and integrating
the cumulative probability distribution to $+\infty$.
For all channels combined, the mean probability is $80\%$ for classification BDT outputs
and $60\%$ for classification BDT training variables in all SRs.
Post-fit distributions of the reconstructed Higgs boson candidate
mass in the three SRs are shown in Figure~\ref{fig:results:mbb}.
 
\begin{figure}[tp]
\centering
\begin{subfigure}[]{0.4\linewidth}
\vspace{0pt}
\includegraphics[width=\linewidth]{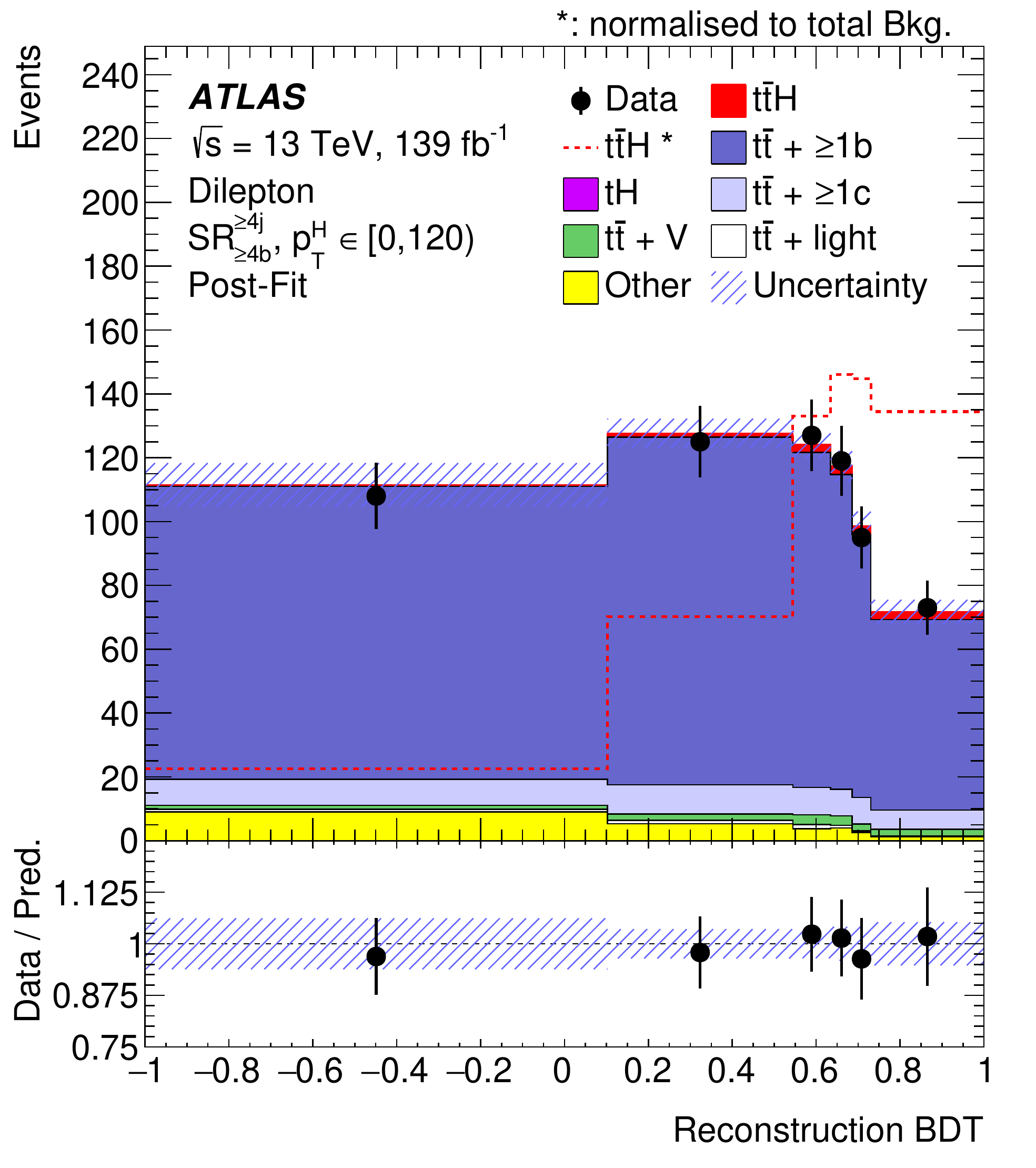}
\caption{}
\end{subfigure}
\begin{subfigure}[]{0.4\linewidth}
\vspace{0pt}
\includegraphics[width=\linewidth]{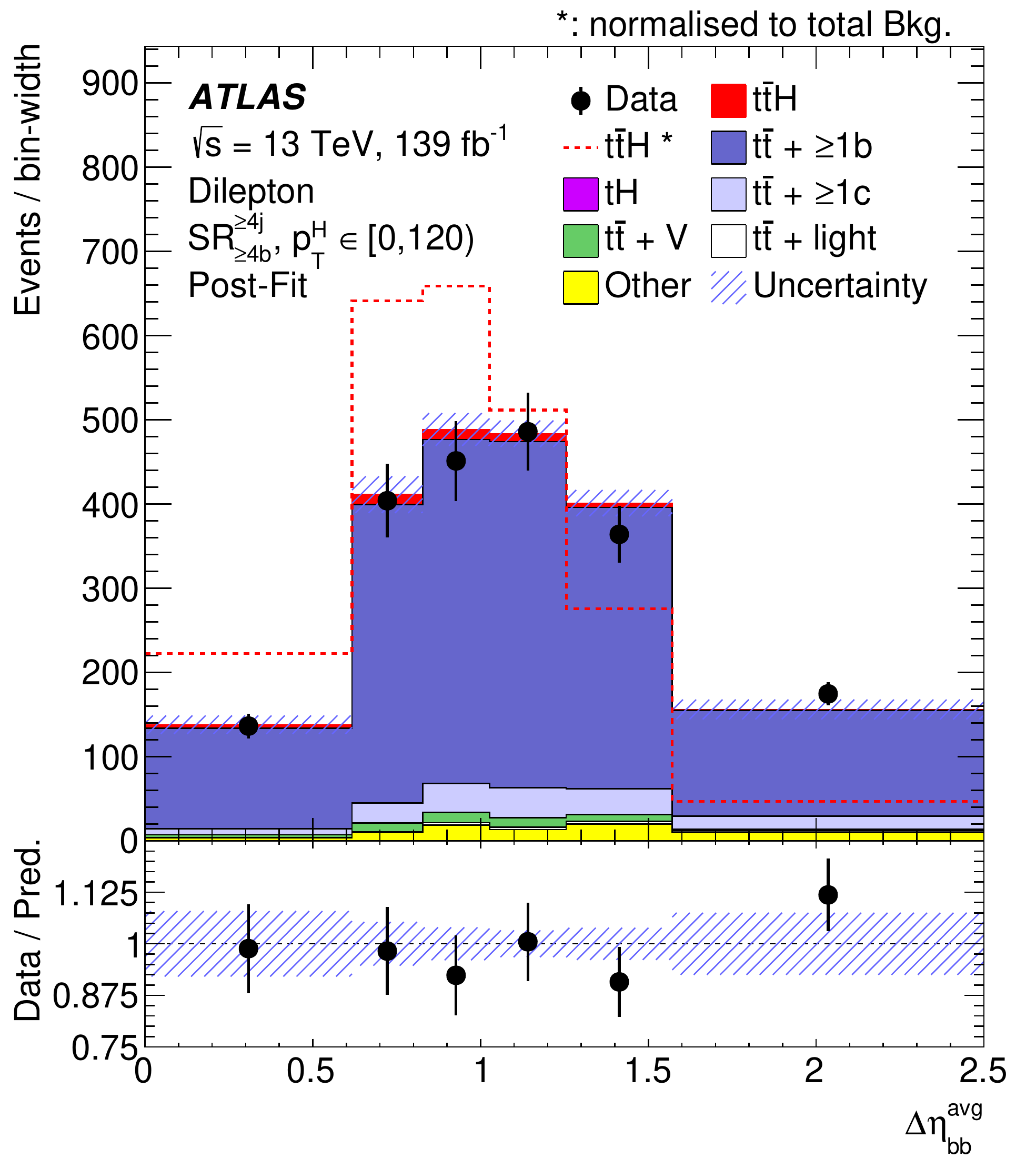}
\caption{}
\end{subfigure}
\caption{Comparison between data and prediction for (a) the reconstruction BDT score for the Higgs boson candidate identified using Higgs boson information, and (b) the average $\Delta\eta$ between $b$-tagged jets, after the inclusive fit to the data in the dilepton resolved channel for $0\le \ptH<120$~\GeV.
The \ttH\ signal yield (solid red) is normalised to the fitted $\mu$ value from the inclusive fit.
The dashed line shows the \ttH\ signal distribution normalised to the total background prediction.
The uncertainty band includes all uncertainties and their correlations.}
\label{fig:results:dilep_PostFit_inputs}
\end{figure}
 
\begin{figure}[tp]
\centering
\begin{subfigure}[]{0.4\linewidth}
\vspace{0pt}
\includegraphics[width=\linewidth]{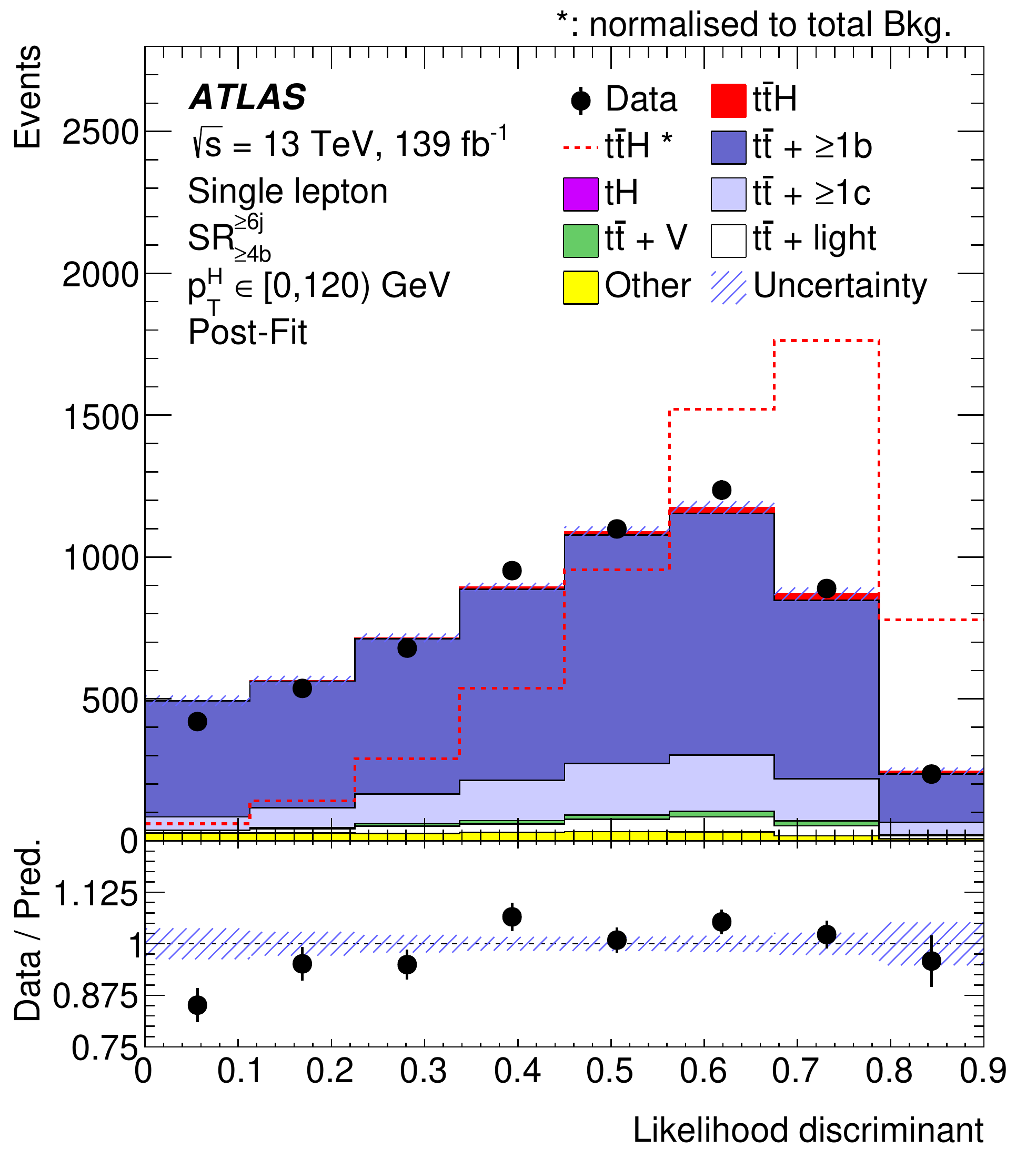}
\caption{}
\end{subfigure}
\begin{subfigure}[]{0.4\linewidth}
\vspace{0pt}
\includegraphics[width=\linewidth]{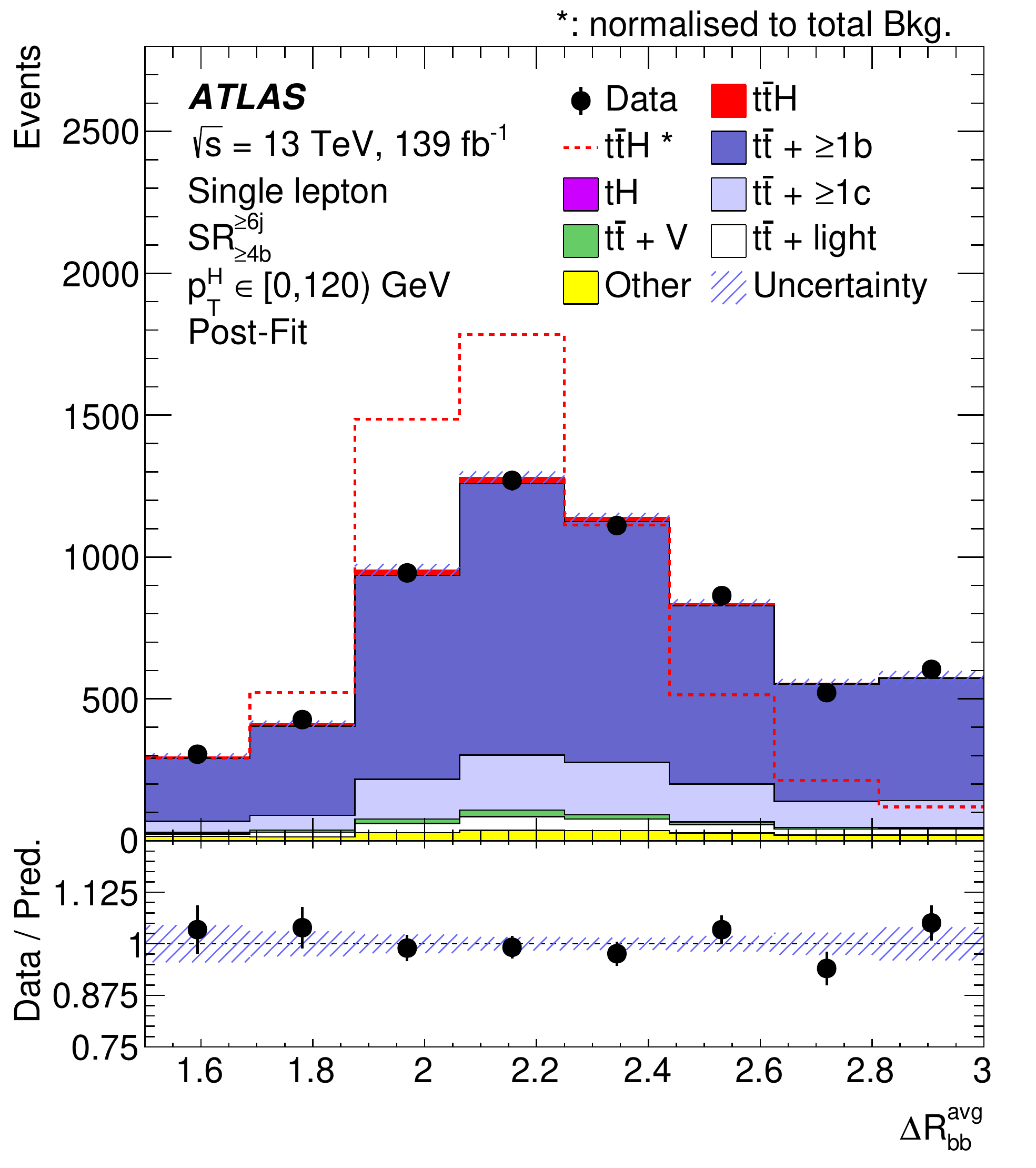}
\caption{}
\end{subfigure}
\caption{Comparison between data and prediction for (a) the likelihood discriminant, and (b) the average $\Delta R$ for all possible combinations of $b$-tagged jet pairs, after the inclusive fit to the data in the single-lepton resolved channel for $0\le \ptH<120$~\GeV.
The \ttH\ signal yield (solid red) is normalised to the fitted $\mu$ value from the inclusive fit.
The dashed line shows the \ttH\ signal distribution normalised to the total background prediction.
The uncertainty band includes all uncertainties and their correlations.}
\label{fig:results:ljets_PostFit_inputs}
\end{figure}
 
\begin{figure}[tp]
\centering
\begin{subfigure}[]{0.4\linewidth}
\vspace{0pt}
\includegraphics[width=\linewidth]{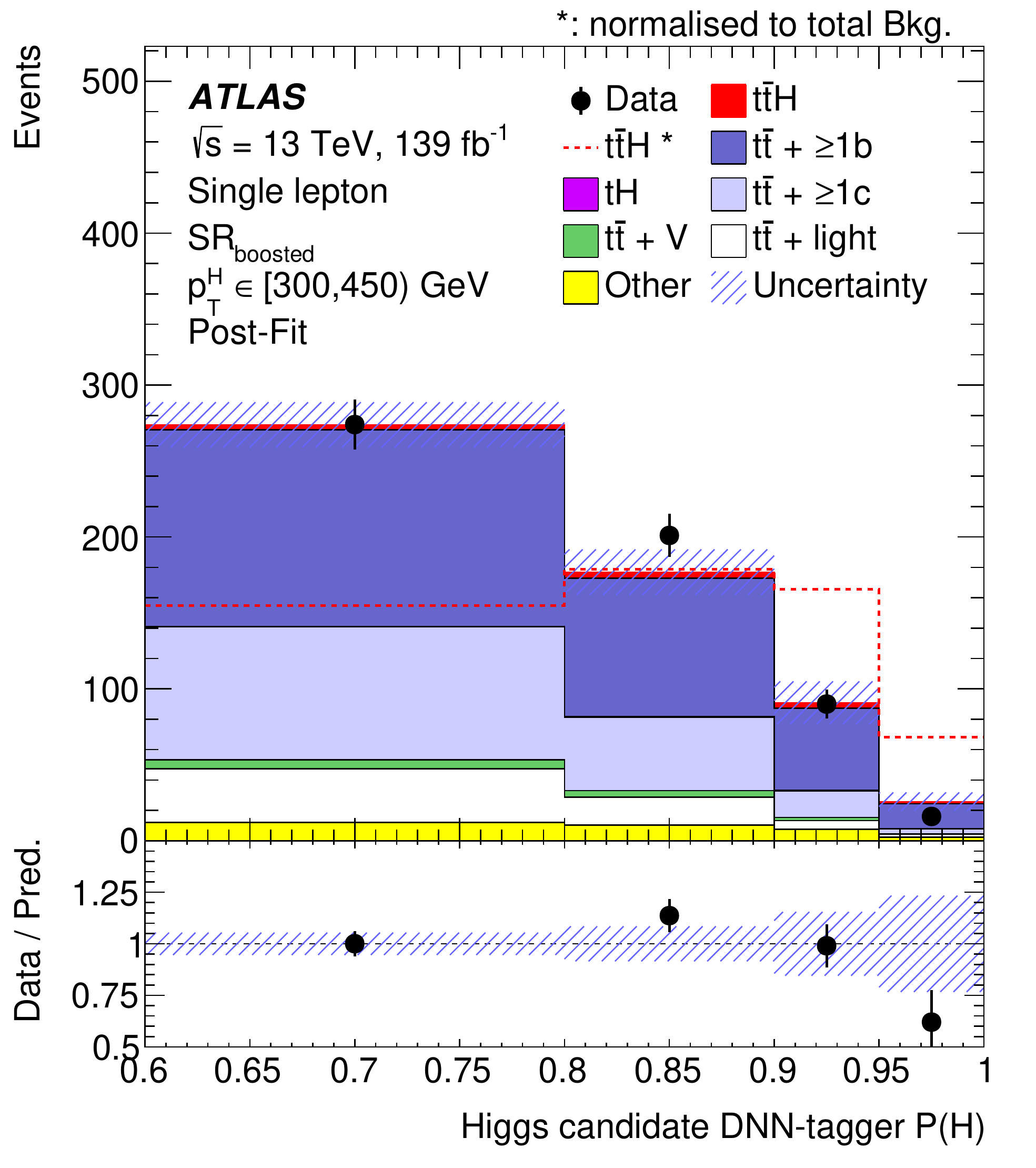}
\caption{}
\end{subfigure}
\hspace{0.5cm}
\begin{subfigure}[]{0.4\linewidth}
\vspace{0pt}
\includegraphics[width=\linewidth]{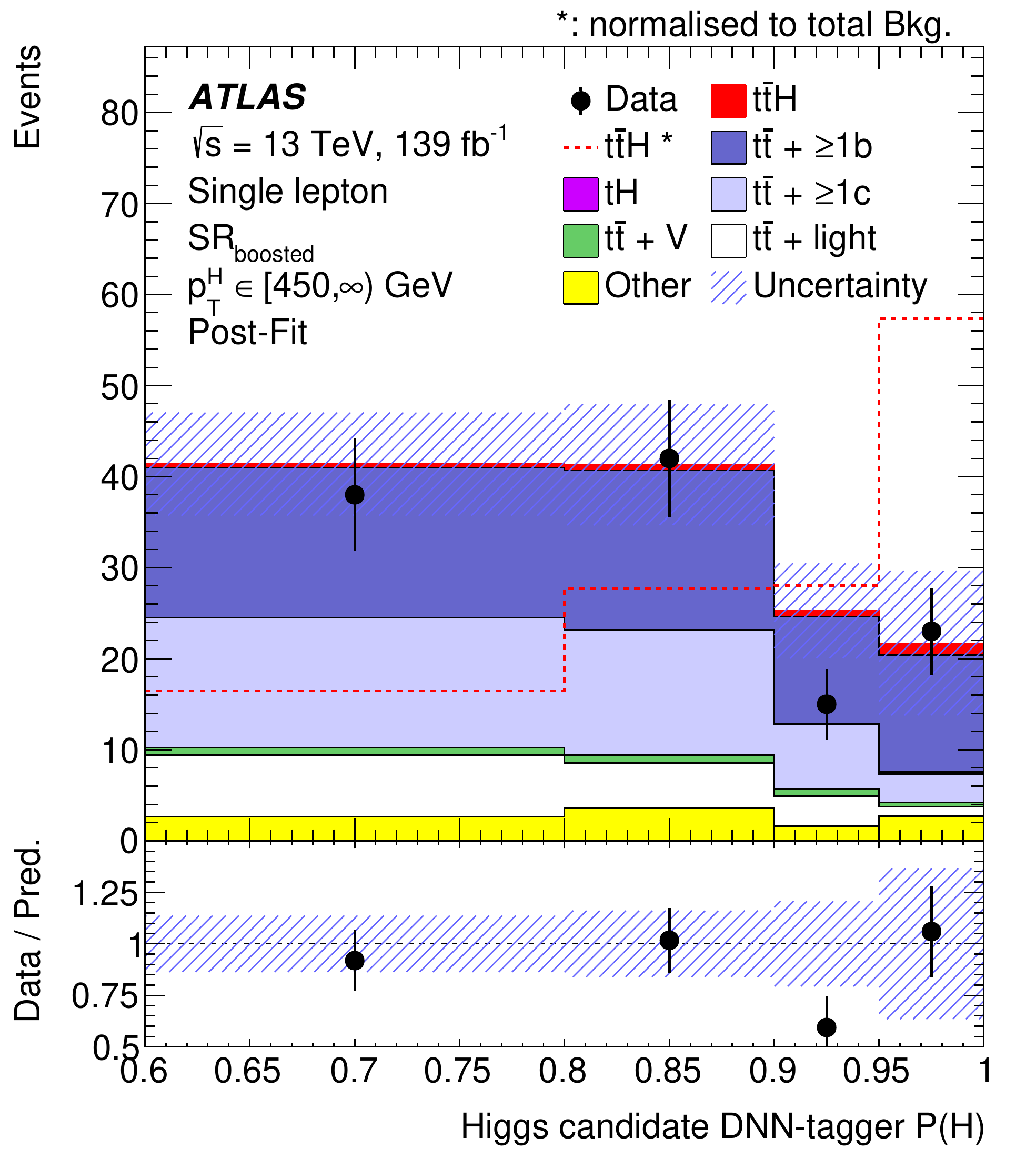}
\caption{}
\end{subfigure}
\caption{Comparison between data and prediction for the DNN $P(H)$ output for the Higgs boson candidate after the inclusive fit to the data in the single-lepton boosted channel for (a) $300\le \ptH<450$~\GeV\ and (b) $\ptH\ge 450$~\GeV. The \ttH\ signal yield
(solid red) is normalised to the fitted $\mu$ value from the inclusive fit. The dashed line shows the \ttH\ signal distribution normalised to the total background prediction.
The uncertainty band includes all uncertainties and their correlations.}
\label{fig:results:boosted_PostFit_pHiggs}
\end{figure}
 
\begin{figure}[tp]
\centering
\begin{subfigure}[]{0.32\linewidth}
\vspace{0pt}
\includegraphics[width=\linewidth]{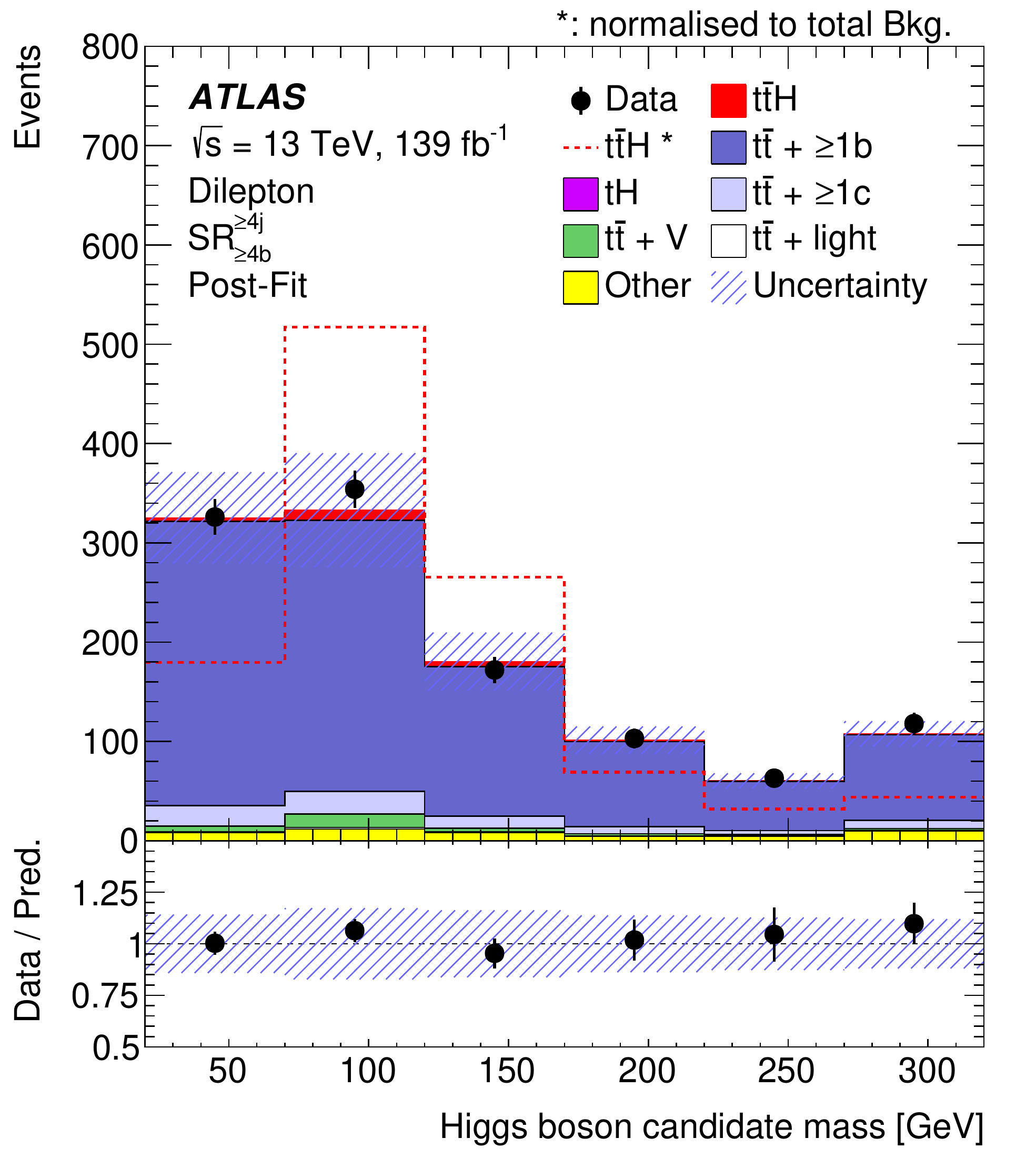}
\caption{}
\end{subfigure}
\begin{subfigure}[]{0.32\linewidth}
\vspace{0pt}
\includegraphics[width=\linewidth]{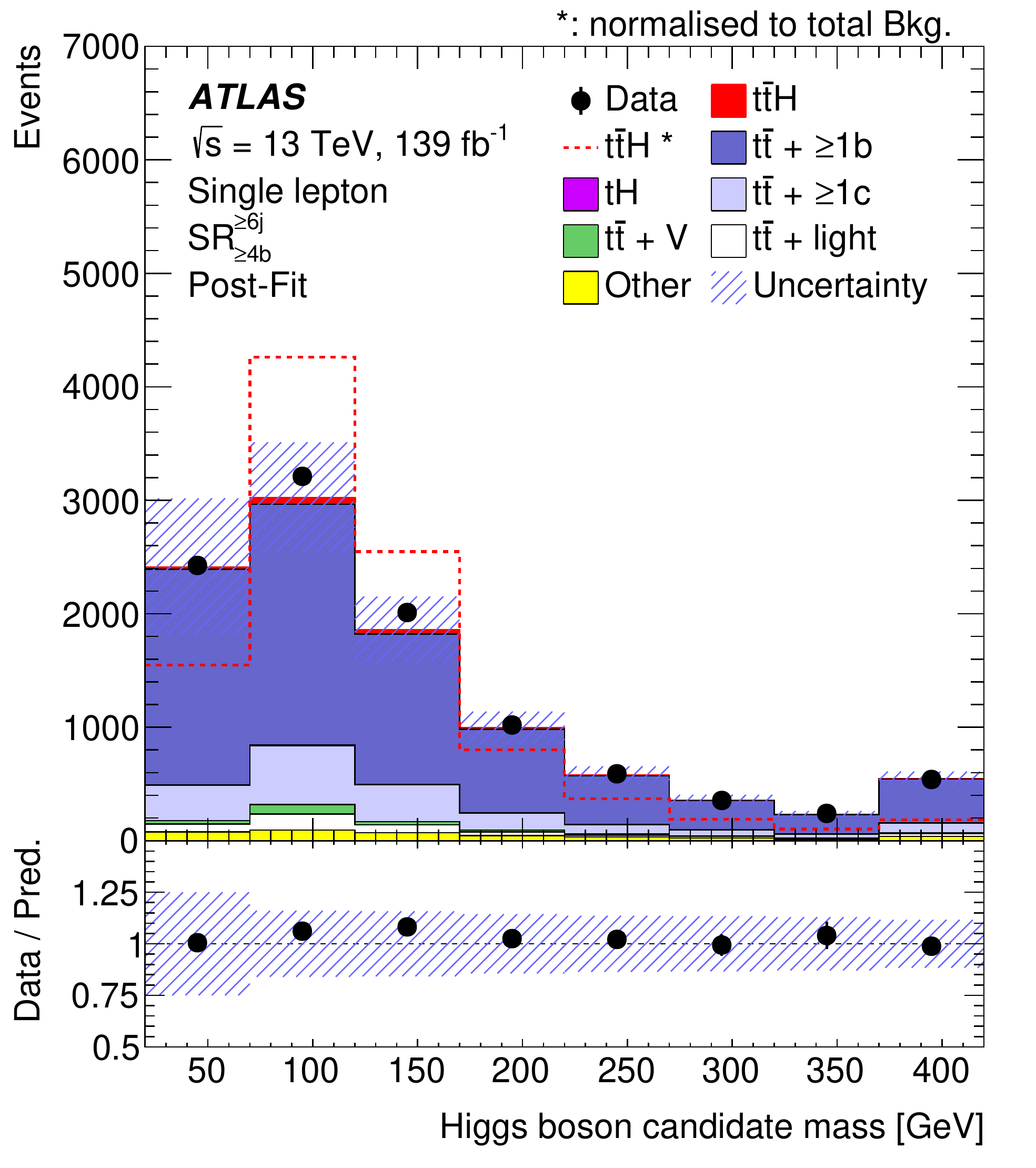}
\caption{}
\end{subfigure}
\begin{subfigure}[]{0.32\linewidth}
\vspace{0pt}
\includegraphics[width=\linewidth]{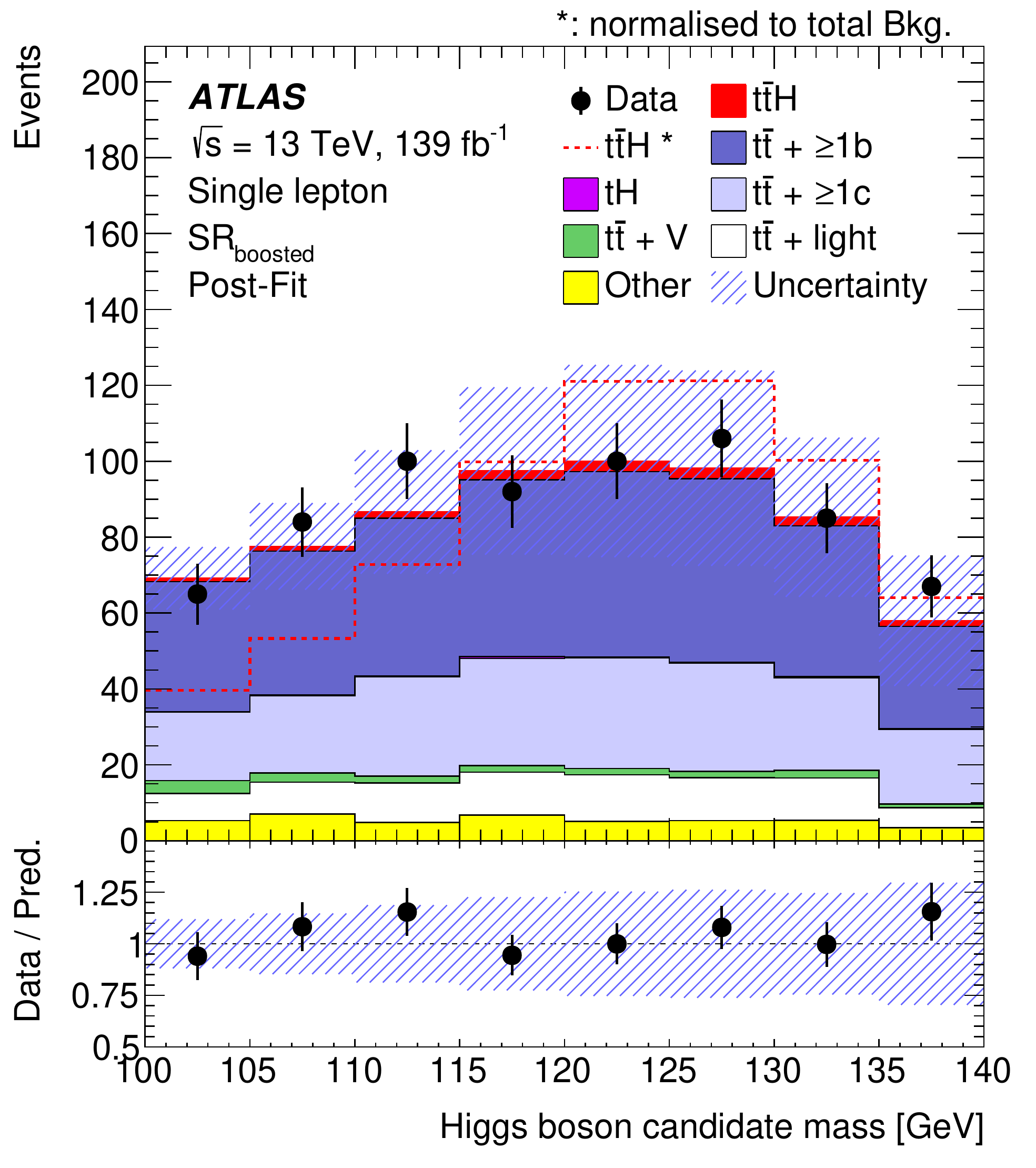}
\caption{}
\end{subfigure}
\caption{Post-fit distributions of the reconstructed Higgs boson candidate mass for the (a) dilepton \SR{4}{4}, (b) single-lepton resolved \SR{6}{4} and (c) single-lepton boosted \SRboosted\ signal regions. The \ttH\ signal yield (solid red) is normalised to the fitted $\mu$ value from the inclusive fit. The dashed line shows the \ttH\ signal distribution normalised to the total background prediction. The uncertainty band includes all uncertainties and their correlations. The first (last) bin includes the underflow (overflow).}
\label{fig:results:mbb}
\end{figure}
 
The probability of the obtained signal strength being compatible with the SM prediction is 8.5\%, estimated by redoing the fit while fixing $\mu=1$. The measured signal strength is compatible with that obtained previously~\cite{HIGG-2017-03} from part of the dataset,
and the impact of systematic uncertainties has been reduced by about a factor of two. The main improvements come from improved theoretical knowledge in \ttbin\ modelling, from the much larger size of simulated event samples for systematic uncertainty estimation as well as from the refined $b$-tagging scale factors and jet energy scale and resolution measurements.
 
The fit is repeated with three independent signal strengths, one each for the dilepton, single-lepton resolved and single-lepton boosted channels. Figure~\ref{fig:results:mu} shows the $\mu$ value obtained for each channel and the single signal strength from the previous fit. In this fit the normalisation factor for the \ttbin\ background is found to be $k(\ttbin) = 1.27 \pm 0.08$, compatible with the value in the single $\mu$ fit. The probability of obtaining a discrepancy between these three $\mu$ values equal to or larger than the one observed is 90\%.
 
\begin{figure}[tp]
\centering
\includegraphics[width=0.6\textwidth]{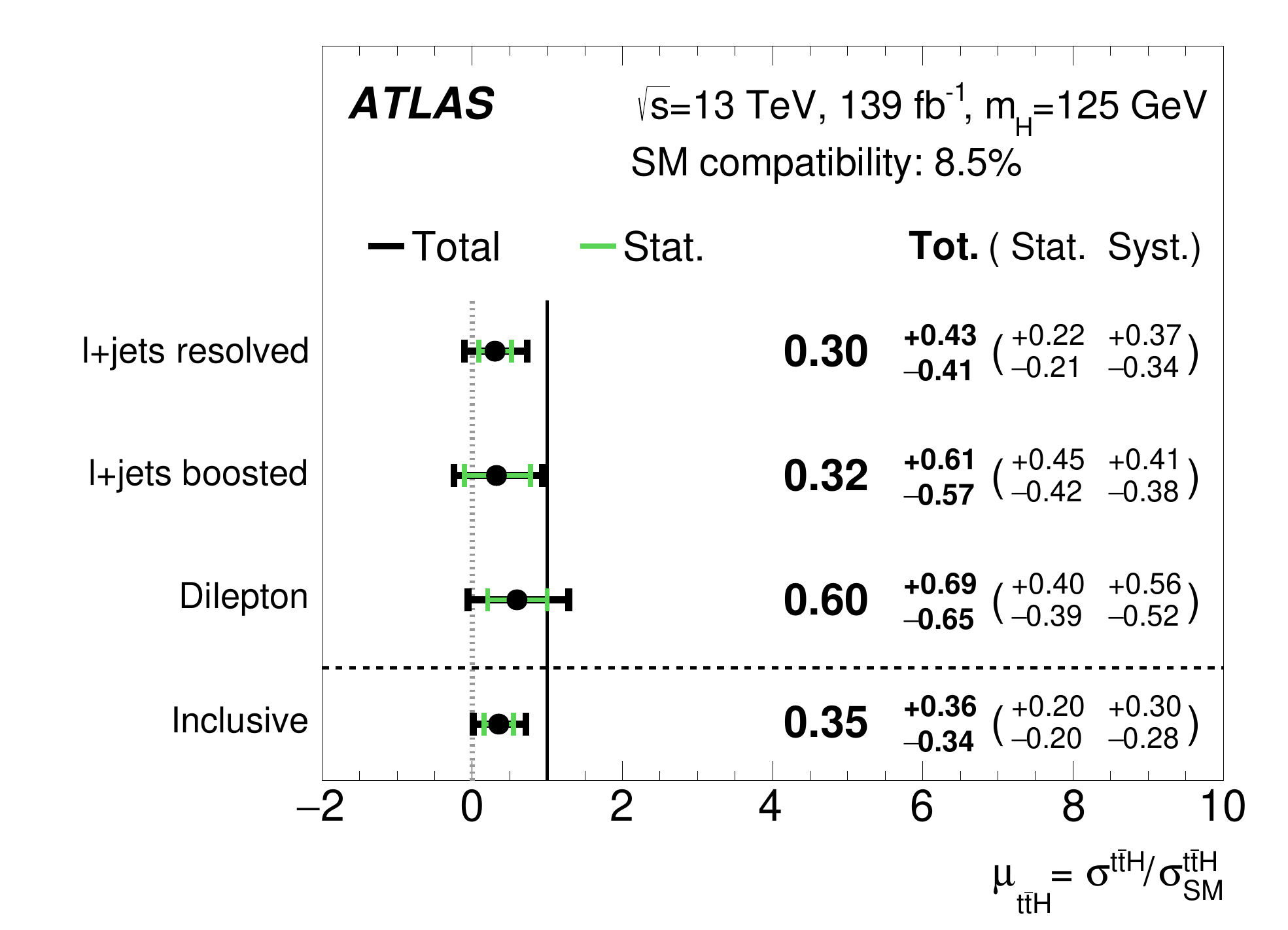}
\caption{Fitted values of the \ttH signal-strength parameter in the individual channels and in the inclusive signal strength measurement.}
\label{fig:results:mu}
\end{figure}

The measurement is largely dominated by systematic
uncertainties. Their contributions to the fit to $\mu$ are
reported in Table~\ref{tab:results:syst}. The dominant impact comes
from the modelling of the \ttbin\ background, followed by the signal modelling, $tW$ modelling and
$b$-tagging efficiency uncertainties. The largest
observed pull on systematic uncertainties, 
as shown in Figure~\ref{fig:Ranking},
is seen in the \ttbin\ ISR uncertainty, and is about $1.2\sigma$, mostly driven by the renormalisation scale. This pull indicates that the data favours a softer renormalisation scale in the ME calculation, as also suggested in Ref.~\cite{Buccioni:2019plc} where a lower scale in $t\bar{t}b\bar{b}$ calculations gives better agreement with $t\bar{t}b\bar{b}j$ calculations.
This effect was shown to not affect the BDT shapes in each individual region, while correcting a mismodelling in the distribution of the number of jets in the event (by adjusting the amount of additional radiation), which affects the categorisation of events. The distributions of the number of jets in the three SRs are shown in Figure~\ref{fig:results:njet} pre-fit and post-fit. Decorrelating this uncertainty between the dilepton and single-lepton channels leads to very similar fitted $\mu$ values and nuisance parameter pulls.
 
Another large pull is on the reconstructed \ptbb\ shape uncertainty in the \ttbin background, as expected from the pre-fit mismodelling (see Figure~\ref{fig:systematics:hpt}) and how this uncertainty is defined: a $+1\sigma$ variation corresponds to correcting the reconstructed \ptH\ shape such that it agrees between data and the background model (see Section~\ref{sec:Systematics:modelling}). The sensitivity of the result to this uncertainty was tested by replacing the data-driven mismodelling with decorrelated free-floating \ttbin normalisation factors across the STXS bins and analysis regions, and no bias was observed on the fitted signal strength. The reconstructed \ptH distributions display good post-fit agreement, as shown in Figure~\ref{fig:results:hpt} (to be compared with the pre-fit discrepancies shown in Figure~\ref{fig:systematics:hpt}).
 
Additionally, a large pull is observed in the \ttbin\ NLO matching uncertainty in the dilepton channel in the $0 \leq \ptH < 120$\,\GeV\ bin, as shown in Figure~\ref{fig:Ranking} where it is ranked sixth for the impact on this measurement.
Due to the large number of nuisance parameters entering into the likelihood fit, there is a non-negligible probability that any one of the \ttbin\ NLO matching uncertainties could be pulled to a value at least as extreme as that observed from a purely statistical standpoint. The probability is calculated to be 17\%. 
Despite the large differences between models (see Section~\ref{sec:Systematics:modelling}), the varying \ttbin composition in the CRs and SRs allows the \ttbin\ subcomponent fraction systematic uncertainty to be constrained. In general the fit mostly constrains the \ttbin modelling uncertainties and the normalisation of the \ttcin background, which is pulled to $0.6\sigma$ and agrees very well with the previous publication~\cite{HIGG-2017-03}.
To further validate the robustness of the fit, a pseudo-data sample was built from simulated events by replacing the nominal \ttbin\ background with the alternative \ttbb\ MC sample generated with \SHERPA[2.2.1]. The fit to this pseudo-data sample did not reveal any significant bias in the signal extraction, the signal strength being compatible with unity within uncertainties.

\begin{table}[tp]
\begin{center}
\sisetup{retain-explicit-plus,round-precision=2,round-mode=places}
\begin{tabular}{l S S}
\toprule
\midrule
Uncertainty source & \multicolumn{2}{c}{$\Delta\mu$} \\
\midrule
Process modelling & & \\
\hspace{1em} \ttH modelling             & +0.130 & -0.054 \\
\hspace{1em} \ttbin modelling & & \\
\hspace{2em} \ttbin NLO matching        & +0.207 & -0.203 \\
\hspace{2em} \ttbin fractions           & +0.115 & -0.123 \\
\hspace{2em} \ttbin FSR                 & +0.104 & -0.107 \\
\hspace{2em} \ttbin PS \& hadronisation & +0.086 & -0.083 \\
\hspace{2em} \ttbin \ptbb\ shape        & +0.040 & -0.037 \\
\hspace{2em} \ttbin ISR                 & +0.036 & -0.039 \\
\hspace{1em} \ttcin modelling           & +0.032 & -0.039 \\
\hspace{1em} \ttlight modelling         & +0.027 & -0.028 \\
\hspace{1em} \tW modelling              & +0.079 & -0.074 \\
\hspace{1em} Background-model statistical uncertainty & +0.044 & -0.045 \\
$b$-tagging efficiency and mis-tag rates & & \\
\hspace{1em} $b$-tagging efficiency & +0.034 &-0.024 \\
\hspace{1em} $c$-mis-tag rates      & +0.034 &-0.031 \\
\hspace{1em} $l$-mis-tag rates      & +0.022 &-0.022 \\
Jet energy scale and resolution & & \\
\hspace{1em} $b$-jet energy scale         & +0.004 & -0.006 \\
\hspace{1em} Jet energy scale (flavour)   & +0.012 & -0.011 \\
\hspace{1em} Jet energy scale (pile-up)    & +0.004 & -0.008 \\
\hspace{1em} Jet energy scale (remaining) & +0.012 & -0.014 \\
\hspace{1em} Jet energy resolution        & +0.020 & -0.019 \\
Luminosity    & +0.011 & -0.002 \\
Other sources & +0.029 & -0.027 \\
\midrule
Total systematic uncertainty & +0.299 & -0.276 \\
\midrule
\midrule
\ttbin normalisation & +0.042 & -0.067 \\
\midrule
Total statistical uncertainty & +0.199  & -0.197 \\
\midrule
\midrule
Total uncertainty & +0.359 & -0.339\\
\midrule
\bottomrule
\end{tabular}
\end{center}
\caption{Breakdown of the contributions to the
uncertainties in $\mu$.  The contributions from the different sources
of uncertainty are evaluated after the fit.
The $\Delta\mu$ values are
obtained by repeating the fit after having fixed a certain set of
nuisance parameters corresponding to a group of systematic
uncertainties, and then evaluating $(\Delta\mu)^2$ by subtracting the resulting
squared uncertainty of $\mu$ from its squared uncertainty found in the full fit.
The same procedure is followed when quoting the effect of the \ttbin\
normalisation. The total uncertainty is different from the sum in
quadrature of the different components due to correlations between
nuisance parameters existing in the fit.}
\label{tab:results:syst}
\end{table}

\begin{figure}[tp]
\centering
\includegraphics[width=0.5\textwidth]{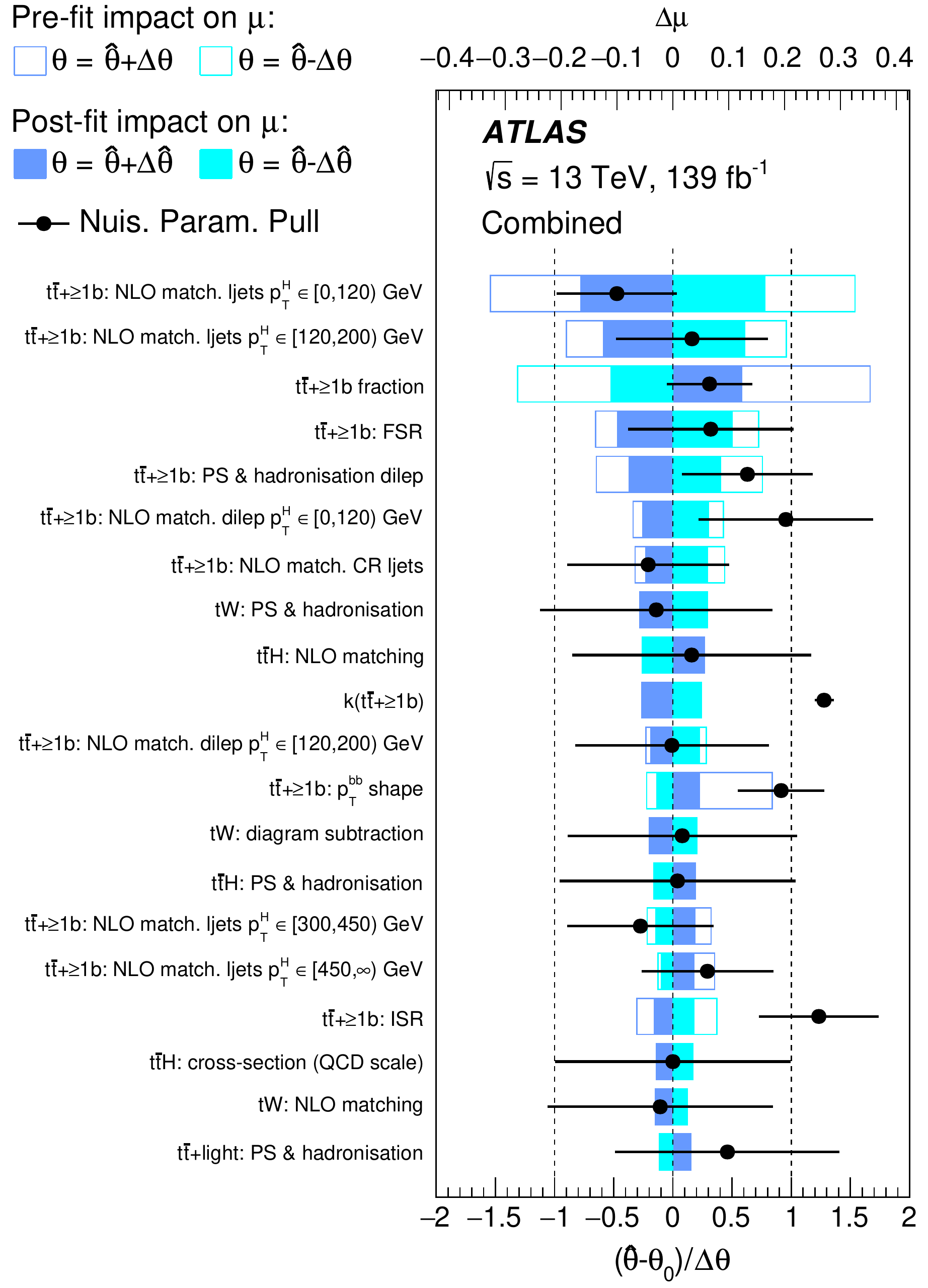}
\caption{Ranking of the 20 nuisance parameters with the largest
post-fit impact on $\mu$ in the fit. Nuisance parameters
corresponding to statistical uncertainties in the simulated event samples are not
included. The empty blue rectangles correspond to the pre-fit
impact on $\mu$ and the filled blue ones to the post-fit impact on
$\mu$, both referring to the upper scale.  The impact of each
nuisance parameter, $\Delta\mu$, is computed by comparing the
nominal best-fit value of $\mu$ with the result of the fit when
fixing the considered nuisance parameter to its best-fit value,
$\hat{\theta}$, shifted by its pre-fit (post-fit) uncertainties
$\pm\Delta\theta$ ($\pm\Delta\hat{\theta}$).  The black points
show the pulls of the nuisance parameters relative to their
nominal values, $\theta_0$.  These pulls and their relative
post-fit errors, $\Delta\hat{\theta}/\Delta\theta$, refer to the
lower scale. 
The `ljets' (`dilep') label refers to the single-lepton (dilepton) channel.}
\label{fig:Ranking}
\end{figure}

\begin{figure}[tp]
\centering
\begin{subfigure}[]{0.32\linewidth}
\vspace{0pt}
\includegraphics[width=\linewidth]{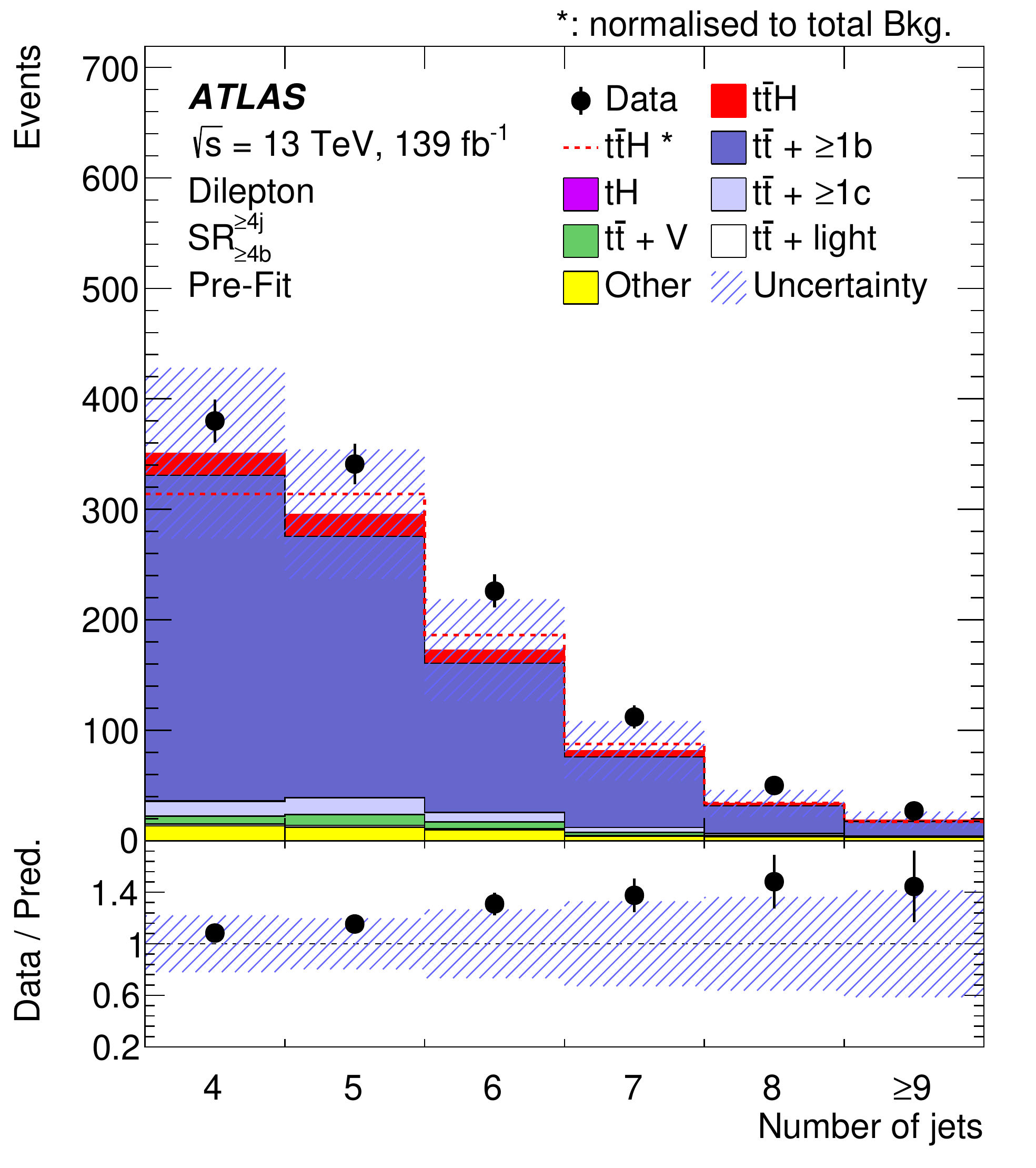}
\caption{}
\end{subfigure}
\begin{subfigure}[]{0.32\linewidth}
\vspace{0pt}
\includegraphics[width=\linewidth]{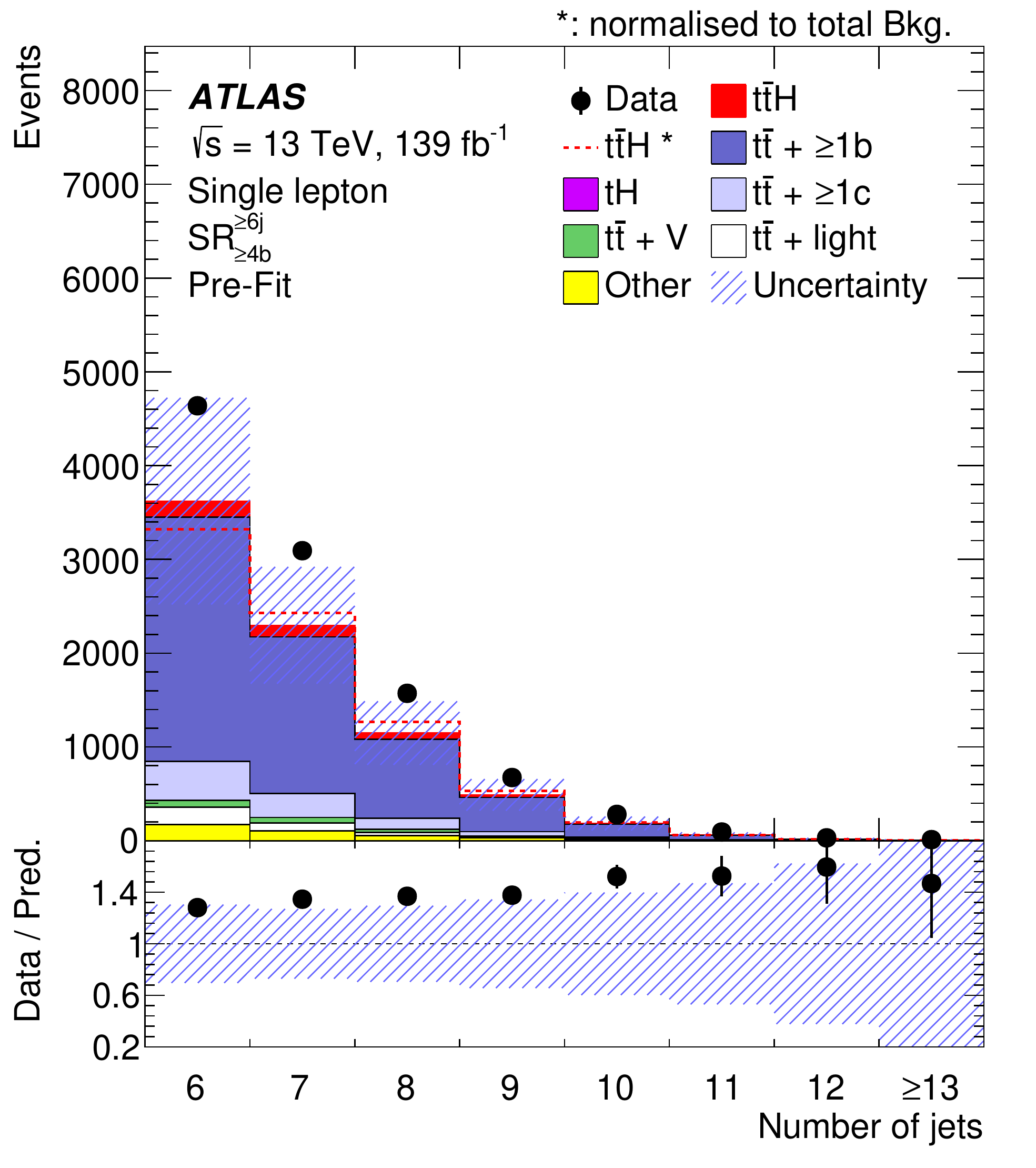}
\caption{}
\end{subfigure}
\begin{subfigure}[]{0.32\linewidth}
\vspace{0pt}
\includegraphics[width=\linewidth]{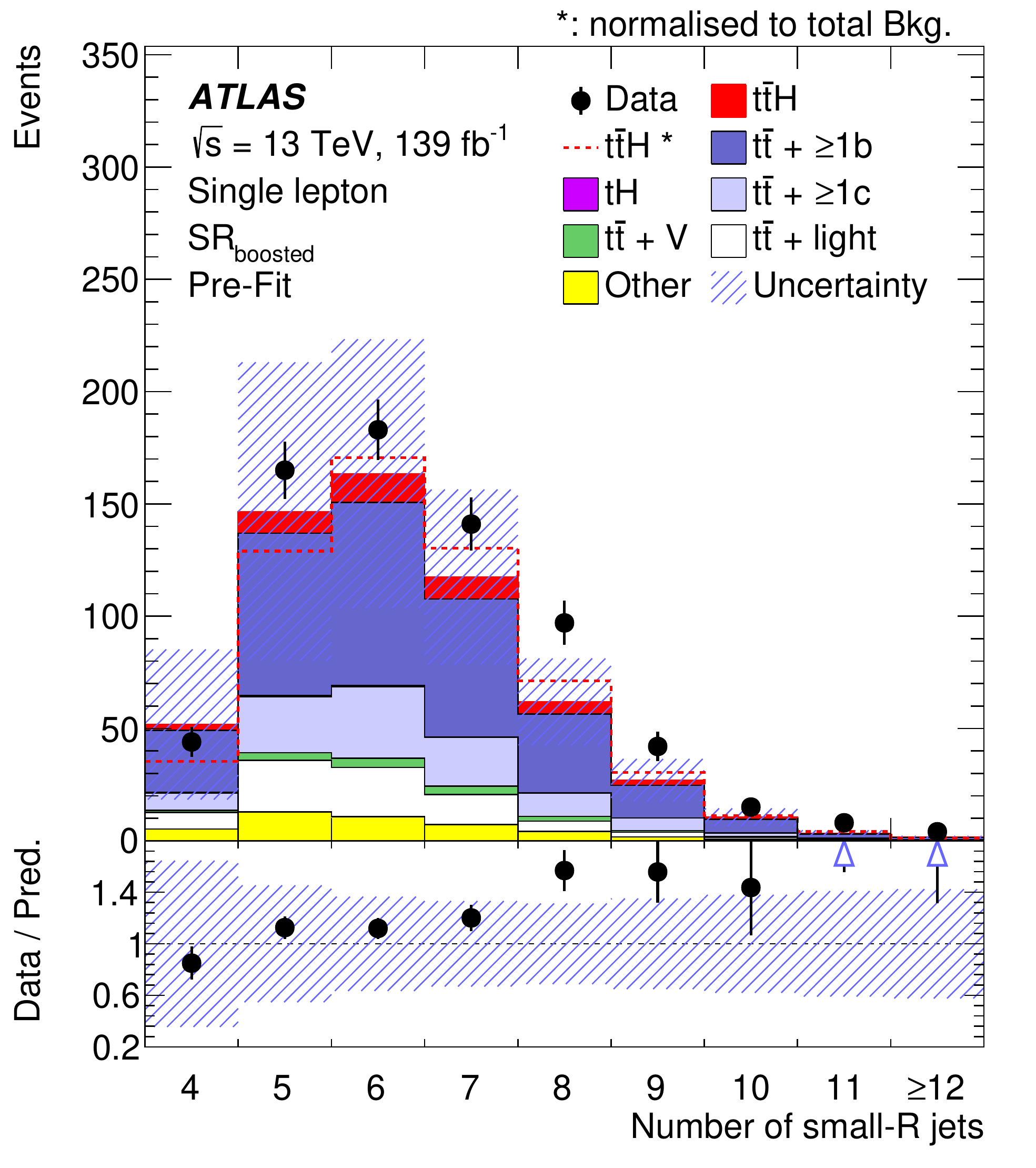}
\caption{}
\end{subfigure}
 
\begin{subfigure}[]{0.32\linewidth}
\vspace{0pt}
\includegraphics[width=\linewidth]{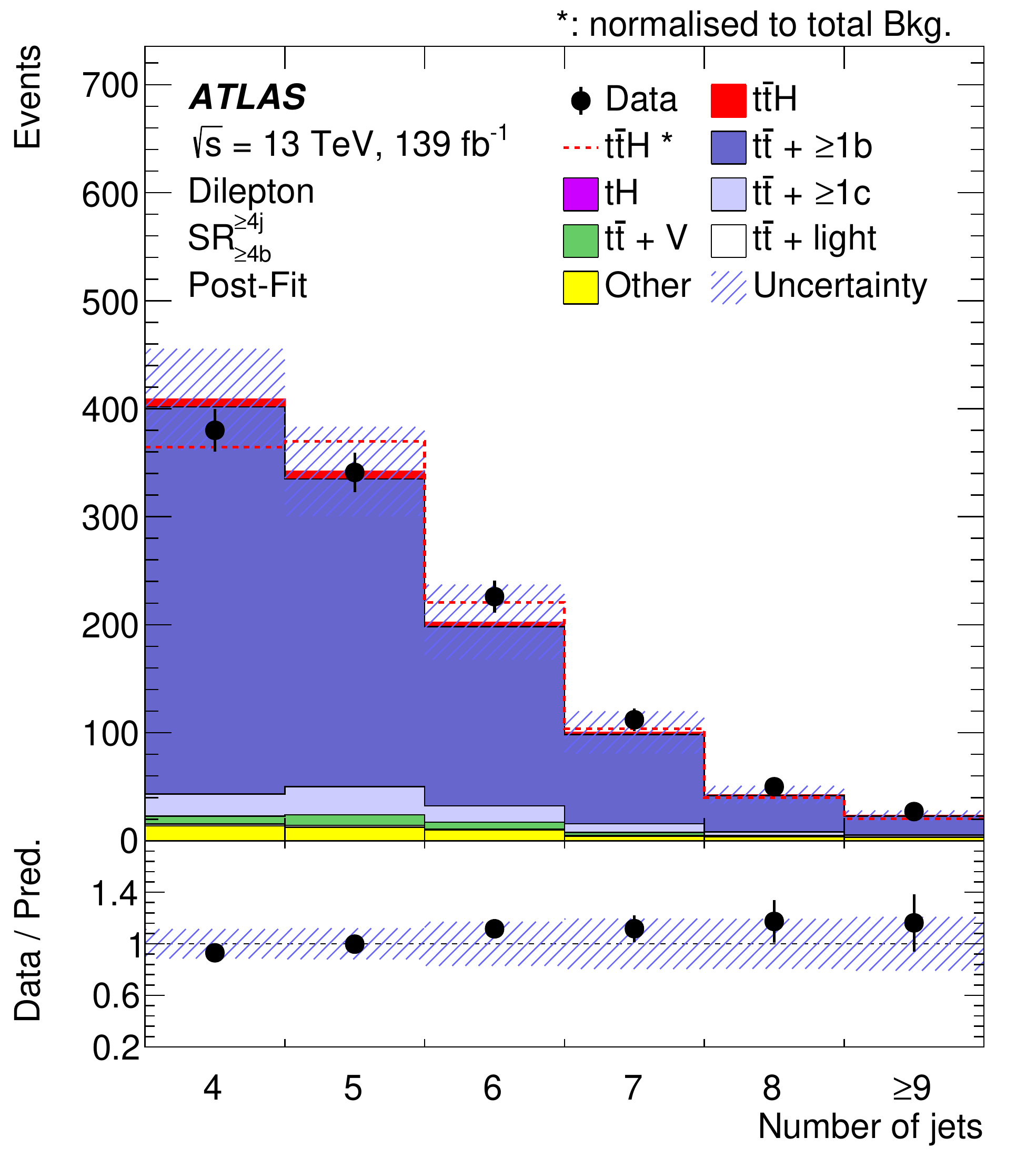}
\caption{}
\end{subfigure}
\begin{subfigure}[]{0.32\linewidth}
\vspace{0pt}
\includegraphics[width=\linewidth]{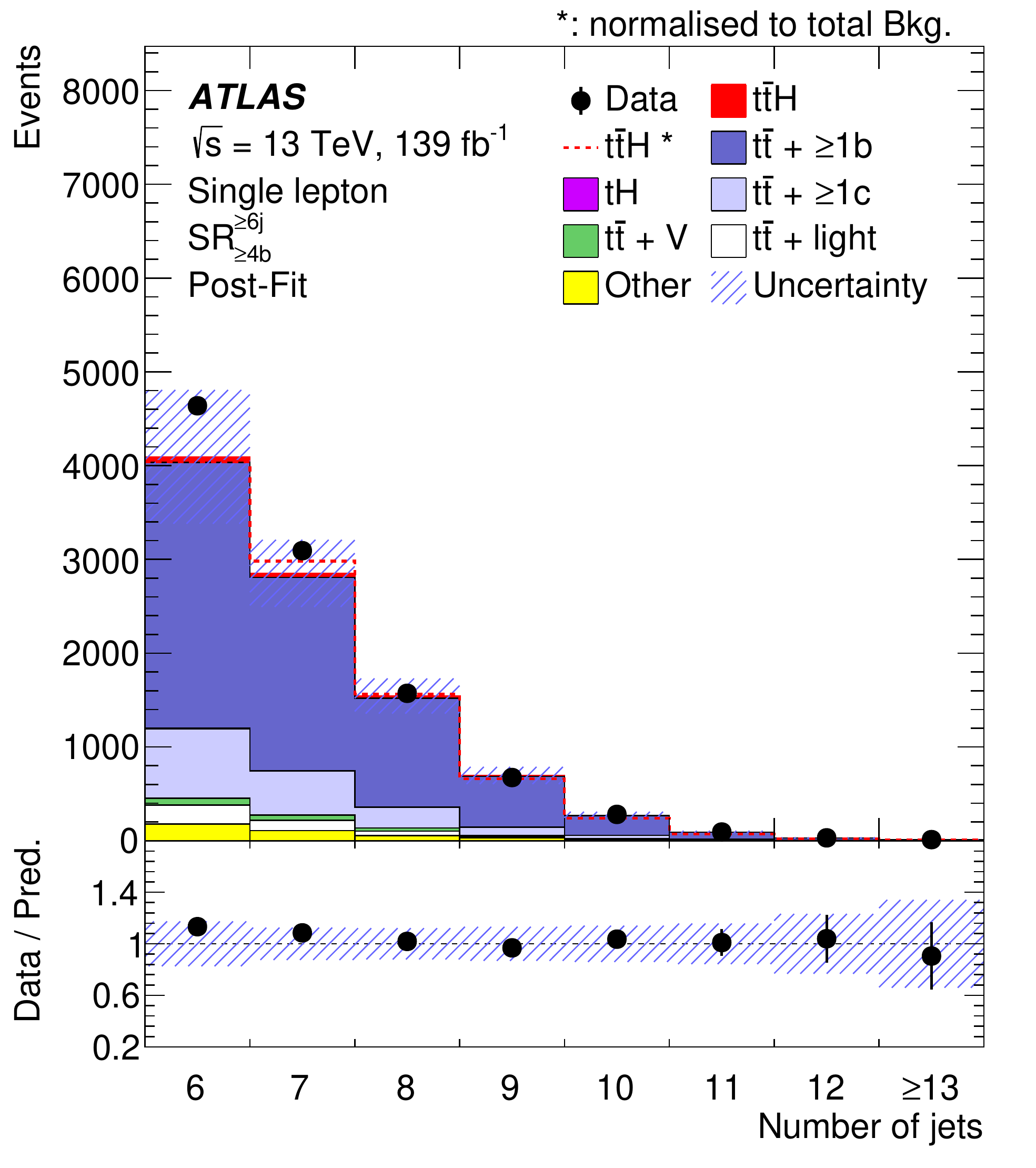}
\caption{}
\end{subfigure}
\begin{subfigure}[]{0.32\linewidth}
\vspace{0pt}
\includegraphics[width=\linewidth]{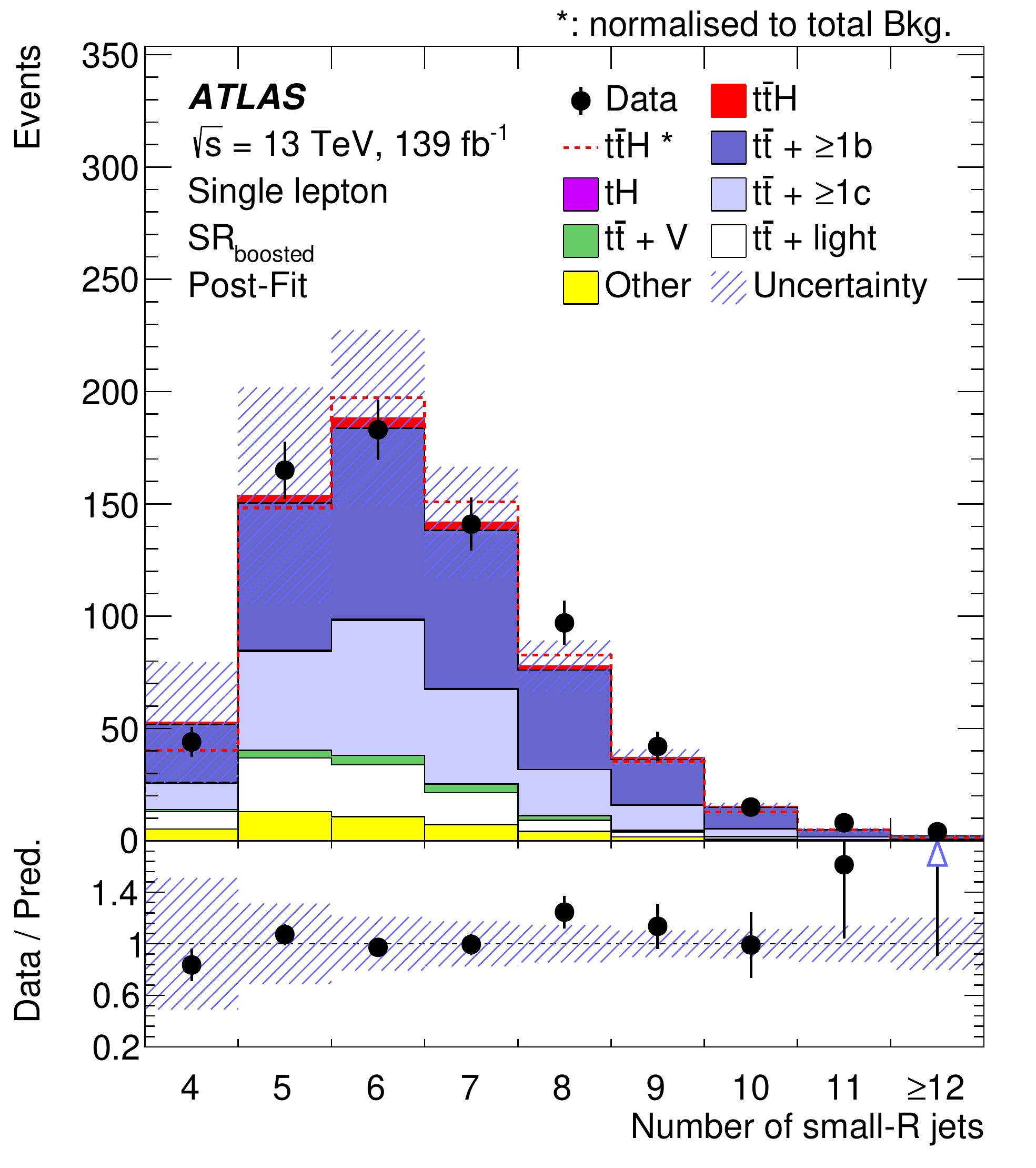}
\caption{}
\end{subfigure}
\caption{Pre-fit (top) and post-fit (bottom) distributions of the number of jets in the (a, d) dilepton \SR{4}{4}, (b, e) single-lepton resolved \SR{6}{4} and (c, f) single-lepton boosted \SRboosted signal regions. The \ttH\ signal yield
(solid red) is normalised to  the Standard Model expectation (the fitted $\mu$ value from the inclusive fit) in the pre-fit (post-fit) distributions. The dashed line shows the \ttH\ signal distribution normalised to the total background prediction. The uncertainty band includes all uncertainties and their correlations, except in the pre-fit distributions where the uncertainty in the $k(\ttbin)$ normalisation factor is not defined.}
\label{fig:results:njet}
\end{figure}

\begin{figure}[tp]
\centering
\begin{subfigure}[]{0.32\linewidth}
\vspace{0pt}
\includegraphics[width=\linewidth]{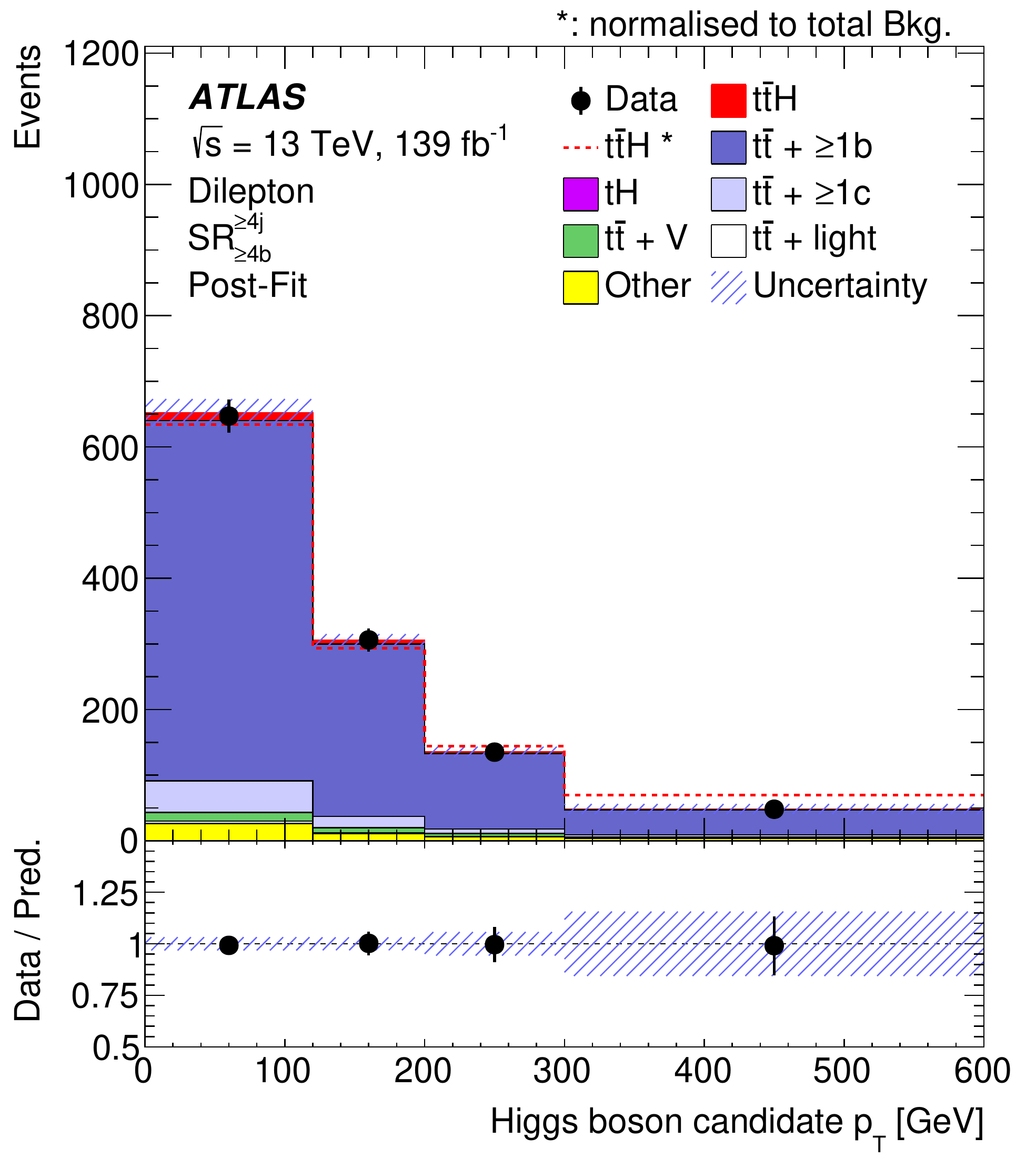}
\caption{}
\end{subfigure}
\begin{subfigure}[]{0.32\linewidth}
\vspace{0pt}
\includegraphics[width=\linewidth]{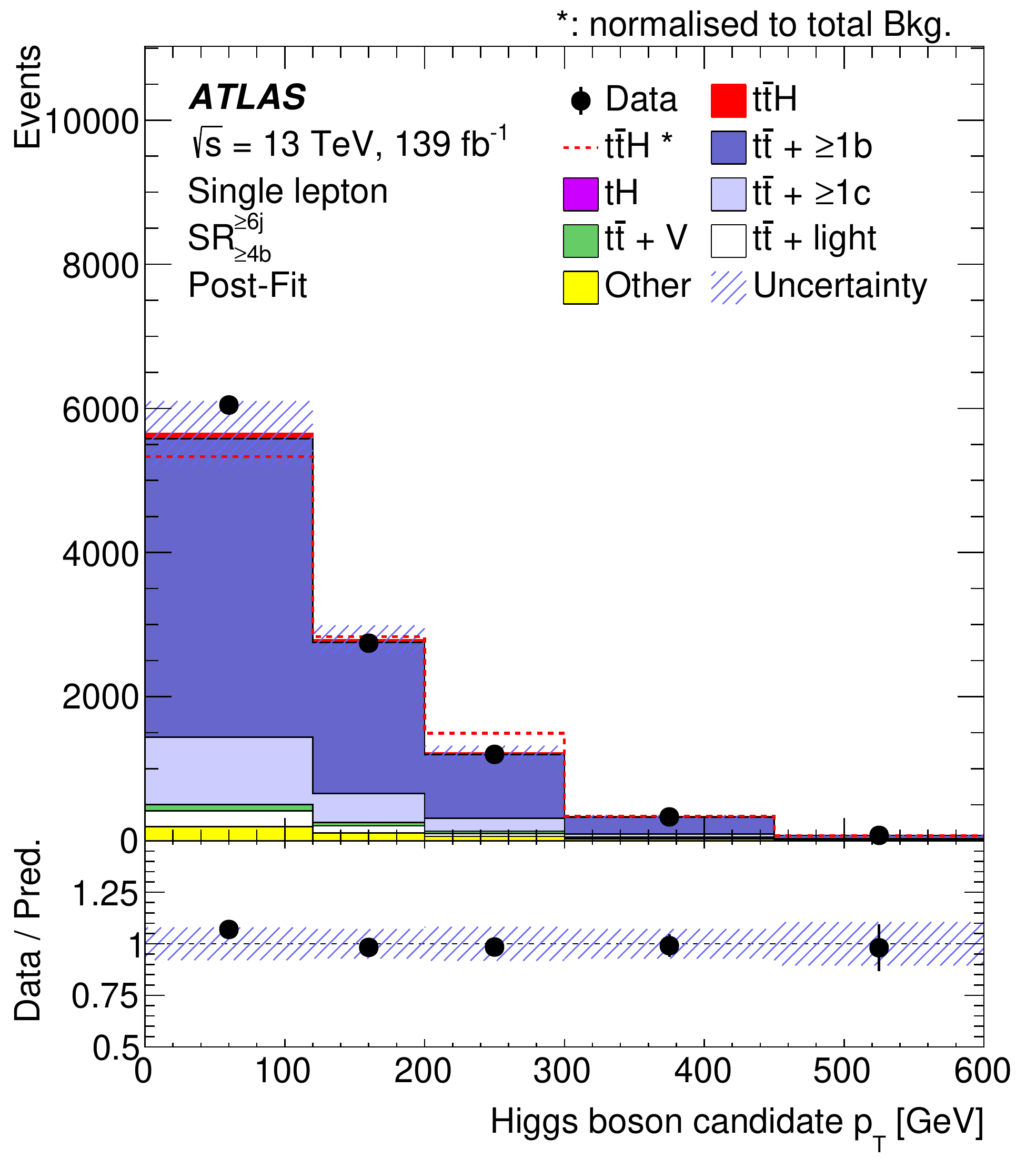}
\caption{}
\end{subfigure}
\begin{subfigure}[]{0.32\linewidth}
\vspace{0pt}
\includegraphics[width=\linewidth]{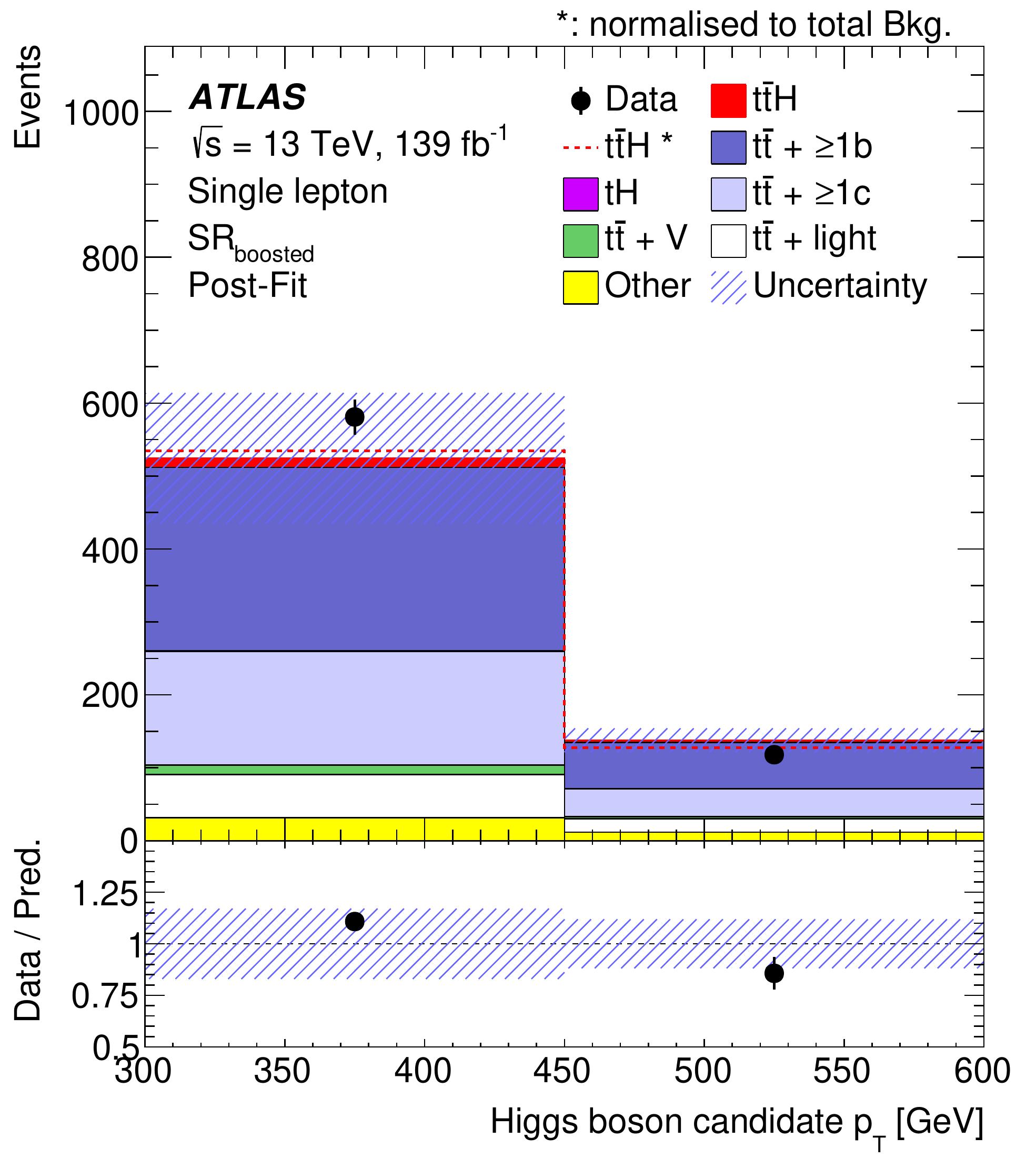}
\caption{}
\end{subfigure}
\caption{Post-fit distributions of the reconstructed Higgs boson candidate \ptH for the (a) dilepton \SR{4}{4}, (b) single-lepton resolved \SR{6}{4} and (c) single-lepton boosted \SRboosted signal regions. The \ttH\ signal yield
(solid red) is normalised to the fitted $\mu$ value from the inclusive fit. The dashed line shows the \ttH\ signal distribution normalised to the total background prediction. The uncertainty band includes all uncertainties and their correlations. The last bin includes the overflow.}
\label{fig:results:hpt}
\end{figure}

The fit is also performed with multiple signal strengths corresponding to the five `truth' \ptHhat bins. 
Figure~\ref{fig:results:muSTXS} shows the values obtained. The normalisation factor is found to be $k(\ttbin) = 1.28 \pm 0.08$, in perfect agreement with the single $\mu$ fit value. The global goodness of fit is 88\%, summarising the
good post-fit modelling obtained. The probability that the obtained signal strengths are compatible with the SM predictions is 45\%, estimated by redoing the fit while fixing the $\mu$ value to 1 in the five bins. Overall, pulls and constraints similar to those in the inclusive measurement are observed. The measurement is dominated by systematic uncertainties in the lowest bin of `truth' \ptHhat (mostly from the \ttbin background modelling), and by statistical uncertainties in the upper three bins.
 
\begin{figure}[tp]
\centering
\includegraphics[width=0.6\textwidth]{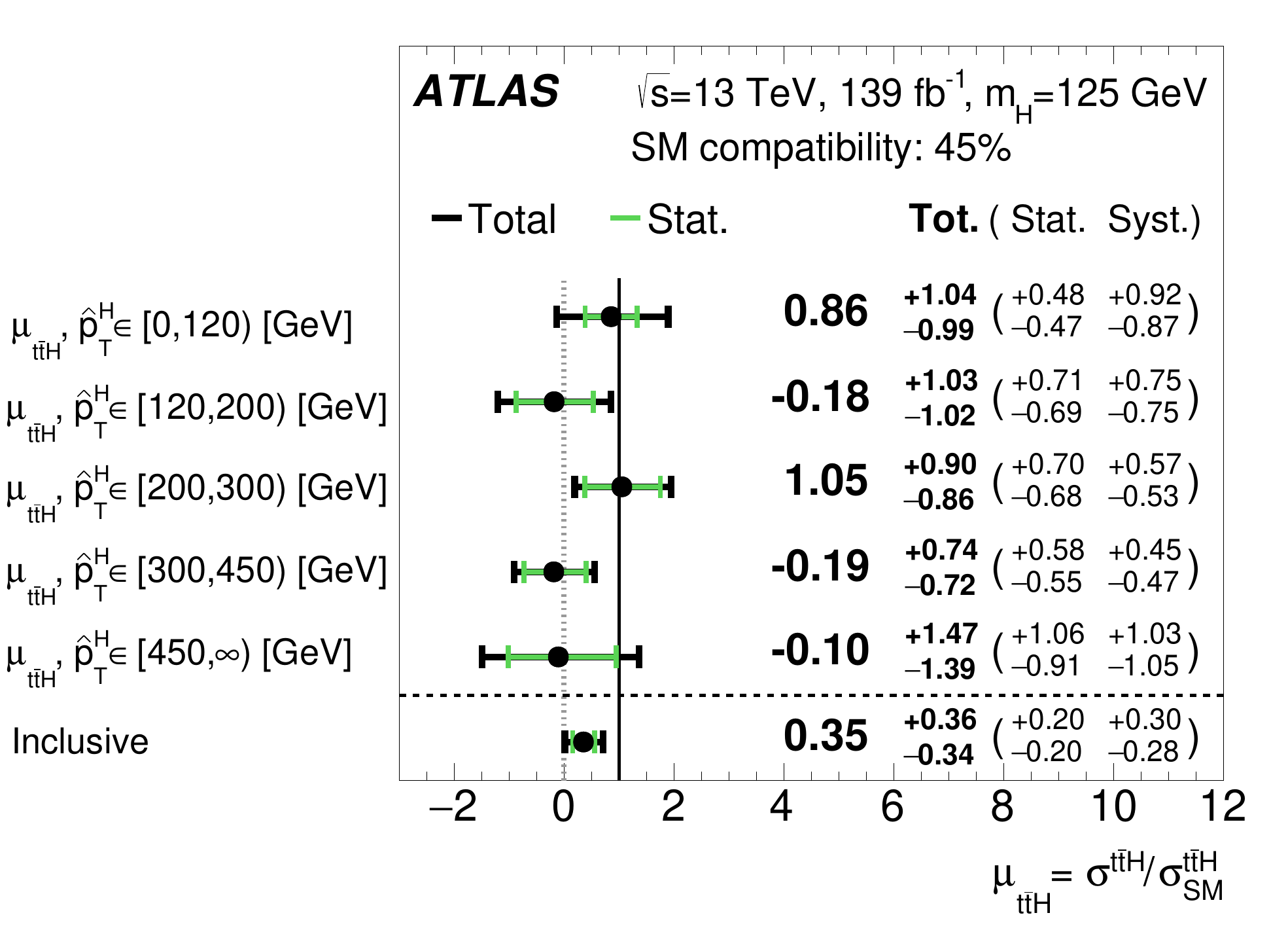}
\caption{Signal-strength measurements in the individual STXS \ptHhat bins, as well as the inclusive signal strength.}
\label{fig:results:muSTXS}
\end{figure}
 
Cross-section upper limits are also derived in the STXS framework. In this case, the
likelihood function is slightly different from the one used to
extract signal strengths: the effects of signal scale and PDF uncertainties on the predicted cross-section are not
included because, while affecting the signal-strength measurements, they do not affect the cross-section measurements.
Scale effects are still present in the statistical model though, via the ISR uncertainty, but with no impact on the overall cross-section.
The inclusive cross-section of 507\,fb is used to calculate these limits, scaled by the fraction of events in each \ptHhat\ bin to establish the fiducial cross-section for each STXS bin.
The measured 95\%
confidence level (CL) cross-section upper limits in each STXS bin are
shown in Figure~\ref{fig:results:sigmaLimitSTXS}, where the
hatched uncertainty bands correspond to the theoretical uncertainty in the
fiducial cross-section prediction in each bin.
 
\begin{figure}[tp]
\centering
\includegraphics[width=0.6\textwidth]{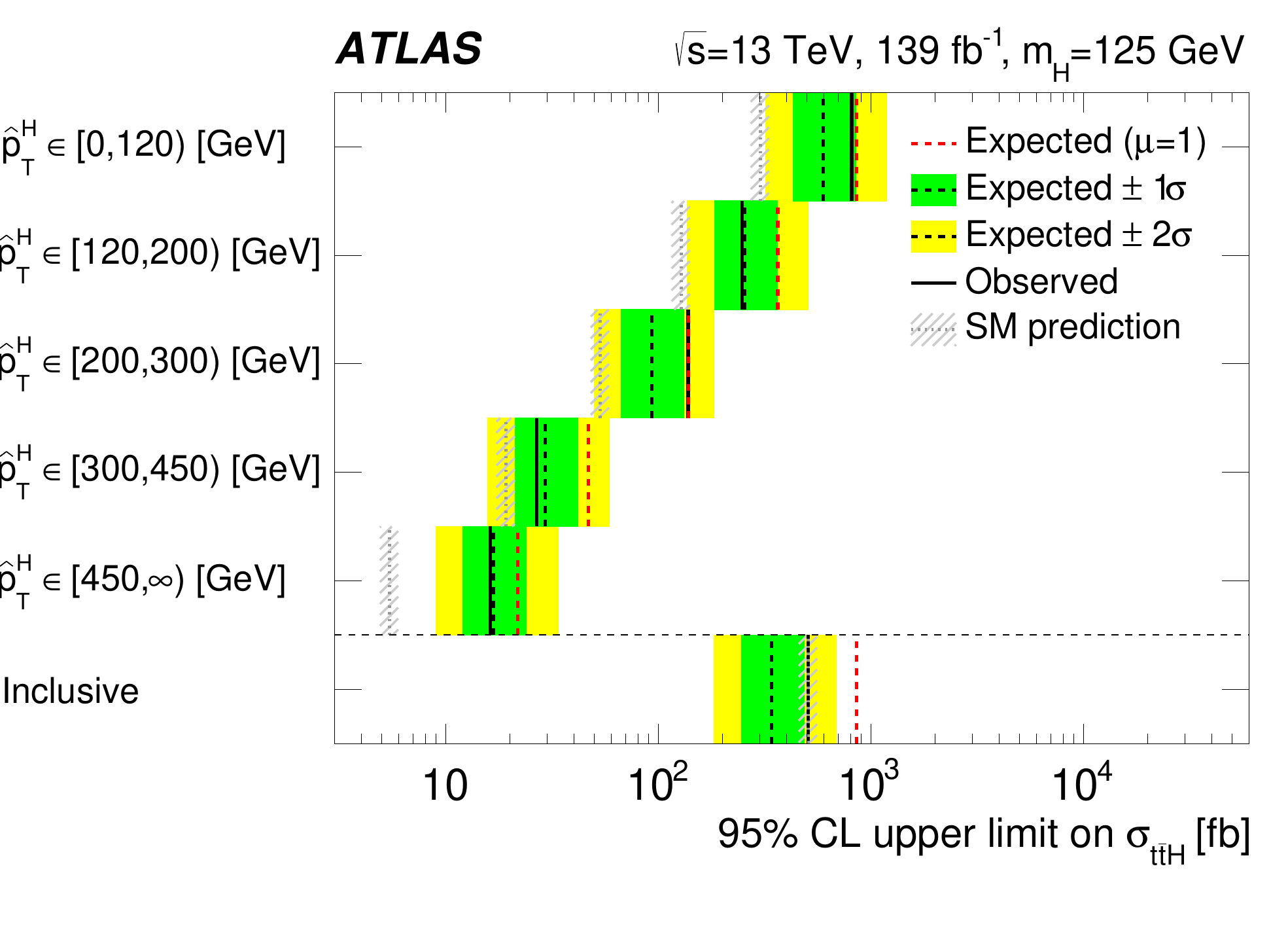}
\caption{95\% CL simplified template
cross-section upper limits in the individual STXS \ptHhat bins, as
well as the inclusive limit. The observed limits are shown (solid
black lines), together with the expected limits both in the
background-only hypothesis (dotted black lines) and in the SM
hypothesis (dotted red lines). In the case of the expected limits
in the background-only hypothesis, one- and two-standard-deviation
uncertainty bands are also shown. The hatched uncertainty bands
correspond to the theory uncertainty in the fiducial
cross-section prediction in each bin.}
\label{fig:results:sigmaLimitSTXS}
\end{figure}
 

\FloatBarrier

\section{Conclusion}
\label{sec:Conclusion}
Measurements of a Standard Model Higgs boson produced in association with a pair of top quarks
and decaying into a pair of $b$-quarks are performed. The results are based on the Run~2 dataset of $pp$
collision data collected at $\sqs=13$~\TeV\ by the ATLAS detector at
the LHC, corresponding to an integrated luminosity of 139~\ifb. The event selection targets top-quark pair decays to a final state containing one or two leptons. Machine-learning techniques are used to discriminate between signal and background events, the latter being dominated by $t\bar{t}$ + jets production.
The measured signal strength is $0.35^{+0.36}_{-0.34}$, corresponding to an observed (expected) significance of 1.0 (2.7) standard deviations.
The probability of the obtained signal strength being compatible with the SM prediction is 8.5\%.
The measurement uncertainty is dominated by systematic uncertainties, despite significant improvement relative to the previous measurement~\cite{HIGG-2017-03}, especially regarding theoretical knowledge of the \ttbin\ background process, which still drives the sensitivity.
To further test the Standard Model, the first differential measurement of the \ttH\ signal strength was performed in five bins of Higgs boson transverse momentum in the STXS framework, including a bin for specially selected boosted Higgs bosons with transverse momentum above 300\,\GeV. Upper limits on the STXS cross-sections are also derived.
Observed results are compatible with Standard Model expectations within uncertainties.


\section*{Acknowledgements}
 

We thank CERN for the very successful operation of the LHC, as well as the
support staff from our institutions without whom ATLAS could not be
operated efficiently.
 
We acknowledge the support of
ANPCyT, Argentina;
YerPhI, Armenia;
ARC, Australia;
BMWFW and FWF, Austria;
ANAS, Azerbaijan;
SSTC, Belarus;
CNPq and FAPESP, Brazil;
NSERC, NRC and CFI, Canada;
CERN;
ANID, Chile;
CAS, MOST and NSFC, China;
Minciencias, Colombia;
MEYS CR, Czech Republic;
DNRF and DNSRC, Denmark;
IN2P3-CNRS and CEA-DRF/IRFU, France;
SRNSFG, Georgia;
BMBF, HGF and MPG, Germany;
GSRI, Greece;
RGC and Hong Kong SAR, China;
ISF and Benoziyo Center, Israel;
INFN, Italy;
MEXT and JSPS, Japan;
CNRST, Morocco;
NWO, Netherlands;
RCN, Norway;
MEiN, Poland;
FCT, Portugal;
MNE/IFA, Romania;
JINR;
MES of Russia and NRC KI, Russian Federation;
MESTD, Serbia;
MSSR, Slovakia;
ARRS and MIZ\v{S}, Slovenia;
DSI/NRF, South Africa;
MICINN, Spain;
SRC and Wallenberg Foundation, Sweden;
SERI, SNSF and Cantons of Bern and Geneva, Switzerland;
MOST, Taiwan;
TAEK, Turkey;
STFC, United Kingdom;
DOE and NSF, United States of America.
In addition, individual groups and members have received support from
BCKDF, CANARIE, Compute Canada and CRC, Canada;
COST, ERC, ERDF, Horizon 2020 and Marie Sk{\l}odowska-Curie Actions, European Union;
Investissements d'Avenir Labex, Investissements d'Avenir Idex and ANR, France;
DFG and AvH Foundation, Germany;
Herakleitos, Thales and Aristeia programmes co-financed by EU-ESF and the Greek NSRF, Greece;
BSF-NSF and GIF, Israel;
Norwegian Financial Mechanism 2014-2021, Norway;
NCN and NAWA, Poland;
La Caixa Banking Foundation, CERCA Programme Generalitat de Catalunya and PROMETEO and GenT Programmes Generalitat Valenciana, Spain;
G\"{o}ran Gustafssons Stiftelse, Sweden;
The Royal Society and Leverhulme Trust, United Kingdom.
 
The crucial computing support from all WLCG partners is acknowledged gratefully, in particular from CERN, the ATLAS Tier-1 facilities at TRIUMF (Canada), NDGF (Denmark, Norway, Sweden), CC-IN2P3 (France), KIT/GridKA (Germany), INFN-CNAF (Italy), NL-T1 (Netherlands), PIC (Spain), ASGC (Taiwan), RAL (UK) and BNL (USA), the Tier-2 facilities worldwide and large non-WLCG resource providers. Major contributors of computing resources are listed in Ref.~\cite{ATL-SOFT-PUB-2021-003}.
 

\appendix
 
\part*{Appendix}
\addcontentsline{toc}{part}{Appendix}
\section{Input variables to the classification BDTs}
\label{sec:Appendix}
In this appendix, the full list of variables used as inputs to the classification BDT, described in Section~\ref{sec:Regions:mva}, in each of the signal regions is reported.
Variables are listed separately in Table~\ref{tab:dilepClassifBDTInputs} for the dilepton channel, in Table~\ref{tab:ljetsClassifBDTInputs} for the resolved single-lepton channel and in Table~\ref{tab:boostedClassifBDTInputs} for the single-lepton boosted channel.
Variables from the reconstruction BDT exploit the chosen jet--parton assignment (see Section~\ref{sec:Regions:mva}).
The $b$-tagging discriminant assigned to each jet and the large-$R$ jets used to build the Higgs-boson and top-quark candidates in the boosted category are defined in Section~\ref{sec:Selection}.
Some kinematic and topological variables are built considering only $b$-tagged jets in the event. In the resolved and boosted single-lepton channels, $b$-tagged jets are defined as the four jets with the largest value of the $b$-tagging discriminant. If two jets have the same $b$-tagging discriminant value, they are ordered by decreasing jet \pt{} value.
In the dilepton channel, jets are required to pass the 70\% working point to be $b$-tagged.
 
\begin{table}[tp]
\centering
\renewcommand{\arraystretch}{1.2}
\begin{tabular}{lll}
\toprule
\midrule
Variable & Definition  \\
\midrule
\midrule
\multicolumn{3}{l}{General kinematic variables} \\
\midrule
\mbbminM           & \begin{minipage}[l]{9.5cm} Minimum invariant mass of a $b$-tagged jet pair                                            \end{minipage} \\[1.2ex]
\mbbmindr          & \begin{minipage}[l]{9.5cm} Invariant mass of the $b$-tagged jet pair with minimum $\Delta R$                          \end{minipage} \\[1.2ex]
\mjjmaxpt          & \begin{minipage}[l]{9.5cm} Invariant mass of the jet pair with maximum \pt                                            \end{minipage} \\[1.2ex]
\mbbmaxpt          & \begin{minipage}[l]{9.5cm} Invariant mass of the $b$-tagged jet pair with maximum \pt                                 \end{minipage} \\[1.2ex]
\detabbav          & \begin{minipage}[l]{9.5cm} Average $\Delta\eta$ for all $b$-tagged jet pairs                                          \end{minipage} \\[1.2ex]
\nhiggsthirtybb    & \begin{minipage}[l]{9.5cm} Number of $b$-tagged jet pairs with invariant mass within 30~\GeV\ of the Higgs boson mass \end{minipage} \\
\midrule
\midrule
\multicolumn{3}{l}{Variables from reconstruction BDT} \\
\midrule
BDT outputs                          & \begin{minipage}[l]{9.5cm}Output of the reco. BDT w/ Higgs info. for the combination selected by the reco. BDTs w/ or w/o Higgs info. \end{minipage} & $\ddagger$ \\[2ex]
$m_{bb}^{\text{Higgs}}$            & \begin{minipage}[l]{9.5cm}Higgs candidate mass                                                  \end{minipage} \\[1.2ex]
$\Delta R_{H\!,\ttbar}$                & \begin{minipage}[l]{9.5cm}$\Delta R$ between Higgs candidate and \ttbar\ candidate system       \end{minipage} & $\dagger$ \\[1.2ex]
$\Delta R_{H\!,\ell}^{\text{min}}$   & \begin{minipage}[l]{9.5cm}Minimum $\Delta R$ between Higgs candidate and lepton                 \end{minipage} \\[1.2ex]
$\Delta R_{H\!,b}^{\text{min}}$      & \begin{minipage}[l]{9.5cm}Minimum $\Delta R$ between Higgs candidate and $b$-jet from top       \end{minipage} \\
\midrule
\bottomrule
\end{tabular}
\caption{Variables used in the classification BDTs in the dilepton channel.
For variables depending on $b$-tagged jets, only jets $b$-tagged using the 70\% working point are considered.
For variables from the reconstruction BDT, those with a $\dagger$ are from the BDT using Higgs boson information,
those with no sign are from the BDT without Higgs boson information
while for those with a $\ddagger$ both versions are used.
}
\label{tab:dilepClassifBDTInputs}
\end{table}
 
\begin{table}[tp]
\centering
\vspace{-0.25cm}
\renewcommand{\arraystretch}{1.2}
\begin{tabular}{lll}
\toprule
\midrule
Variable & Definition  \\
\midrule
\midrule
\multicolumn{3}{l}{General kinematic variables} \\
\midrule
\drbbav          & \begin{minipage}[l]{10.3cm} Average $\Delta R$ for all $b$-tagged jet pairs                                                                             \end{minipage} \\[1ex]
\drbbmaxpt       & \begin{minipage}[l]{10.3cm} $\Delta R$ between the two $b$-tagged jets with the largest vector sum $\pt$                                                \end{minipage} \\[1ex]
\maxdeta         & \begin{minipage}[l]{10.3cm} Maximum $\Delta\eta$ between any two jets                                                                                   \end{minipage} \\[1ex]
\mbbmindr        & \begin{minipage}[l]{10.3cm} Mass of the combination of two $b$-tagged jets with the smallest $\Delta R$                                                 \end{minipage} \\[2ex]
\nhiggsthirtybb  & \begin{minipage}[l]{10.3cm} Number of $b$-tagged jet pairs with invariant mass within 30~\GeV\ of the Higgs boson mass                                  \end{minipage} \\[2ex]
\apla            & \begin{minipage}[l]{10.3cm} $1.5 \lambda_2$, where $\lambda_2$ is the second eigenvalue of the momentum tensor [\href{https://dx.doi.org/10.1103/PhysRevD.48.R3953}{Phys. Rev. D \textbf{48} (1993) R3953}] built with all jets \end{minipage} \\[2ex]
$H_1$            & \begin{minipage}[l]{10.3cm} Second Fox--Wolfram moment computed using all jets and the lepton                                                           \end{minipage} \\
\midrule
\midrule
\multicolumn{3}{l}{Variables from reconstruction BDT} \\
\midrule
BDT output \rule{0pt}{3ex}           & \begin{minipage}[l]{10.3cm} Output of the reconstruction BDT                                              \end{minipage} & $\dagger$ \\[1ex]
$m_{bb}^{\text{Higgs}}$            & \begin{minipage}[l]{10.3cm} Higgs candidate mass                                                          \end{minipage}     \\[1ex]
$m_{{H}\!,b_{\text{lep top}}}$       & \begin{minipage}[l]{10.3cm} Mass of Higgs candidate and $b$-jet from leptonic top candidate               \end{minipage}     \\[1ex]
$\Delta R^{\text{Higgs}}_{bb}$     & \begin{minipage}[l]{10.3cm} $\Delta R$ between $b$-jets from the Higgs candidate                          \end{minipage}     \\[1ex]
$\Delta R_{H\!,\ttbar}$                & \begin{minipage}[l]{10.3cm} $\Delta R$ between Higgs candidate and \ttbar{} candidate system              \end{minipage} & $\dagger$ \\[1ex]
$\Delta R_{H\!,\text{lep top}}$      & \begin{minipage}[l]{10.3cm} $\Delta R$ between Higgs candidate and leptonic top candidate                 \end{minipage}     \\
\midrule
\midrule
\multicolumn{3}{l}{Variables from likelihood calculations} \\
\midrule
LHD  & \begin{minipage}[l]{10.3cm}Likelihood discriminant     \end{minipage}\\[1ex]
\midrule
\midrule
\multicolumn{3}{l}{Variables from $b$-tagging}\\[0.1ex]
\midrule
$w_{b\textrm{-tag}}^{\text{Higgs}}$ & \begin{minipage}[l]{10.3cm}Sum of $b$-tagging discriminants of jets from best Higgs candidate from the reconstruction BDT  \end{minipage} \\[2ex]
$B_{\text{jet}}^{3}$                & \begin{minipage}[l]{10.3cm}3$^\text{rd}$ largest jet $b$-tagging discriminant                                            \end{minipage} \\[1ex]
$B_{\text{jet}}^{4}$                & \begin{minipage}[l]{10.3cm}4$^\text{th}$ largest jet $b$-tagging discriminant                                            \end{minipage} \\[1ex]
$B_{\text{jet}}^{5}$                & \begin{minipage}[l]{10.3cm}5$^\text{th}$ largest jet $b$-tagging discriminant                                            \end{minipage} \\
\midrule
\bottomrule
\end{tabular}
\caption{Input variables to the classification BDTs in the single-lepton resolved channel.
For variables depending on $b$-tagged jets, jets are sorted by their pseudo-continuous $b$-tag score, and by their \pt\ when they have the same pseudo-continuous $b$-tag score.
For variables from the reconstruction BDT, those with a $\dagger$ are from the BDT using Higgs boson information,
those with no $\dagger$ are from the BDT without Higgs boson information.
}
\label{tab:ljetsClassifBDTInputs}
\end{table}
 
\begin{table}[tp]
\centering
\renewcommand{\arraystretch}{1.2}
\begin{tabular}{ll}
\toprule
\midrule
Variable & Definition \\
\midrule
$m_{bb}^{\text{Higgs}}$ & Higgs candidate mass\\[1ex]
\ptH & Higgs candidate transverse momentum\\[1ex]
$\eta^{\text{Higgs}}_{\text{lep}}$ & $\eta$ of the Higgs candidate relative to the lepton\\[1ex]
$P(H)$ & DNN Higgs probability for the Higgs candidate\\[1ex]
$m_{\text{had top}}$ & Hadronic top candidate mass\\[1ex]
$\pt^{\text{had top}}$ & Hadronic top candidate transverse momentum\\[1ex]
$\eta_{\text{had top}}^{\text{lep}}$ & $\eta$ of the hadronic top candidate relative to the lepton\\[1ex]
$B_{\text{had top}}^{i}$ & $i^\text{th}$ largest jet $b$-tagging discriminant associated to the hadronic top candidate\\[1ex]
$m_{\text{lep top}}$ & Leptonic top candidate mass\\[1ex]
$\pt^{\text{lep top}}$ & Leptonic top candidate transverse momentum\\[1ex]
$B_{\text{lep top}}$ & $b$-tagging discriminant of the jet associated to the leptonic top candidate\\[1ex]
$n_{\text{jets}}$ & Small-$R$ jets multiplicity\\[1ex]
$\Delta R_{H\!,\text{had top}}$ & $\Delta R$ between the Higgs and the hadronic top candidates\\[1ex]
$\Delta R_{H\!,\text{lep top}}$ & $\Delta R$ between the Higgs and the leptonic top candidates\\[1ex]
$\Delta R_\text{had top,lep top}$ & $\Delta R$ between the hadronic top and the leptonic top candidates\\[1ex]
$\pt^{\ttH}$ & \ttH system transverse momentum\\[1ex]
$\pt^{\ttbar}$ & \ttbar system transverse momentum \\[2ex]
$w_{b\text{-tag}}^{\text{sum}}$ & \begin{minipage}[l]{12.4cm}Sum of $b$-tagging discriminants of jets from Higgs, hadronic and leptonic top candidates\end{minipage}\\[3ex]
$w_{b\text{-tag}}^{\text{add jet}}$ & \begin{minipage}[l]{12.4cm}Fraction of the sum of $b$-tagging discriminants of all jets not associated to Higgs or hadronic top candidates\end{minipage}\\
\midrule
\bottomrule
\end{tabular}
\caption{Input variables to the classification BDT in the single-lepton boosted channel.
For variables depending on $b$-tagged jets, jets are sorted by their pseudo-continuous $b$-tag score, and by their \pt\ when they have the same pseudo-continuous $b$-tag score.
The $i$ index runs from zero to two.
}
\label{tab:boostedClassifBDTInputs}
\end{table}

\FloatBarrier

\printbibliography
 
\clearpage
 
\begin{flushleft}
\hypersetup{urlcolor=black}
{\Large The ATLAS Collaboration}

\bigskip

\AtlasOrcid[0000-0002-6665-4934]{G.~Aad}$^\textrm{\scriptsize 98}$,    
\AtlasOrcid[0000-0002-5888-2734]{B.~Abbott}$^\textrm{\scriptsize 124}$,    
\AtlasOrcid[0000-0002-7248-3203]{D.C.~Abbott}$^\textrm{\scriptsize 99}$,    
\AtlasOrcid[0000-0002-2788-3822]{A.~Abed~Abud}$^\textrm{\scriptsize 34}$,    
\AtlasOrcid[0000-0002-1002-1652]{K.~Abeling}$^\textrm{\scriptsize 51}$,    
\AtlasOrcid[0000-0002-2987-4006]{D.K.~Abhayasinghe}$^\textrm{\scriptsize 91}$,    
\AtlasOrcid[0000-0002-8496-9294]{S.H.~Abidi}$^\textrm{\scriptsize 27}$,    
\AtlasOrcid[0000-0002-9987-2292]{A.~Aboulhorma}$^\textrm{\scriptsize 33e}$,    
\AtlasOrcid[0000-0001-5329-6640]{H.~Abramowicz}$^\textrm{\scriptsize 157}$,    
\AtlasOrcid[0000-0002-1599-2896]{H.~Abreu}$^\textrm{\scriptsize 156}$
% ...
\bigskip
\\

% ...
$^{27}$Physics Department, Brookhaven National Laboratory, Upton NY; United States of America.\\
% ...
$^{33}$$^{(a)}$Facult\'e des Sciences Ain Chock, R\'eseau Universitaire de Physique des Hautes Energies - Universit\'e Hassan II, Casablanca;$^{(b)}$Facult\'{e} des Sciences, Universit\'{e} Ibn-Tofail, K\'{e}nitra;$^{(c)}$Facult\'e des Sciences Semlalia, Universit\'e Cadi Ayyad, LPHEA-Marrakech;$^{(d)}$LPMR, Facult\'e des Sciences, Universit\'e Mohamed Premier, Oujda;$^{(e)}$Facult\'e des sciences, Universit\'e Mohammed V, Rabat;$^{(f)}$Mohammed VI Polytechnic University, Ben Guerir; Morocco.\\
$^{34}$CERN, Geneva; Switzerland.\\
% ...
$^{51}$II. Physikalisches Institut, Georg-August-Universit\"{a}t G\"ottingen, G\"ottingen; Germany.\\
% ...
$^{91}$Department of Physics, Royal Holloway University of London, Egham; United Kingdom.\\
% ...
$^{98}$CPPM, Aix-Marseille Universit\'e, CNRS/IN2P3, Marseille; France.\\
$^{99}$Department of Physics, University of Massachusetts, Amherst MA; United States of America.\\
% ...
$^{124}$Homer L. Dodge Department of Physics and Astronomy, University of Oklahoma, Norman OK; United States of America.\\
% ...
$^{156}$Department of Physics, Technion, Israel Institute of Technology, Haifa; Israel.\\
$^{157}$Raymond and Beverly Sackler School of Physics and Astronomy, Tel Aviv University, Tel Aviv; Israel.\\
% ...
$^{a}$ Also at Borough of Manhattan Community College, City University of New York, New York NY; United States of America.\\
$^{b}$ Also at Bruno Kessler Foundation, Trento; Italy.\\
$^{c}$ Also at Center for High Energy Physics, Peking University; China.\\
$^{d}$ Also at Centro Studi e Ricerche Enrico Fermi; Italy.\\
$^{e}$ Also at CERN, Geneva; Switzerland.\\
$^{f}$ Also at D\'epartement de Physique Nucl\'eaire et Corpusculaire, Universit\'e de Gen\`eve, Gen\`eve; Switzerland.\\
% ...

\end{flushleft}


\end{document}